%% file: main.tex
\def\isarxiv{1} 

\ifdefined\isarxiv
\documentclass[11pt]{article}

\usepackage[numbers]{natbib}

\usepackage{subfig}

\usepackage{graphicx}

\else

\documentclass{article} 
\usepackage{microtype} 
\usepackage{graphicx}
\usepackage{subfig}
\usepackage{hyperref}
\usepackage{icml2024}

\fi

\usepackage{amsmath}
\usepackage{amsthm}
\usepackage{amssymb}
\usepackage{algorithm}
\usepackage{algpseudocode}
\usepackage{grffile}
\usepackage{wrapfig,epsfig}
\usepackage{url}
\usepackage{xcolor}
\usepackage{epstopdf}

\usepackage{bbm}
\usepackage{dsfont}

\allowdisplaybreaks

\ifdefined\isarxiv

\usepackage{tikz}
\usepackage{hyperref}  
\hypersetup{colorlinks=true,citecolor=blue,linkcolor=blue} 
\usetikzlibrary{arrows}
\usepackage[margin=1in]{geometry}

\else

\usepackage{microtype}
\usepackage{hyperref}
\definecolor{mydarkblue}{rgb}{0,0.08,0.45}
\hypersetup{colorlinks=true, citecolor=mydarkblue,linkcolor=mydarkblue}

\fi
\graphicspath{{./figs/}}

\theoremstyle{plain}
\newtheorem{theorem}{Theorem}[section]
\newtheorem{lemma}[theorem]{Lemma}
\newtheorem{definition}[theorem]{Definition}

\newcommand{\wh}{\widehat}
\newcommand{\wt}{\widetilde}

\newcommand{\R}{\mathbb{R}}

\renewcommand{\tilde}{\wt}
\renewcommand{\hat}{\wh}

\DeclareMathOperator*{\E}{{\mathbb{E}}}

\DeclareMathOperator{\OPT}{OPT}

\DeclareMathOperator{\poly}{poly}

\DeclareMathOperator{\cost}{cost}

\makeatletter
\newcommand*{\RN}[1]{\expandafter\@slowromancap\romannumeral #1@}
\makeatother

\usepackage{lineno}

\ifdefined\isarxiv

\else

\usepackage[textsize=tiny]{todonotes}

\icmltitlerunning{A Faster $k$-means++ Algorithm}

\fi

\begin{document}

\ifdefined\isarxiv


\title{A Faster $k$-means++ Algorithm}

\author{
Jiehao Liang\thanks{\texttt{liangjiehao127@gmail.com}. University of California, Berkeley.}
\and 
Somdeb Sarkhel\thanks{\texttt{sarkhel@adobe.com}. Adobe Research.}
\and 
Zhao Song\thanks{\texttt{zsong@adobe.com}. Adobe Research.}
\and 
Chenbo Yin\thanks{\texttt{chenboyin1@gmail.com}. University of Washington. }
\and 
Junze Yin\thanks{\texttt{junze@bu.edu}. Boston University. }
\and 
Danyang Zhuo\thanks{\texttt{danyang@cs.duke.edu}. Duke University.}
}

\else

\twocolumn[
\icmltitle{A Faster $k$-means++ Algorithm}



\icmlsetsymbol{equal}{*}

\begin{icmlauthorlist}
\icmlauthor{Firstname1 Lastname1}{equal,yyy}
\icmlauthor{Firstname2 Lastname2}{equal,yyy,comp}
\icmlauthor{Firstname3 Lastname3}{comp}
\icmlauthor{Firstname4 Lastname4}{sch}
\icmlauthor{Firstname5 Lastname5}{yyy}
\icmlauthor{Firstname6 Lastname6}{sch,yyy,comp}
\icmlauthor{Firstname7 Lastname7}{comp}
\icmlauthor{Firstname8 Lastname8}{sch}
\icmlauthor{Firstname8 Lastname8}{yyy,comp}
\end{icmlauthorlist}

\icmlaffiliation{yyy}{Department of XXX, University of YYY, Location, Country}
\icmlaffiliation{comp}{Company Name, Location, Country}
\icmlaffiliation{sch}{School of ZZZ, Institute of WWW, Location, Country}

\icmlcorrespondingauthor{Firstname1 Lastname1}{first1.last1@xxx.edu}
\icmlcorrespondingauthor{Firstname2 Lastname2}{first2.last2@www.uk}

\icmlkeywords{Machine Learning, ICML}

\vskip 0.3in
]



\printAffiliationsAndNotice{\icmlEqualContribution} 

\fi

\ifdefined\isarxiv
\date{\empty}
\begin{titlepage}
  \maketitle
  \begin{abstract}
\input{abstract}

  \end{abstract}
  \thispagestyle{empty}
\end{titlepage}

{\hypersetup{linkcolor=black}
}
\newpage

\else

\begin{abstract}
\input{abstract}
\end{abstract}

\fi

\input{intro} 
\input{related}

\input{tech}
\input{preli}

\input{data}

\input{time}

\input{exp}

\input{concl}

\ifdefined\isarxiv
\bibliographystyle{alpha}
\bibliography{ref}
\else
\bibliography{ref}
\bibliographystyle{icml2024}
\fi

\newpage
\onecolumn
\appendix
\section*{Appendix}

\input{correct}

\input{app_exp}





\end{document}

%% file: abstract.tex
$k$-means++ is an important algorithm for choosing initial cluster centers for the $k$-means clustering algorithm. In this work, we present a new algorithm that can solve the $k$-means++ problem with nearly optimal running time. Given $n$ data points in $\mathbb{R}^d$, the current state-of-the-art algorithm runs in $\widetilde{O}(k )$ iterations, and each iteration takes $\widetilde{O}(nd k)$ time. The overall running time is thus $\widetilde{O}(n d k^2)$. We propose a new algorithm \textsc{FastKmeans++} that only takes in $\widetilde{O}(nd + nk^2)$ time, in total.

%% file: intro.tex
\section{Introduction}
$k$-means clustering aims to cluster $n$ data points from $\R^d$ into $k$ clusters, such that the average distance between a data point and the center of the cluster it belongs to is minimized. $k$-means has many important real-world applications in signal processing \cite{gg12, dmc15} and unsupervised machine learning \cite{dh04, cn12}.

A central problem in $k$-means is how to pick the initial locations of the $k$ clusters to avoid $k$-means from finding poor clusters. Much work focuses on finding a ``good'' initialization for the $k$-means algorithm. The well-known heuristic Lloyd algorithm \cite{llo82} performs well in practice while there is no theoretical guarantee for its approximation at initialization. There are already constant approximation algorithms with theoretical guarantee \cite{jv01,kmn+04}. Unfortunately, $(1+\epsilon)$-approximation algorithm for arbitrary small $\epsilon>0$ does not exist if $k$ and $d$ can be large \cite{lsw17}. In recent works \cite{bv15,ca18,cakm19,frs19}, one achieved $(1 + \epsilon)$  approximation algorithm. However, the running time of these algorithms depends on $d$ double exponentially. For fixed $k$ and $\epsilon$, one may reduce input size or number of potential solutions to achieve constant approximation \cite{hpm04,as05,fms07,che09,kss10}.

$k$-means++~\cite{av06} is an algorithm that can find a center set that achieves $O(\log k)$ approximation for the optimal center set. $k$-means++ has thus been integrated into many standard machine learning libraries and is used as the default initialization method, replacing the traditional random initialization method. If allowed to sample $O(k)$ centers, \cite{adk09} shows that $k$-means++ achieves constant bi-criteria approximation with constant probability. The more accurate trade-off between the number of centers to be sampled and the expected cost can be found in \cite{wei16}. \cite{ajos19} provides a $6.357$ ratio approximation guarantee using primal-dual algorithm.

We note that applying a local search algorithm is one feasible way to achieve constant approximation for $k$-means clustering \cite{kmn+04,fs08,ca18,frs19}. However, these algorithms suffer from a potentially large number of iterations. A more efficient $k$-means++ algorithm via local search was proposed in \cite{ls19} to achieve constant approximation. It is an iterative method that requires $\tilde{O}(k)$ iterations, and each iteration has to take $\tilde{O}(nd k)$ time. The overall running time is thus $\tilde{O}(n d k^2)$. This running time is fine for small and low-dimensional data, but it is increasingly insufficient for today's explosion of dataset sizes $(n)$, data dimensionality $(d)$, and the number of clusters $(k)$.

In this paper, we propose a faster version of the $k$-means++ algorithm, called \textsc{FastKMeans++}. We show that choosing initial cluster locations can be much faster for constant approximation. We observe that the major computation cost originates from distance calculation. To accelerate the algorithm, especially for high dimensional situations, we design a distance oracle using JL lemmas \cite{jl84} to approximate distance. We can approximate the $k$-means++ algorithm in $\tilde{O}(nd + nk^2)$ time and maintain constant approximation. When $k = O(\sqrt{d})$, we obtain optimal running time. In experiments, we use the $k$-means++ algorithm \cite{ls19} and \textsc{FastKMeans++} algorithm to cluster the same synthetic point set respectively. We use $k$-means++ as a baseline to evaluate the performance of \textsc{FastKMeans++}. Our experimental results demonstrate that \textsc{FastKMeans++} is faster than $k$-means++ in practice.

\subsection{Our Results}

We state an informal version of our results as follows. 

\begin{theorem}[Informal Version of Theorem~\ref{thm:k_means_formal}]\label{thm:k_means_informal}

Given point set $P \subset \R^d$ and $Z = \tilde{O}(k)$, the running time of Algorithm~\ref{alg:k_means} is 
$\tilde{O}(n(d + k^2))$ 
which uses $O(n(d + k))$ space.

We use $C$ to denote the result of Algorithm~\ref{alg:k_means}. Let $C^*$ be the set of optimum centers. Then we have
\begin{align*}
\E[\cost(P, C)] = O(\cost(P, C^*)).
\end{align*}
\end{theorem}

Our result shows that our algorithm using a distance oracle still achieves constant approximation in expectation and runs in nearly optimal running time.

%% file: related.tex
\section{Related Work}

\paragraph{Clustering}
In \cite{hpm04}, they showed the existence of small coresets for the problems of computing $k$-means clustering for points in low dimensions. In \cite{as05}, when they got clusterings for many different values of k, they used a quality measure of clusterings that is independent of k to determine a good choice of k, like the average silhouette coefficient.  In \cite{fms07}, they use a weak ($\epsilon$, k)-coreset to obtain a PTAS for the $k$-means clustering problem with running time $O(nkd + d \cdot Poly(k/\epsilon) + 2\tilde{O}(k/\epsilon))$. In \cite{che09}, they use coresets whose size is with polynomial dependency on the dimension d to maintain a $(1+\epsilon)$-approximate $k$-means clustering of a stream of points in $\R ^d$. In \cite{kss10}, they provide simple randomized algorithms for the $k$-means that yield $(1+\epsilon)$ approximations with probability $\geq 1/2$ and running times of $O(2^{(k/\epsilon)O(1)} dn)$. In \cite{ls19}, they provide a $O(1)$-approximation algorithm and runs in $O(ndk^2)$ time. In \cite{cln+20}, they provide a $\log(k)$-approximation and runs in $(n d + d k^3)$ time. Our algorithm achieves $O(1)$-approximation and runs in $O(nd + nk^2)$.
The approximation factor of \cite{ls19} is the same as ours, but their running time is always slower than ours. 
The approximation factor of \cite{cln+20} is much worse than ours. Also when $d \approx n$, their algorithm \cite{cln+20} takes $O(nk^3)$ while ours takes only $O(nk^2)$.

\paragraph{Sketching for Iterative Algorithm}

Sketching is an effective method to increase the speed of machine learning algorithms and optimization techniques. It involves compressing the large input matrix down to a much smaller sketching matrix that retains most of the key features of the original. As the algorithm now processes this smaller representation rather than the full high-dimensional input, its runtime can be substantially reduced. The core benefit is that the key information is preserved while removing redundant data that would otherwise slow computations. By working with this concise sketch version, large performance gains are possible without significantly impacting output quality.

In this work, we apply the sketching technique to an iterative algorithm. It is well-known that the sketching technique can be applied to many fundamental problems that are solved by the iterative type of algorithm. For example,
linear programming \cite{cls19,jswz21,sy21,gs22}, empirical risk minimization \cite{lsz19,qszz23}, cutting plane method \cite{jlsw20}, computing John Ellipsoid \cite{ccly19,syyz22}, low-rank approximation \cite{syyz23_weighted}, tensor decomposition \cite{dsy23}, integral minimization problem \cite{jlsz23}, federated learning \cite{bsy23}, linear regression problem \cite{cw13,nn13,syyz23_ellinf,syz23}, matrix completion \cite{gsyz23}, training over-parameterized neural tangent kernel regression \cite{bpsw21,szz21,z22,als+22}, attention computation problem \cite{syz23_atten,gswy23,gsy23_hyper}, and matrix sensing \cite{dls23}.

%% file: tech.tex
\paragraph{Roadmap} In Section~\ref{sec:preliminary}, we introduce preliminary knowledge including notation and folklore lemmas. In Section~\ref{sec:data_structure}, we present our data structure of \textsc{DistanceOracle}. In Section~\ref{sec:main_result}, we present the main results for \textsc{FastKMeans++} algorithm. In Section~\ref{sec:running_time}, we demonstrate the running time and space storage for the algorithm. 
We provide the pseudocode of our algorithms and use a series of experiments to justify the advantages of our \textsc{FastKMeans++} algorithm in Section~\ref{sec:experiments}. In Section~\ref{sec:conclusion}, we draw our conclusion. 

%% file: preli.tex
\section{Preliminary}\label{sec:preliminary}

In Section~\ref{sec:prel_notation}, we first introduce several basic notations, such as the number of elements in a set, the distance between two nodes, and mathematical expectation. We introduce related definitions throughout the paper in Section~\ref{sec:definition}, such as the sum of the distance between each node and its closest center and the mass center of multiple points. We present useful lemmas in Section~\ref{sec:prel_folklore}, such as the method to calculate or estimate the sum of the distance between each node in a point set and a specific center or between a given node and each node in a center point set. 

\subsection{Notation}\label{sec:prel_notation}

We use $[n]$ to denote set $\{1, 2, \cdots, n\}$. Given finite set $S$, we use $|S|$ to denote the number of elements in set $S$. We use $\|\cdot\|_2$ to denote $\ell_2$ norm in Euclidean space. We use $| \cdot |$ to denote absolute value. Given random variable $X$, we use $\E[X]$ to denote its expectation. 
For any function $f$, we use $\wt{O}(f)$ to denote $f \cdot \poly(\log f)$.

\subsection{Related Definitions}\label{sec:definition}

We define the cost given data set $P$ and center set $C$.

\begin{definition}[Cost]\label{def:cost}
Given set $P \subset \R^d$ and set $C \subset \R^d$, we define  
$
    \cost(P,C):= \sum_{i=1}^n \min_{c \in C}\|p_i - c\|_2^2.
$
\end{definition}

Also, we define the mean or center of gravity for a point set.
\begin{definition}[Mean]\label{def:mean_center}
Given data set $P \subset \R^d$, we define the mean of this set
$
    \mu(P):= \frac{1}{|P|} \sum_{p \in P} p.
$
\end{definition}

\subsection{Useful Lemmas}\label{sec:prel_folklore}

The following JL Lemma provides guidance for achieving distance oracle data structure.

\begin{lemma}[JL Lemma, \cite{jl84}]\label{lem:jl_lemma}
For any $X \subset \R^d$ of size $n$, there exists an embedding $f: \R^d \mapsto \R^m$ where $m = O(\epsilon^{-2} \log n)$ such that \[ (1-\epsilon) \cdot \| x - y \|_2^2 \leq \| f(x) - f(y) \|_2^2 \leq (1+\epsilon)\cdot \| x - y \|_2^2\]
\end{lemma}


The following lemma is folklore which bridges the cost (Definition~\ref{def:cost}) and mean (Definition~\ref{def:mean_center})

\begin{lemma}\label{lem:cost_mu}
Let $P \subset \mathbb{R}^{d}$ be a point set. Let $c \in \mathbb{R}^{d}$ be a center. Let $\mu(P)$ be defined as $\mu(P) := \frac{1}{|P|} \sum_{p \in P} p$ (Definition~\ref{def:mean_center}). Then we have 
\begin{align*}
    \cost(P,\{c\})=|P| \cdot\|c-\mu(P)\|_2^{2}+\cost(P, \mu(P))
\end{align*}

\end{lemma}

We present a variation of Corollary 21 in \cite{cdm20} which gives an upper bound on the cost difference.

\begin{lemma}\label{lem:two_point_cost_error}
For arbitrary $\epsilon>0$, let $p, q$ be two different points in $\R^{d}$ and $C \subset \R^{d}$ be the center set. Then 
\begin{align*}
     & ~  |\cost(\{p\}, C)-\cost(\{q\}, C)| 
    \leq ~ \epsilon \cdot \cost(\{p\}, C) \\
    + & ~  (1+\frac{1}{\epsilon})\|p-q\|_2^{2}
\end{align*}

\end{lemma}

%% file: data.tex
\section{Data Structure}\label{sec:data_structure}

 In Section~\ref{sec:distance_oracle}, we present the data structure of \textsc{DistanceOracle}. In Section~\ref{sec:running_time_distance_oracle}, we state the running time results for \textsc{DistanceOracle}.

\subsection{Distance Oracle}\label{sec:distance_oracle}

In this section, we state the result for the Distance Oracle data structure. In \textsc{Init} procedure, we preprocess the original data set $P \subset \R^d$ using JL transform and store the transformed sketches. When we need to compute distance, we only use these sketches so that we can get rid of factor $d$. We also use index set $S \subset [n]$ to denote current centers. We maintain a balanced search tree array $v$ of size $n$ that records the closet center for each data point and supports \textsc{Insert}, \textsc{Delete}, and \textsc{GetMin} operations. Using tree array $v$, we are able to compute the current cost (Definition~\ref{def:cost}) with \textsc{Cost} operation and also the cost after adding a center with \textsc{Query} operation. 

\begin{theorem}[Distance Oracle]\label{thm:distance_oracle}
There is a data structure that uses space 
\begin{align*}
   O(n (d + k +\epsilon^{-2}\log n \log (n/\delta)) )
\end{align*}
 with the following procedures:

\begin{itemize}
    \item \textsc{Init}$(P = \{x_1, \cdots, x_n\} \subset \R^d, S \subset [n], \epsilon \in (0, 0.1), \delta \in (0, 0.1))$ Given data $P \subset \R^d$, center $S \subset [n]$, precision parameter $\epsilon \in (0, 0.1)$ and failure probability $\delta \in (0, 0.1)$, it takes $O(\epsilon^{-2}nd\log(n/\delta))$ time to initialize.
    \item \textsc{Cost}$()$ 
    It returns an approximate cost $\tilde{d}$ using $O(n)$ time such that
    \begin{align*}
        (1 - \epsilon) \cost(P,S) \leq \tilde{d} \leq (1 + \epsilon) \cost(P,S)
    \end{align*}
    \item \textsc{Insert}$(j \in [n])$ Given index $j \in [n]$, it inserts distances between $y_j$ and all sketches to $v$ using $O(\epsilon^{-2}n \log (n/\delta))$ time.
    \item \textsc{Delete}$(j \in [n])$ Given index $j \in [n]$, it deletes all distances between $y_j$ and all sketches using $O(n \log n)$ time
    \item \textsc{Sample}$()$ It samples an index $j$ using $O(n)$ time
    \item \textsc{Query}$( j \in [n])$ Given an index $j \in [n]$, it returns a value $\tilde{q}$ that satisfies
    \begin{align*}
        (1 - \epsilon) \cost(P, \{x_j\}) \leq \tilde{q} \leq (1 + \epsilon) \cost(P, \{x_j\})
    \end{align*}
    using $O(\epsilon^{-2}n \log (n/\delta))$ time
\end{itemize}
\end{theorem}
\begin{proof}

 By Lemma~\ref{lem:running_time_init}, Lemma~\ref{lem:running_time_cost}, Lemma~\ref{lem:running_time_insert}, Lemma~\ref{lem:running_time_delete}, Lemma~\ref{lem:running_time_sample} and Lemma~\ref{lem:running_time_query}, we obtain the running time for procedure \textsc{Init}, \textsc{Cost}, \textsc{Insert}, \textsc{Delete}, \textsc{Sample} and \textsc{Query}, respectively. In each of these lemmas, we calculate and prove the time complexity of each procedure.

In \textsc{DistancOracle}, it stores $n$ data points in $\R^d$, $n$ sketches in $\R^m$ and maintain an $n$-array. Each entry of the $n$-array is a balanced search tree with $O(k)$ nodes. Also, we have $m = O(\epsilon^{-2}  \log (n/\delta))$. The total space storage is 
\begin{align*}
    O(nd) + O(nm) + O(nk)
    = O(n (d + k +\epsilon^{-2} \log (n/\delta)))
\end{align*}
\end{proof}

\begin{algorithm}[!ht]\caption{Distance Oralce}\label{alg:sample_oracle}
    \begin{algorithmic}[1]
        \State {\bf data structure} \textsc{DistanceOracle} \Comment{Theorem~\ref{thm:distance_oracle}}
        \State {\bf members}
        \State \hspace{4mm} $n , m , d, k\in \mathbb{N}_+$
        \State \hspace{4mm} $P = \{ x_1, \cdots, x_n \}$ \Comment{$|P|=n, P \subset \R^d$}
        \State \hspace{4mm} $Q = \{ y_1, \cdots, y_n \}$ \Comment{$|Q|=n, Q \subset \R^m$}
        \State \hspace{4mm} $S \subset [n]$
        \State \hspace{4mm} $v$ \Comment{This is a length-$n$ array and each entry in the array is a balanced search tree where the root has the minimum value.}
        \State \hspace{4mm} $\mathrm{sum}$
        \State {\bf end members}
        
        \Procedure{Init}{$P \subset \R^d, S \subset [n], \epsilon \in (0, 0.1), \delta \in (0, 0.1)$}
        \State $P \gets P$
        \State $m \gets O( \epsilon^{-2} \log(n/\delta) )$ 

        \For{$i=1 \to n$} \label{lin:do_init_forloop}
            \State $y_i \gets \Pi x_i$ \label{lin:do_init_sketch}
        \EndFor 
        \State $S \gets S$
        \EndProcedure
        \Procedure{Cost}{$ $}
            \State $\mathrm{sum} \gets 0$
            \For{$i=1 \to n$}\label{lin:do_cost_forloop}
                \State $\mathrm{sum} \gets \mathrm{sum} + v[i].\textsc{GetMin}()$
                \label{lin:do_cost_sum_getmin}
            \EndFor 
            \State \Return sum
        \EndProcedure 
        \Procedure{Insert}{$j \in [n]$}
            \State $S \gets S \cup j$
            \For{$i=1 \to n$}\label{lin:do_insert_forloop}
                \State $v[i].\textsc{Insert}(j, \| y_i - y_j \|_2^2 )$\label{lin:do_insert_insert}
                
            \EndFor 
            \State \textsc{Cost}$()$
        \EndProcedure
        \Procedure{Delete}{$j \in [n]$}
            \State $S \gets S \setminus j$
            \For{$i=1 \to n$}\label{lin:do_delete_forloop}
                \State $v[i].\textsc{Delete}(j)$\label{lin:do_delete_delete}
            \EndFor 
             \State \textsc{Cost}$()$\label{lin:do_delete_cost}
        \EndProcedure 
        \State {\bf end data structure}
       \end{algorithmic}
\end{algorithm}     
 \begin{algorithm}[!ht]\caption{Distance Oralce}\label{alg:sample_oracle_second}
    \begin{algorithmic}[1]
        \State {\bf data structure} \textsc{DistanceOracle} \Comment{Theorem~\ref{thm:distance_oracle}}   
        \Procedure{Sample}{$ $}
            \State $\mathrm{sum} \gets 0$
            \For{$i = 1 \to n$}\label{lin:do_sample_forloop}
                \State $u_i \gets v[i].\textsc{GetMin}()$\label{lin:do_sample_getmin}
                \State $\mathrm{sum} \gets \mathrm{sum} + u_i$\label{lin:do_sample_sum}
            \EndFor
            \State Sample an index $j$ via probability equal to $\frac{u_j}{\mathrm{sum}}$\label{lin:do_sample_sample}
            \State \Return $j$
        \EndProcedure
        \Procedure{Query}{$j $}
            \State $\mathrm{tmp} \gets 0$
            \For{$i=1 \to n$}\label{lin:do_query_forloop}
                \State $u_i \gets \min \{ v[i].\textsc{GetMin}(), \|y_i - y_j\|_2^2 \}$ \label{lin:do_query_getmin_sketch}
                \State $\mathrm{tmp} \gets \mathrm{tmp} + u_i $
            \EndFor 
            \State \Return $\mathrm{tmp}$
        \EndProcedure 
        \State {\bf end data structure}
    \end{algorithmic}
\end{algorithm}
\subsection{Running Time of Distance Oracle}\label{sec:running_time_distance_oracle}

We start to prove the running time for \textsc{Init}.

\begin{lemma}[Running Time of \textsc{Init}]\label{lem:running_time_init}
Given data $P \subset \R^d$, index set $S \subset [n]$, precision parameter $\epsilon \in (0, 0.1)$ and failure probability $\delta \in (0, 0.1)$, procedure \textsc{Init} takes $O(\epsilon^{-2}nd\log(n/\delta))$ time to initialize.
\end{lemma}

\begin{proof}
The running time is dominated by the for-loop in Line~\ref{lin:do_init_forloop} with $n$ iterations. In each iterations, it samples a sketch matrix $\Pi$ from $\R^{m \times d}$ and obtains a sketch $\Pi x$ (Line~\ref{lin:do_init_sketch}), which takes $O(md)$ time. Thus, the total running time is 
\begin{align*}
    n \cdot O(md) = O(\epsilon^{-2}nd \log (n/\delta))
\end{align*}
which follows from $m = O(\epsilon^{-2}\log (n / \delta))$
\end{proof}

Next, we turn to prove the running time for \textsc{Cost}.

\begin{lemma}[Running Time of \textsc{Cost}]\label{lem:running_time_cost}
Procedure \textsc{Cost} returns the current cost using $O(n)$ time.
\end{lemma}
\begin{proof}
Procedure \textsc{Cost} runs a for-loop in Line~\ref{lin:do_cost_forloop} with $n$ iterations. In each iteration, it runs in $O(1)$ time (Line~\ref{lin:do_cost_sum_getmin}). Thus the total running time is $O(n)$.
\end{proof}

We show the running time of \textsc{Insert} in the following lemmas.

\begin{lemma}[Running Time of \textsc{Insert}]\label{lem:running_time_insert}
Given index $j \in [n]$, it inserts distances between $y_j$ and all sketches to $v$  using $O(\epsilon^{-2}n \log (n/\delta))$ time.
\end{lemma}

\begin{proof}
Procedure \textsc{Insert} runs a for-loop (Line~\ref{lin:do_insert_forloop}) with $n$ iterations then calls procedure \textsc{Cost}. In each iteration $i$, it first calculate $\|y_i - y_j\|_2^2$ which takes $O(m)$ time and insert into balanced tree $v[i]$ using $O(\log n)$ time (Line~\ref{lin:do_insert_insert}). After the for-loop, it calls \textsc{Cost} which takes $O(n)$ time. Thus, the total running time is $n \cdot O(m + \log n) + O(n) = O(\epsilon^{-2}n\log (n/\delta))$ which follows from $m = O(\epsilon^{-2}\log(n/\delta))$
\end{proof}

We turn to prove the running time of \textsc{Delete}.

\begin{lemma}[Running Time of \textsc{Delete}]\label{lem:running_time_delete}
Given index $j \in [n]$, procedure \textsc{Delete} deletes all distances between $y_j$ and all sketches using $O(n \log n)$ time
\end{lemma}
\begin{proof}
Procedure \textsc{Delete} runs a for-loop (Line~\ref{lin:do_delete_forloop}) with $n$ iterations then calls procedure \textsc{Cost}. In each iteration $i$, it delete node $j$ in balanced tree $v[i]$ (Line~\ref{lin:do_delete_delete}) using $O(\log n)$ time. After the for-loop, it calls \textsc{Cost} (Line~\ref{lin:do_delete_cost}) which takes $O(n)$ time. Thus, the total running time is $n \cdot O(\log n) + O(n) = O(\epsilon^{-2} \log n \log (n/\delta))$ which follows from $m = O(\epsilon^{-2}\log(n/\delta))$
\end{proof}

We present the running time of \textsc{Sample}

\begin{lemma}[Running Time of \textsc{Sample}]\label{lem:running_time_sample}
Procedure \textsc{Sample} samples an index $j$ according to probability 
$
\frac{\cost(\{x_j\}, C)}{\sum_{i = 1}^n \cost(\{x_i\}, C)}
$
using $O(n)$ time.
\end{lemma}

\begin{proof}
Procedure \textsc{Sample} first runs a for-loop (Line~\ref{lin:do_sample_forloop}) with $n$ iterations then sample an element from $[n]$. In each iteration, Line~\ref{lin:do_sample_getmin} and Line~\ref{lin:do_sample_sum} runs in $O(1)$ time. The sampling step in Line~\ref{lin:do_sample_sample} runs in $O(n)$ time. Thus, the total running time is $O(n)$.
\end{proof}

We show the running time of \textsc{Query} procedure.

\begin{lemma}[Running Time of \textsc{Query}]\label{lem:running_time_query}
Given an index $j \in [n]$, procedure \textsc{Query} returns the sum of distances between $y_j$ and all sketches using $O(\epsilon^{-2}n \log (n/\delta))$ time 
\end{lemma}

\begin{proof}
Procedure \textsc{Query} runs a for-loop (Line~\ref{lin:do_query_forloop}) with $n$ iterations. In each iteration $i$, it calculates $\|y_i - y_j\|_2^2$ which takes $O(m)$ time and \textsc{GetMin} procedure for balanced search tree (Line~\ref{lin:do_query_getmin_sketch}) runs in $O(\log n)$ time. Thus, the total running time is $n \cdot O(m + \log n) = O(\epsilon^{-2}n \log (n/\delta))$.
\end{proof}

\section{Main Result}\label{sec:main_result}

We state the main result for our algorithm including running time, space storage, and constant approximation guarantee.

\begin{theorem}[Formal Version of Theorem~\ref{thm:k_means_informal}]\label{thm:k_means_formal}
 
Given data set $P \subset \R^d$, number of centers $k \in \mathbb{N}$, precision parameter $\epsilon \in (0, 0.1)$, failure probability $\delta \in (0, 0.1)$ and set $Z = O(k \log \log k)$,  
the running time of the \textsc{FastKMeans++} algorithm is 
\begin{align*} 
O(\epsilon^{-2}n(d + k^2 \log \log k ) \log (n/\delta)).
\end{align*}

Let $C$ be the output of our \textsc{FastKMeans++} Algorithm~\ref{alg:k_means}. Let $C^*$ be the set of optimal centers. Then, we have  
$
\E[\cost(P, C)] = O(\cost(P, C^*)).
$


In addition this algorithm requires $O(n ( d+k + \epsilon^{-2} \log(n/\delta)))$ space.
\end{theorem}
\begin{proof}
It follows from Lemma~\ref{lem:space_kmeans++_alg} (Space part), Lemma~\ref{lem:fast_kmeans++_time} (Running time part), and Lemma~\ref{lem:correctness_k_means} (Correctness part).
\end{proof}

%% file: time.tex
\section{Running Time and Space}\label{sec:running_time}

Section~\ref{sec:running_time_k_means} presents the running time for \textsc{FastKMeans++} algorithm. Section~\ref{sec:running_time_local_search} shows the running time for \textsc{LocalSearch++}. Section~\ref{sec:space_storage} states the space storage for \textsc{FastKMeans++}.

\subsection{Running Time of \textsc{FastKMeans++}}\label{sec:running_time_k_means}

In this section, we present the running time of the \textsc{FastKMeans++} algorithm.

\begin{lemma}[Running Time of \textsc{FastKMeans++}, Running tme of Theorem~\ref{thm:k_means_formal}]\label{lem:fast_kmeans++_time}
Given data set $P \subset \R^d$, number of cluster centers $k \in \mathbb{N}$, precision $\epsilon \in (0, 0.1)$, failure probability $\delta \in (0, 0.1)$ and $Z = O(k \log \log k)$, the running time of Algorithm~\ref{alg:k_means} is 
\begin{align*} 
O(\epsilon^{-2}n(d + k^2 \log \log k ) \log (n/\delta)).
\end{align*}
\end{lemma}

\begin{proof}

The running time consists of three parts.

\begin{itemize}
    \item By Theorem~\ref{thm:distance_oracle}, Line~\ref{lin:k_means_do_init} takes $O(\epsilon^{-2}nd\log(n/\delta))$ time and Line~\ref{lin:k_means_do_insert} takes $O(\epsilon^{-2}n \log (n/\delta))$ time. 
    \item In Line~\ref{lin:k_means_k_forloop}, there is a for-loop with $O(k)$ iterations. In each iteration, Line~\ref{lin:k_means_k_forloop_do_sample} takes $O(n)$ time and Line~\ref{lin:k_means_k_forloop_do_insert} runs in $O(\epsilon^{-2}n \log (n/\delta))$ time. 
    \item In Line~\ref{lin:k_means_z_forloop}, there is another for-loop with $Z$ iterations. In each iteration, it calls \textsc{LocalSearch++} which takes $O(\epsilon^{-2}nk \log(n/\delta))$ time by Lemma~\ref{lem:running_time_local_search}. 
\end{itemize}

Thus, the total running time for \textsc{FastKMeans++} is
\begin{align*}
    & ~ O(\epsilon^{-2}nd\log(n/\delta)) + O(\epsilon^{-2}n\log n \log (n/\delta)) \\
    + & ~  k \cdot (O(n) + O(\epsilon^{-2}n\log n \log (n/\delta))) \\
    + & ~ Z \cdot O((\epsilon^{-2}n k \log(n/\delta)) 
    \\
    = & ~ O(\epsilon^{-2}n(d + \log n) \log(n/\delta)) \\
    + & ~  O(\epsilon^{-2}nk  \log(n / \delta)) \\
    + & ~  O(\epsilon^{-2}n k^2 \log (\log k) \log (n) \log (n/\delta))
    \\
    = & ~ O(\epsilon^{-2}n(d + k^2 \log \log k ) \log (n/\delta))
\end{align*}
where the first step follows from $Z = O(k\log \log k)$, and the last step follows from summing all terms in the first step.

Thus, we complete the proof. 
\end{proof}

\subsection{Running Time of LocalSearch++}\label{sec:running_time_local_search}

In this section, we prove the running time for \textsc{LocalSearch++} (Algorithm~\ref{alg:local_search})

\begin{lemma}[Running Time of \textsc{LocalSearch++}]\label{lem:running_time_local_search}
Given data set $P \subset \R^d$ and center set $C \subset \R^d$, the running time of \textsc{LocalSearch++} is $O(\epsilon^{-2} n k \log (n/\delta))$.
\end{lemma}

\begin{proof}
In Line~\ref{lin:local_sample}, it calls \textsc{Sample} procedure which takes $O(n)$ time. Line~\ref{lin:local_cost} takes $O(n)$ time. Line~\ref{lin:local_forloop} is a for-loop with $k$ iterations. In each iteration, it takes $O(n\log n)$ time to run \textsc{Delete} (Line~\ref{lin:local_forloop_delete}), $O(\epsilon^{-2}n \log (n/\delta))$ to run \textsc{Query} procedure (Line~\ref{lin:local_forloop_query}) and $O(\epsilon^{-2} n \log (n/\delta))$ to run \textsc{Insert} (Line~\ref{lin:local_forloop_insert}). The running time for one step iteration is $O(\epsilon^{-2}n \log (n/\delta))$. Line~\ref{lin:local_delete} and Line~\ref{lin:local_insert} runs in $O(\epsilon^{-2}n \log (n/\delta))$ time. 

Thus, the total running time for \textsc{LocalSearch++} is 
\begin{align*}
     & ~  O(n) + k \cdot O(\epsilon^{-2}n \log (n/\delta)) + O(\epsilon^{-2}n \log (n/\delta)) \\
   =  & ~  O(\epsilon^{-2}nk\log(n/\delta))
\end{align*}

Thus, we complete the proof.
\end{proof}

 \subsection{Space Storage}\label{sec:space_storage}
In this section, we prove the space storage of \textsc{FastKMeans++} algorithm.

\begin{lemma}[Space Storage For Algorithm~\ref{alg:k_means}, Space part of Theorem~\ref{thm:k_means_formal}]\label{lem:space_kmeans++_alg}
The space storage of Algorithm~\ref{alg:k_means} is $O(n (d + k +\epsilon^{-2} \log (n/\delta)) )$. 
\end{lemma}

\begin{proof}

The space storage in Algorithm~\ref{alg:k_means} comes from \textsc{DistanceOracle} and center set $C \subset [n]$ (Line~\ref{lin:center_set}) .

We have the total space storage for \textsc{FastKMeans++}
\begin{align*}
    & ~ O(n (d + k +\epsilon^{-2} \log (n/\delta))) + O(n)\\
    = & ~ O(n (d + k +\epsilon^{-2} \log (n/\delta)))
\end{align*}

Thus, we complete the proof.
\end{proof}

%% file: exp.tex
\begin{figure*}[!ht]
    \centering
    \subfloat[time vs $n$]
    {\includegraphics[width=0.25\textwidth]{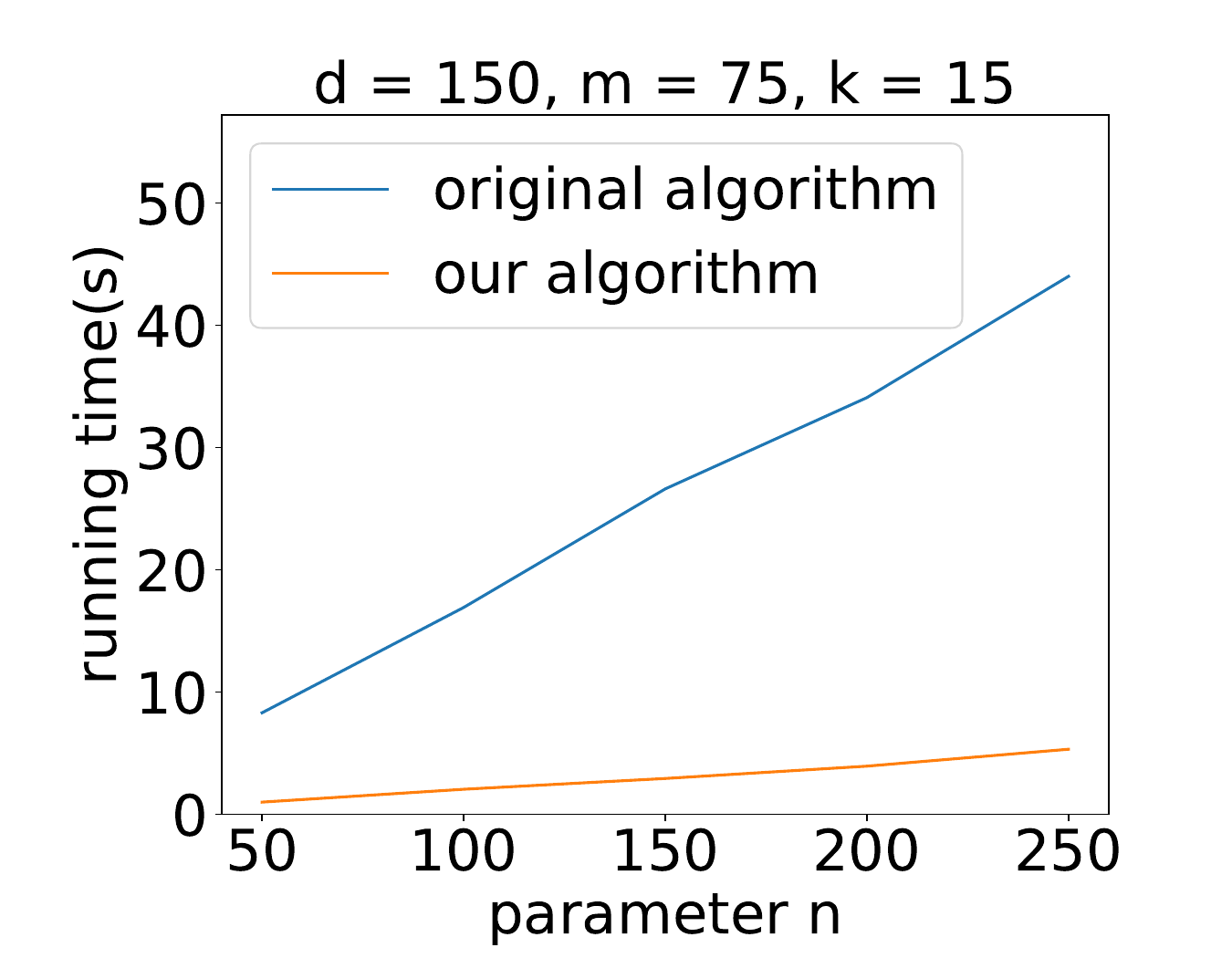}
    \label{fig:param_n_running_time}
    }
    \subfloat[time vs $d$]
    {
    \includegraphics[width=0.25\textwidth]{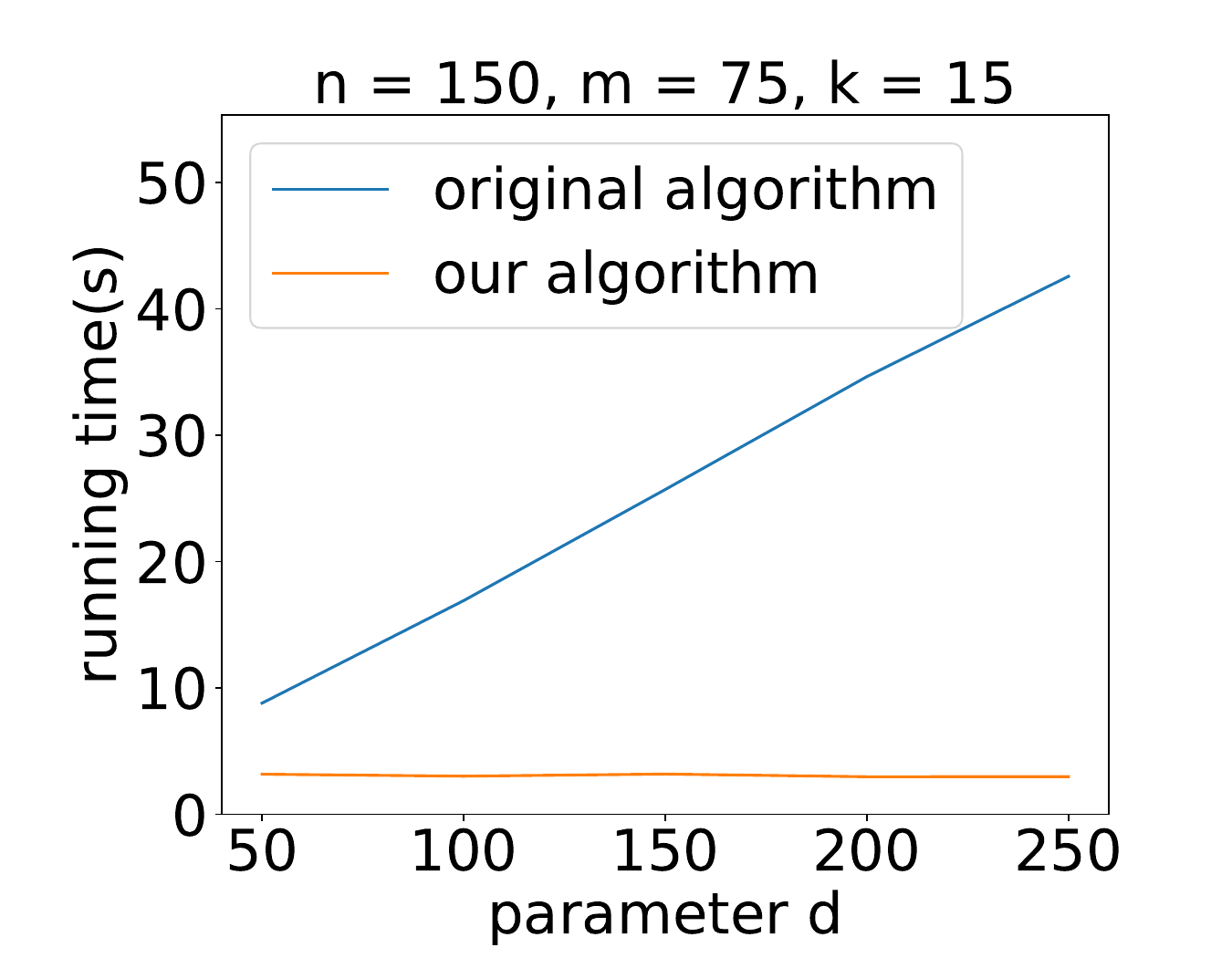}
   \label{fig:param_d_running_time}
    }
    \subfloat[time vs $m$]
    {\includegraphics[width=0.25\textwidth]{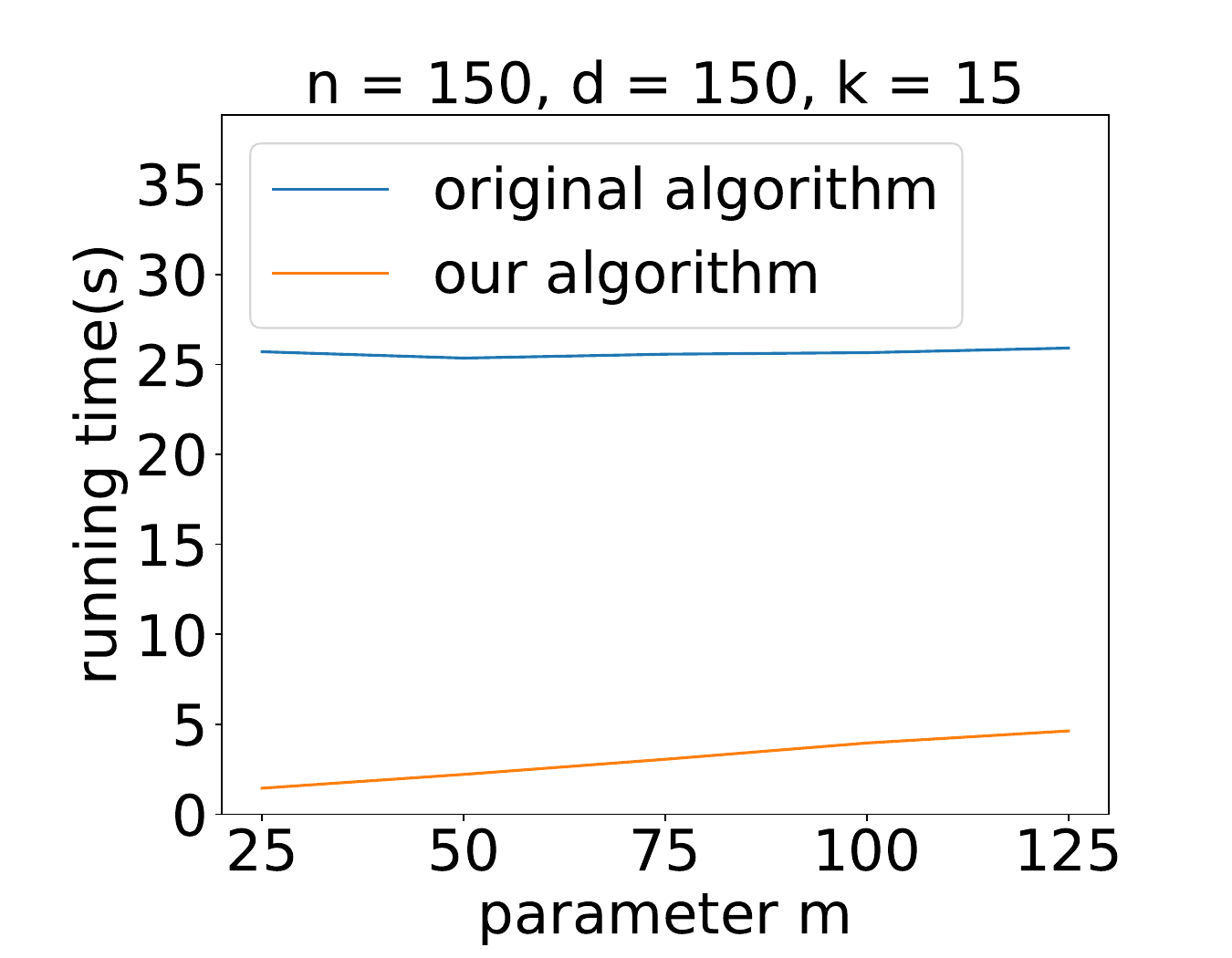}
    \label{fig:param_m_running_time}
    }
    \subfloat[time vs $k$]
    {
    \includegraphics[width=0.25\textwidth]{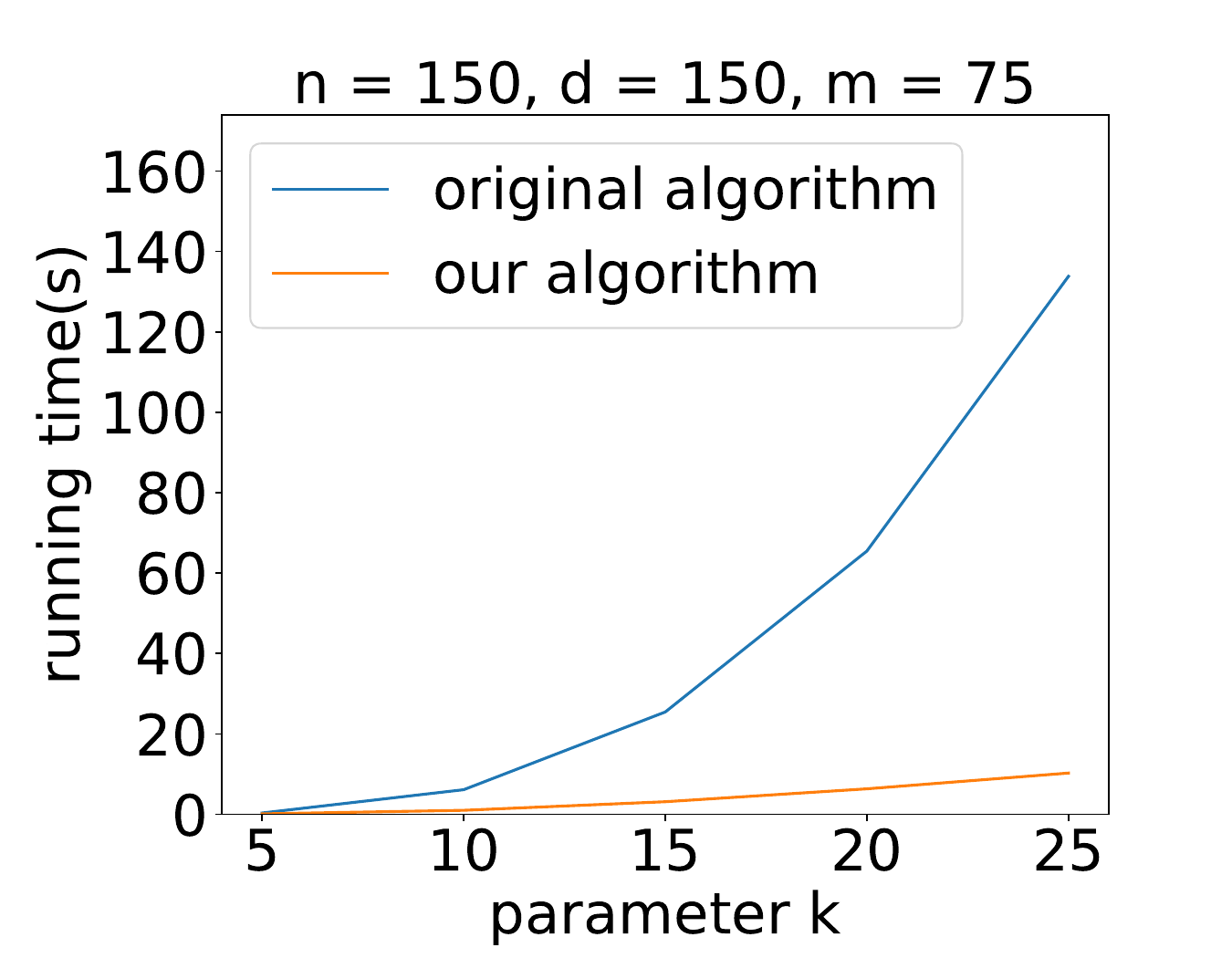}
    \label{fig:param_k_running_time}
    }
    \caption{The relationship between each parameter and the running time, where original algorithm denotes Algorithm in \cite{ls19}, and our algorithm denotes \textsc{FastKMeans++} in Theorem \ref{thm:k_means_formal}. Let $n$ be the number of points in the point set. Let $d$ denote the dimension of each node. Let $m$ denote the dimension of each node after we process them with a sketching matrix. Let $k$ be the number of clusters and centers. 
    }
\end{figure*}

 \begin{figure*}[!ht]

    \subfloat[SCADI data set]{
    \includegraphics[width=0.25\textwidth]{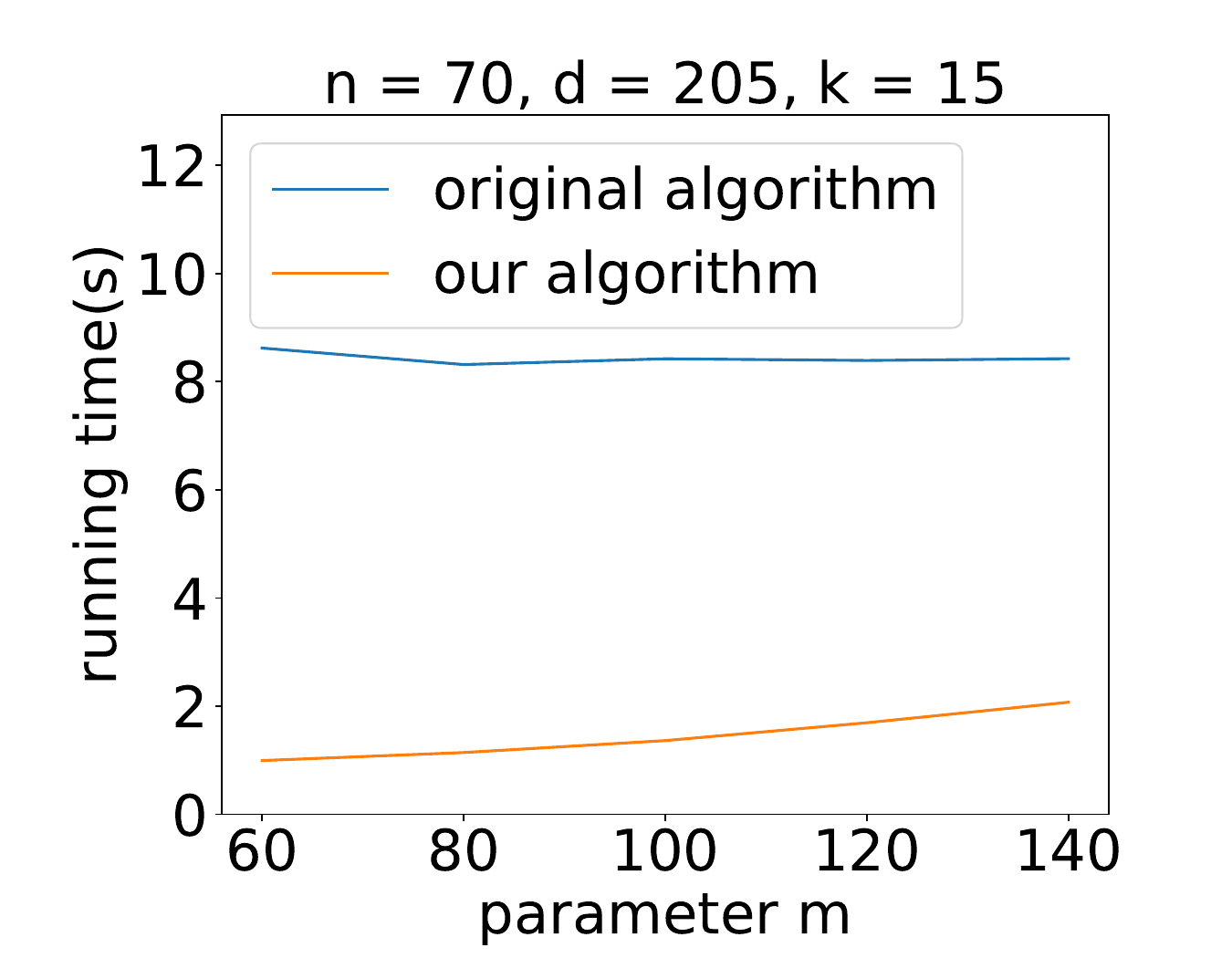}
    \includegraphics[width=0.25\textwidth]{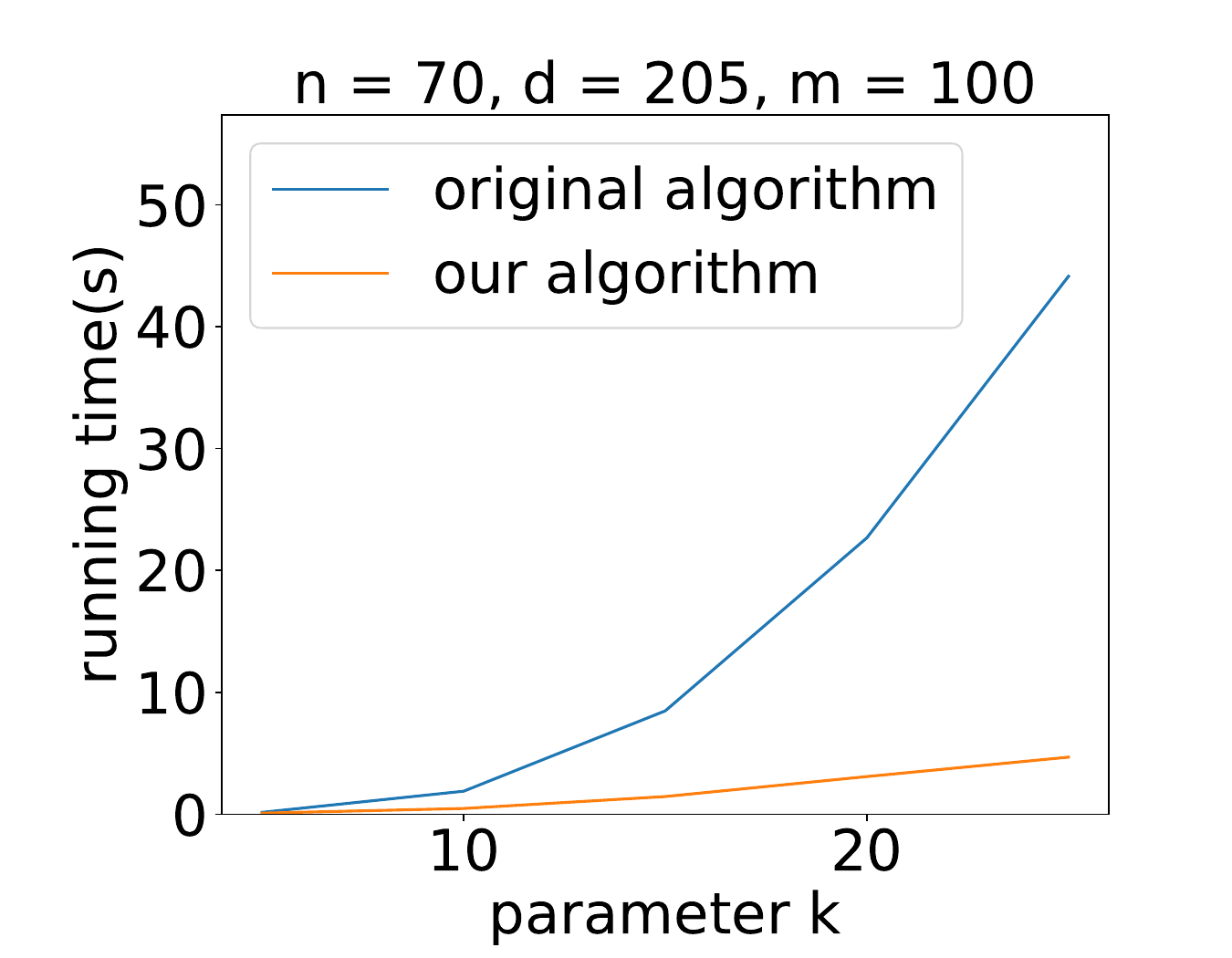}
    \label{fig:scadi_running_time}
}
   \subfloat[MUPCI data set]{
    \includegraphics[width=0.25\textwidth]{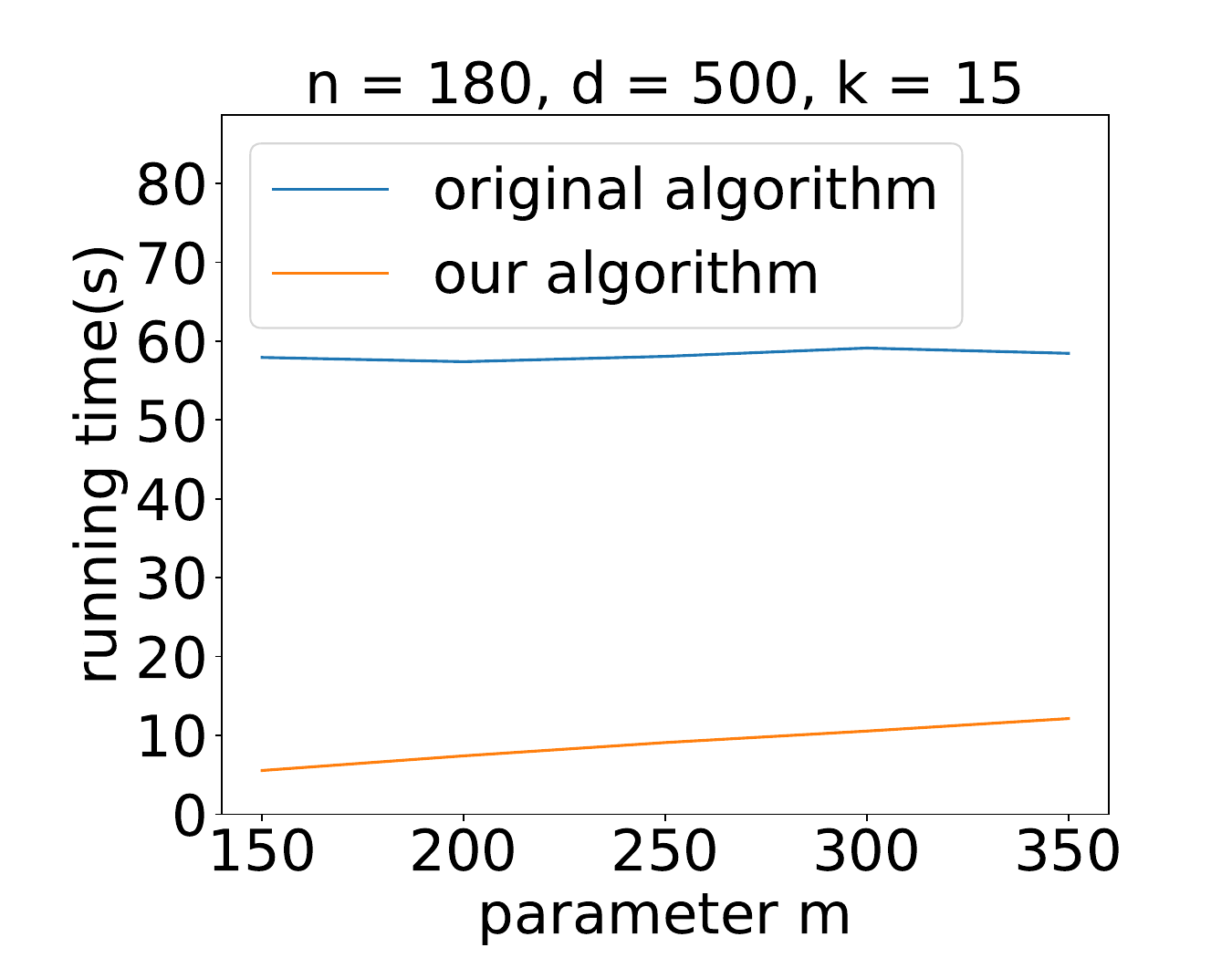}
    \includegraphics[width=0.25\textwidth]{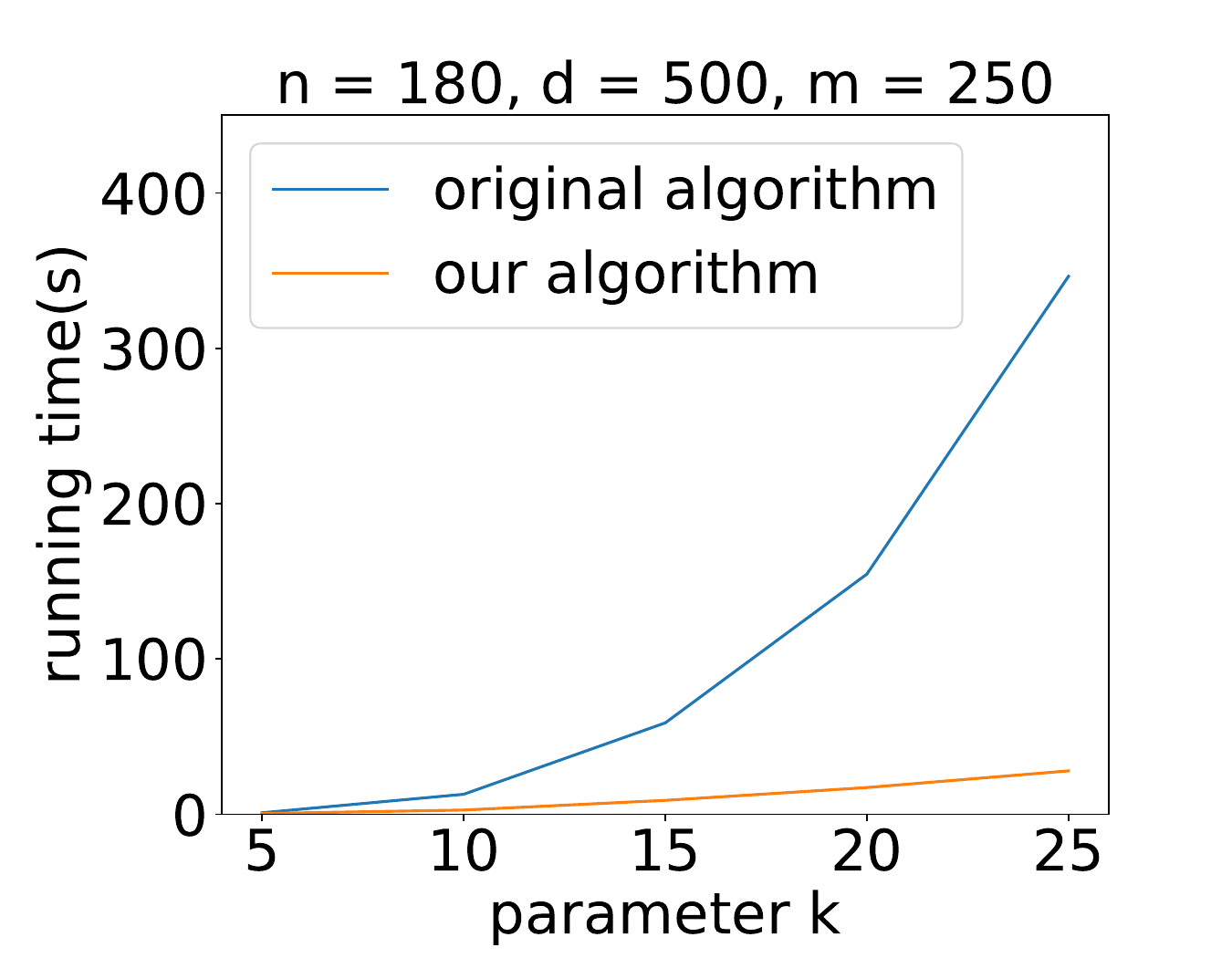}
    \label{fig:mupci_running_time}
}
    \caption{The running time of $k$-means++ and \textsc{FastKMeans++} algorithm on SCADI and MUPCI data set. $m$ denotes the dimension of each node after we transformed all of them, and $k$ denotes the number of clusters. There are two figures for each data set. The left one shows the relationship between the running time of two algorithms and $m$, and the right one shows the relationship between the running time of two algorithms and $k$.}
    \label{fig:scadi_mupci_running_time}
\end{figure*}

\begin{figure*}[!ht]

    \subfloat[Libras Movement data set]{
    \includegraphics[width=0.25\textwidth]{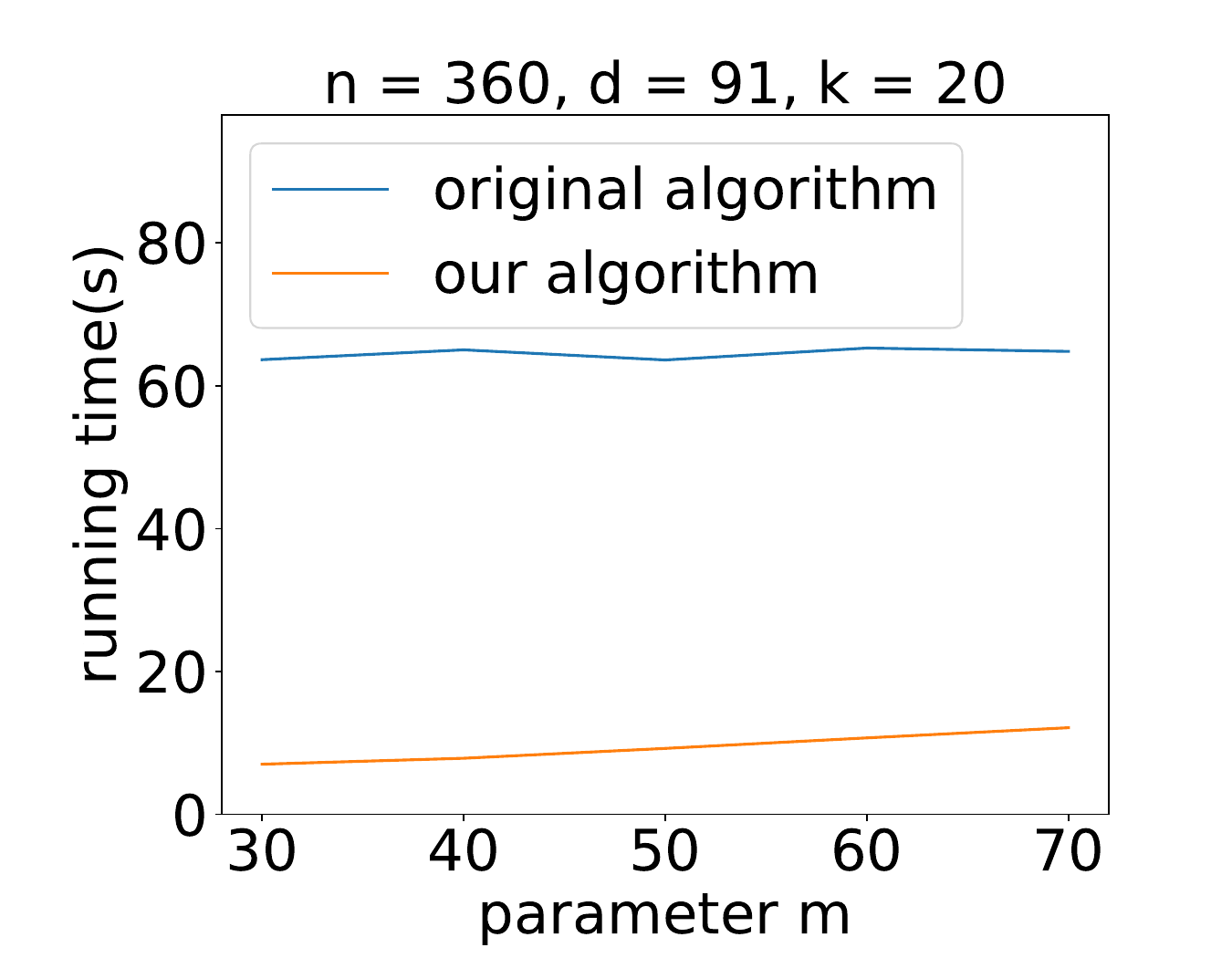}
    \includegraphics[width=0.25\textwidth]{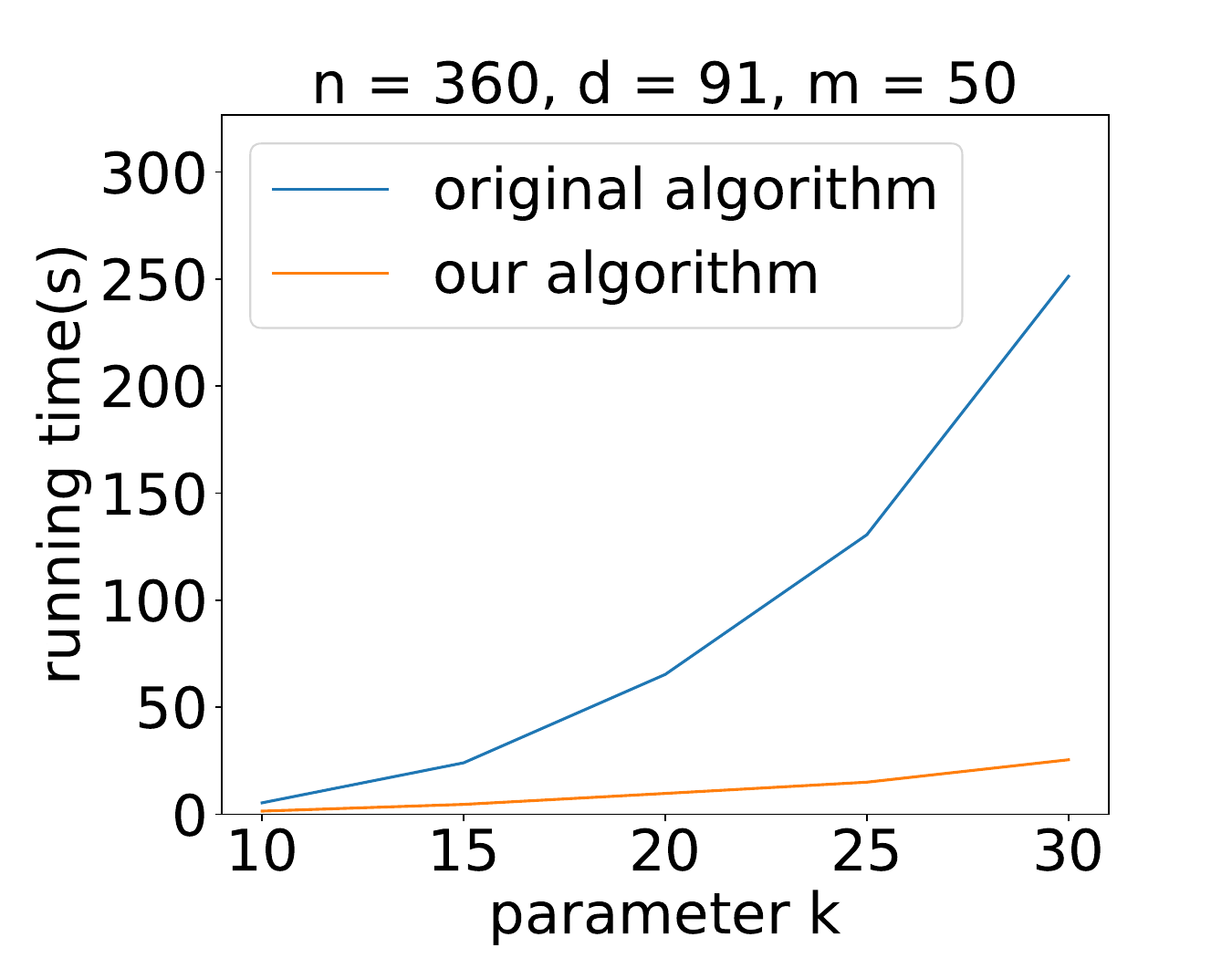}
    \label{fig:libras_movement_running_time}
}
   \subfloat[STDW data set]{
    \includegraphics[width=0.25\textwidth]{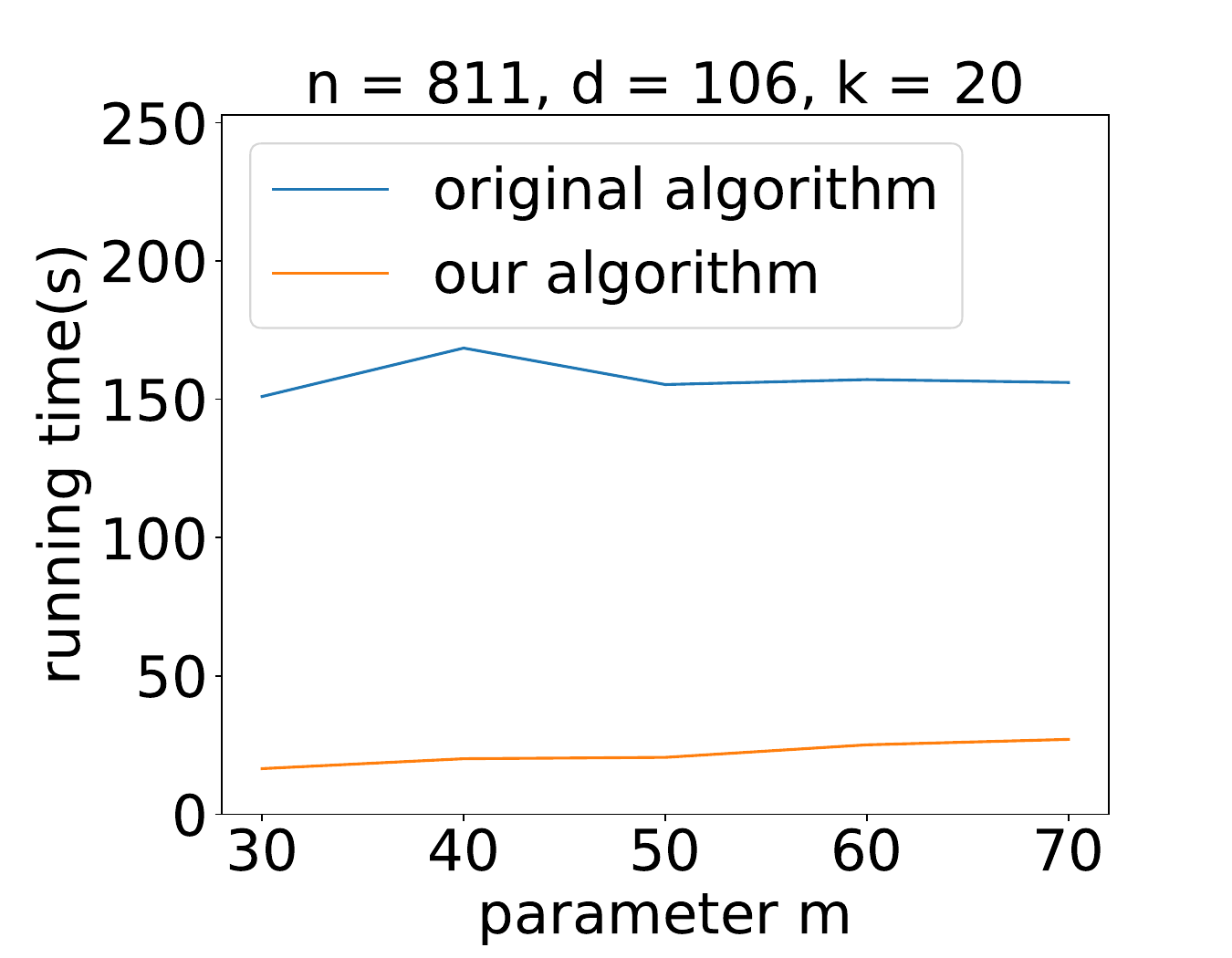}
    \includegraphics[width=0.25\textwidth]{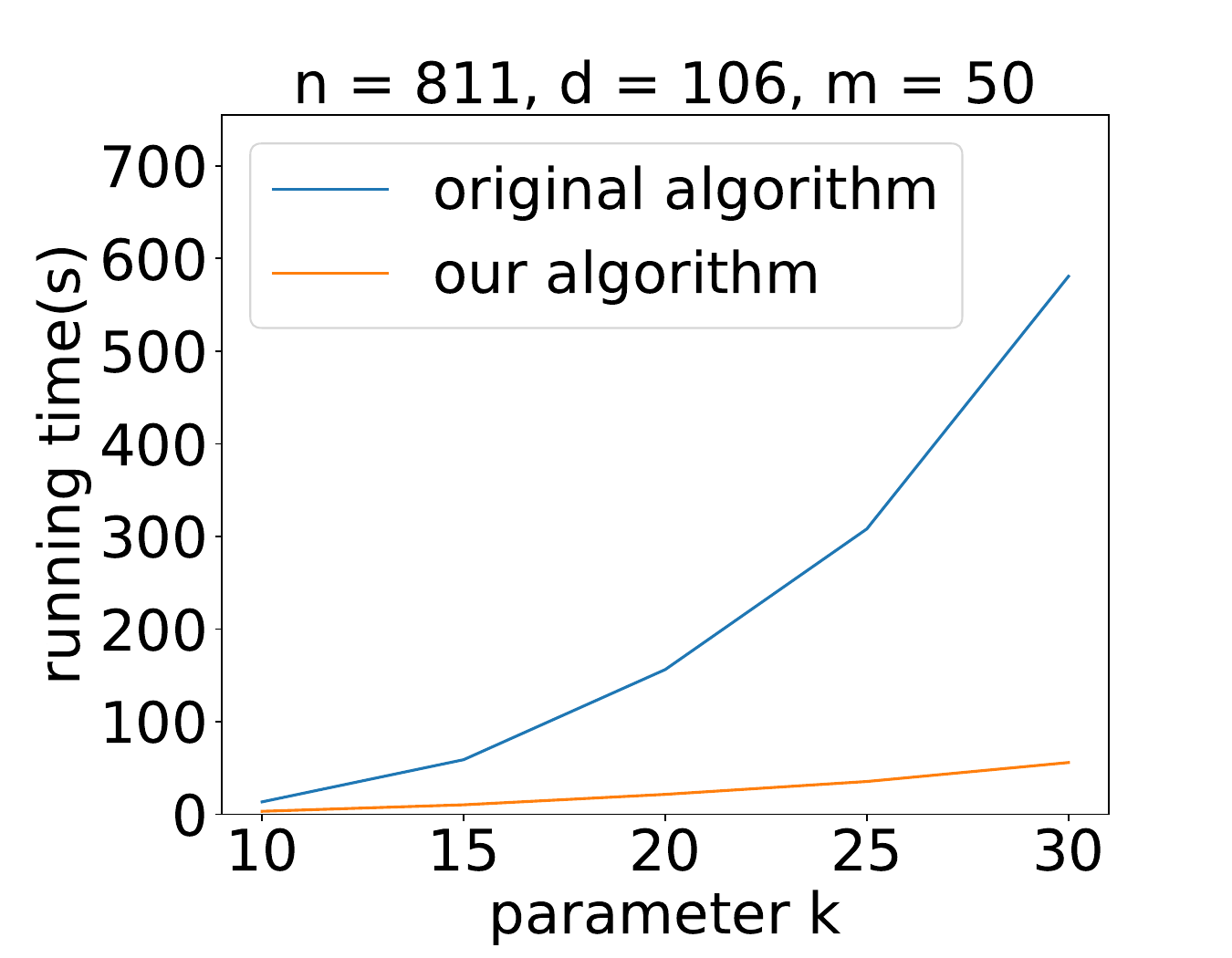}
    \label{fig:stdw_running_time}
}
    \caption{The running time of $k$-means++ and \textsc{FastKMeans++} algorithm on Libras Movement and STDW data set. $m$ denotes the dimension of each node after we transformed all of them, and $k$ denotes the number of clusters. There are two figures for each data set. The left one shows the relationship between the running time of two algorithms and $m$, and the right one shows the relationship between the running time of two algorithms and $k$.}
    \label{fig:libras_movement_stdw_running_time}
\end{figure*}

\section{Experiments} \label{sec:experiments}

\textbf{Purpose.} In this section, we use $k$-means++ algorithm in \cite{ls19} and \textsc{FastKMeans++} algorithm in Theorem \ref{thm:k_means_formal} to cluster the same synthetic point set respectively. We use $k$-means++ as a baseline to evaluate the performance of \textsc{FastKMeans++}.  
We designed a new data structure \textsc{DistanceOracle} in Theorem \ref{thm:distance_oracle}. In \textsc{DistanceOracle}, we use a balanced search tree to store the nodes of each cluster to make the searching process faster. In order to shorten the time of calculating the distance between two nodes, we also create a sketching matrix that decreases the number of dimensions of each node. We adjust the parameters of our \textsc{FastKMeans++} and original $k$-means++ algorithm spontaneously to observe the influence of each parameter on their running time. Our summary of the result is as follows:
\begin{itemize}
    \item Our \textsc{FastKMeans++} algorithm is much faster than original $k$-means++  algorithm if $m$ is small enough. 
    \item The running time of $k$-means++ and \textsc{FastKMeans++} algorithm will increase linearly as $n$ increases.  
    \item The running time of original $k$-means++ algorithm will increase linearly as $d$ increases, but $d$ doesn't affect the running time of our \textsc{FastKMeans++} algorithm very much. 
    \item The running time of our \textsc{FastKMeans++} algorithm will increase linearly as $m$ increases, but $m$ is unrelated with the running time of original $k$-means++ algorithm. 
    \item The running time of $k$-means++ and \textsc{FastKMeans++} algorithm will increase squarely as $k$ increases. 
\end{itemize}

\textbf{Setup.} We use a computer of which the CPU is AMD Ryzen 7 4800H, GPU is RTX 2060 laptop. The operating system is Windows 11 Pro and we use Python as the code language.  
Let $\ell_2$ denote distance weights. Let $n$ denote the number of points in the point set. Let $d$ denote the dimension of each node. Let $m$ denote the dimension of each node after we process them with a sketching matrix. Let $k$ denote the number of clusters and centers. Let $Z = O (k \log \log k)$ denote the number of iterations.  

\begin{table*}[!ht]\caption{
Let $n$ denote the number of nodes and $d$ denote the dimension of each node. Here MUPCI denotes Mturk User-Perceived Clusters over Images. Let STDW denote the Sales Transactions Dataset Weekly Data Set. Let LM denote the Libras Movement. 
$m$ denotes the dimension of each node after we transformed all of them, and $k$ denotes the number of clusters.}
\begin{center}
\begin{tabular}{ |l|l|l|l|l| } \hline 
 Dataset Names &  $n$ & $d$ &  $m$ & $k$\\ \hline
 SCADI &  $70$ & $206$ & $[60, 80, 100, 120, 140]$ & $[5, 10, 15, 20, 25]$\\ \hline
MUPCI  & $180$ & $500$ & $[150, 200, 250, 300, 350]$ & $[5, 10, 15, 20, 25]$ \\ \hline
LM  & $360$ & $91$ & $[30, 40, 50, 60, 70]$ & $ [10, 15, 20, 25, 30]$ \\ \hline
STDW  & $811$ & $53$  & $[30, 40, 50, 60, 70]$ & $[10, 15, 20, 25, 30]$ \\ \hline

\end{tabular}
\end{center}
\end{table*}

\textbf{Sythetic Data Generation.} We generate $n$ random nodes of $x_1, \cdots, x_n \in \R^d$ and each node is generated as follows
\begin{itemize}
    \item Pick each coordinate of node from $[-1,1]$.
    \item Normalize each vector so that its $\ell_2$ norm is $1$.
\end{itemize}

In the original $k$-means++ algorithm, for node $x_i$ and node $x_j$, we use  $\| x_i - x_j\|_2^2$ to denote the distance between them. This takes $O(d)$ time. As per update, it will take $O(n k d)$ time since there are $k$ clusters containing $n$ points. For $Z = O (k \log \log k)$ iterations, it takes $O(n d k^2 \log \log k)$ time.

In our \textsc{FastKMeans++} algorithm, we compute $y_i = S x_i$ for $\forall i \in [n]$. It takes $O(nd)$ time since there are $n$ nodes in total. So when we initialize our data structure, it will take $O(n m d)$ time. In per update, we use $\| y_i - y_j\|_2^2$ to denote the distance between them. This takes $O(m)$ time. So it will take $O(n k m)$ time since we need to iterate over all clusters and calculate the distance to all other clusters.

For $Z = O (k \log \log k)$ iterations, it will take $O(n d k^2 \log \log k)$ time during the update process. Therefore, it will take 
\begin{align*}
    O(n m d + n m k^2 \log \log k)
\end{align*}
time in total. 

\textbf{Parameter Setting.}  In our experiments, we choose $n=150$, $d=150$, $m = 75$ and $k = 15$ as primary condition.

\textbf{Real Datasets.} In this part, we run $k$-means++ and our \textsc{FastKMeans++} algorithm on real data sets from UCI
library \cite{dg19} to observe if our algorithm is better than the original one in a real-world setting.
\begin{itemize}
    \item SCADI Data Set \cite{zbd18,bz19} : This dataset contains $206$ attributes of $70$ children with physical and motor disability based on ICF-CY.
    \item Mturk User-Perceived Clusters over Images Data set \cite{clg+16}: This dataset random sampled $180$ images from the NUS-WIDE image database. Each image has $500$ features consisting of a bag of words based on SIFT descriptions. 
    \item Libras Movement Data set \cite{dmr+09}: The dataset contains $15$ classes of $24$ instances each. In the video pre-processing, a time normalization is carried out selecting 45 frames from each video, according to a uniform distribution. In each frame, the centroid pixels of the segmented objects (the hand) are found, which compose the discrete version of the curve F with $45$ points. Each instance contains $90$ features of this curve F.
   \item Sales Transactions Dataset Weekly Data set \cite{ts14}: This data set contains $811$ sales transactions weekly. Each record has one product code and data for $52$ weeks.
\end{itemize}

\textbf{Results}.
    To evaluate the influence of $n$, we vary the value of $n \in [50,250]$ while keeping other parameters the same. Fig.~\ref{fig:param_n_running_time} indicates the running time of original $k$-means++ and our \textsc{FastKMeans++} algorithm increases linearly while $n$ increases.  We note that our \textsc{FastKMeans++} algorithm is faster than the original $k$-means++ algorithm while $n$ changes. The ratio of the running time of the original $k$-means++ to that of \textsc{FastKMeans++} also increases from $8.26$ to $8.33$ as $n$ increases.

    Then we estimate the influence of $d$, we vary the value of $d \in [50,250]$ while making sure $n = 15, m = 75, k = 15$ during the whole testing process. We compare the performance of the original $k$-means++ algorithm and our \textsc{FastKMeans++} algorithm in Fig.~\ref{fig:param_d_running_time}. It shows the running time of the original $k$-means++ algorithm increases linearly while $d$ increases. 
    The ratio of the running time of the original $k$-means++ to that of \textsc{FastKMeans++} also increases from $2.77$ to $14.30$ as $d$ increases. The running time of our \textsc{FastKMeans++} seems unrelated with $d$ because in this case, $d$ is too small compared with other parameters like $n$ and $m$. To show the influence of $d$ on our \textsc{FastKMeans++} algorithm more clearly, we will explain it with more figures in the appendix.
    
    To evaluate the influence of $m \in [25,125] $, we vary the value of $m$ while keeping other parameters the same. Fig.~\ref{fig:param_m_running_time} shows the running time of the original $k$-means++ algorithm doesn't change but that of our algorithm increases linearly while $m$ increases. And our \textsc{FastKMeans++} algorithm runs faster compared to the original algorithm if $m$ is small enough. The ratio of the running time of the original  $k$-means++ to that of \textsc{FastKMeans++} decreases from $17.73$ to $5.59$ as $m$ increases. 
    We vary the value of $k \in [5,25]$ and run $k$-means++ and \textsc{FastKMeans++}. From Fig.~\ref{fig:param_k_running_time} we can see that the running time of original $k$-means++ and our algorithm increases squarely while $k$ increases. We note that our \textsc{FastKMeans++} algorithm is faster than the original $k$-means++ algorithm while $k$ changes. The ratio of the running time of the original $k$-means++ to that of \textsc{FastKMeans++} also increases from $2.95$ to $13.04$ as $k$ increases. This is consistent with our original thought. 

    \textbf{Results for real data set.} Fig.~\ref{fig:scadi_running_time}, Fig.~\ref{fig:mupci_running_time}, Fig.~\ref{fig:libras_movement_running_time} and Fig.~\ref{fig:stdw_running_time} show the relationship between the running time of $k$-means++ algorithm and \textsc{FastKmeans++} and parameter $m$ and $k$ on SCADI, MUPCI, Libras Movement and STDW data sets respectively. They show that our \textsc{FastKmeans++} is much faster than the original $k$-means++ algorithm on real data sets.

%% file: concl.tex
\section{Conclusion}\label{sec:conclusion}
In this paper, we accelerate the $k$-means++ algorithm using a distance oracle. We carefully design the data structure to support initialize, query, cost, etc. operations. We remove the factor $d$ of running time in each iteration of the local search. Finally, we run our \textsc{FastKMeans++} algorithm and the original $k$-means++ algorithm. In addition, we also implement our algorithm and obtain the experimental results which support our theoretical result. 
In this paper, we primarily focus on designing data structures and algorithms. We believe that proving a certain lower bound for our problem is also an interesting future direction. Our proposed \textsc{FastKMeans++} algorithm not only improves computational efficiency but also maintains the quality of the clustering solution. We believe that our contributions will significantly advance the current state of scalable machine learning methods, opening up new possibilities for handling large-scale datasets.

\section*{Impact Statement}

In this paper, we present the algorithm, theoretical analysis, and experiment. To the best of our knowledge, we do not foresee any potential negative societal consequences of our work.

%% file: correct.tex
\paragraph{Roadmap.}

In Section~\ref{sec:correctness}, we provide the pseudocode of our algorithms and proof of the correctness result. In Section~\ref{sec:app_exp}, we provide more experimental results.

\section{Correctness}\label{sec:correctness}

In Section \ref{sec:capture_and_assign}, we defined two operations capture and reassign respectively. Then in Section \ref{sec:upper_bound_for_reassign_Operation}, we prove the upper bound for reassign operation. In Section \ref{sec:good_cluster}, we defined a good cluster whose index is in $H$. $H$ is the subset of center set $C=\{c_{1}, \cdots, c_{k}\}$ that capture (Definition~\ref{def:capture}) exactly one cluster center from optimal center set $C^* = \{c_1^*, \cdots, c_k^*\}$. Then we calculate a lower bound for the proportion of good clusters under certain conditions. In Section \ref{sec:lower_bound_for_cost}, we calculate a lower bound for the cost related to the mean center(Definition~\ref{def:mean_center}). In Section \ref{sec:good_cluster_for_general_cluster}, similar to Section \ref{sec:good_cluster}, we provide a definition of a good cluster for general cases. And we calculate a lower bound for the proportion of such good clusters under certain conditions. With lemmas and definitions in these sections, we prove that in each iteration of \textsc{LocalSearch++}, we can reduce the cost by $(1 - \frac{1}{100k})$ with constant probability in Section \ref{sec:cost_reduction}. Then in Section \ref{sec:correctness_k_means}, we prove the correctness of \textsc{FastKMeans++}.

\begin{algorithm}[!ht]\caption{Fast $k$-means++}\label{alg:k_means}
    \begin{algorithmic}[1] 
        \State {\bf data structure} \textsc{FastKMeans++}
\Comment{Theorem~\ref{thm:k_means_formal}}
        \State {\bf members}
            \State \hspace{4mm} \textsc{DistanceOracle} \textsc{do}

            \State \hspace{4mm} $C \subset [n]$ \label{lin:center_set}
        \State {\bf end members}
        \Procedure{Seeding}{$P \subset \R^d, k \in \mathbb{N}_+, Z \in \mathbb{N}_+, \epsilon \in (0, 0.1), \delta \in (0, 0.1)$}
        
        \State \textsc{do}.\textsc{Init}($P, \epsilon, \delta$)\label{lin:k_means_do_init}  
        \State Sample index $j$ from $[n]$ uniformly at random
        \State $C \gets \{j\}$
        \State \textsc{do}.\textsc{Insert}$(j)$\label{lin:k_means_do_insert}
        \For{$i = 2 \to k$}\label{lin:k_means_k_forloop}
            \State  $j \gets \textsc{do}.\textsc{Sample}()$ \label{lin:k_means_k_forloop_do_sample}
            \State $C \gets C \cup \{j\}$
            \State $\textsc{do}.\textsc{Insert}(j)$ \label{lin:k_means_k_forloop_do_insert}
        \EndFor
        \For{$ i = 2 \to Z$}\label{lin:k_means_z_forloop}
            \State  \textsc{LocalSearch++}$()$\label{lin:k_means_local_search}
        \EndFor
        \State \Return $C$
        \EndProcedure
        \State {\bf end data structure}
    \end{algorithmic}
\end{algorithm}

\begin{algorithm}[!ht]\caption{LocalSearch++}\label{alg:local_search}
    \begin{algorithmic}[1]
        \State {\bf data structure} \textsc{FastKMeans++}    
        \Procedure{LocalSearch++}{$ $}  \Comment{Lemma~\ref{lem:running_time_local_search}}
        \State $j \gets \textsc{do}.\textsc{Sample}()$ \label{lin:local_sample}
        \State $a \gets j$
        \State $q \gets a$
        \State $\mathrm{tmpb} \gets $ \textsc{do}.\textsc{Cost}$()$ \label{lin:local_cost}
        \For{$b \in C$} \label{lin:local_forloop} 
            \State \textsc{do}.\textsc{Delete}$(b)$ \label{lin:local_forloop_delete}
            \State $\mathrm{tmpa} \gets \textsc{do}.\textsc{Query}( a )$ \label{lin:local_forloop_query}
            \If{$\mathrm{tmpa}< \mathrm{tmpb}$}
                \State $q \gets b$
                \State tmpb $\gets$ tmpa
            \EndIf
            \State \textsc{do}.\textsc{Insert}$(b)$ \label{lin:local_forloop_insert}
        \EndFor
        \State \textsc{do}.\textsc{Delete}$(q)$ \label{lin:local_delete}
        \State \textsc{do}.\textsc{Insert}$(a)$ \label{lin:local_insert}
        \State $C \gets (C \setminus q) \cup a$
        \EndProcedure
        \State {\bf end data structure}
    \end{algorithmic}
\end{algorithm}

\subsection{Capture and Assign}
\label{sec:capture_and_assign}
In this section, we introduce two operations capture and assign.
\begin{definition}[Capture]\label{def:capture}
Let $C^* = \{c_1^*, \cdots, c_k^*\}$ be optimal centers and $C = \{c_1, \cdots, c_k\}$ is current cluster centers. An optimal center $c^*$ is captured by a center $c \in C$ if 
\begin{align*}
    c=\arg\min_{c \in C} \|c^* - c\|_2^2.
\end{align*}
 
\end{definition}

Then, we state the definition of reassign operation as follows.

\begin{definition}[Reassign]\label{def:reassign}
Let $P \subset \R^{d}$ be a data set. Let $C \subset \R^{d}$ be cluster centers set and $C^*=\{c_{1}^{*}, \cdots, c_{k}^{*}\}$ be the optimal cluster centers set. Let $P_{i}, P_{i}^{*}, 1 \leq i \leq k$ be the corresponding clusters. Let $H$ denote the subset of center set $C=\{c_{1}, \cdots, c_{k}\}$ that capture (Definition~\ref{def:capture}) exactly one cluster center from $C^*$. Let $L$ be the subset of cluster centers $C=\{c_1, \cdots, c_k\}$ that doesn't capture (Definition~\ref{def:capture}) any optimal centers. We use $P_{h}$ to denote the cluster with index $h \in H$ and use $P_{h}^{*}$ to denote the cluster in $C^*$ captured by $c_{h}$. We use a similar notation for the index $\ell \in L$.

For $h \in H$, we define the reassignment cost of $c_{h}$ as
\begin{align*} 
& ~ \mathrm{reassign}(P, C, c_{h}) \\
:= & ~ \cost(P \setminus P_{h}^{*}, C \setminus\{c_{h}\})-\cost(P \setminus P_{h}^{*}, C)
\end{align*}

For $\ell \in L$ we define the reassignment cost of $c_{\ell}$ as 
\begin{align*}
\mathrm{reassign}(P, C, c_{\ell}) := \cost(P, C \setminus\{c_{\ell}\})-\cost(P, C)    
\end{align*}
\end{definition}

\subsection{Upper Bound for Reassign Operation}
\label{sec:upper_bound_for_reassign_Operation}
In this section, we calculate an upper bound for reassign operation.
\begin{lemma}\label{lem:reassign_upper_bound}
For $r \in H \cup L$ we have
\begin{align*}
\mathrm{reassign}(P, C, c_{r}) \leq \frac{21}{100} \cost(P_{r}, C)+24 \cost(P_{r}, C^{*})    
\end{align*}
\end{lemma}

\begin{proof}
If a center $c \in C$ does not capture any optimal center, we call it a lonely center. If a center is lonely, we think of it as a center that can be moved to a different cluster. We use $Q_{r}$ to denote points obtained from $P_{r} \setminus P_{r}^{*}$ by moving each point in $P_{i}^{*} \cap P_{r}, i \neq r$, to $c_{i}^{*}$. By Lemma~\ref{lem:two_point_cost_error} 
with $\epsilon=1 / 10$, we obtain an upper bound for the change of cost with respect to $C$, which results from moving the points to $Q_{r}$. For $p \in P_{r} \setminus P_{r}^{*}$, we use $q_{p} \in Q_{r}$  to denote the point of $Q_{r}$ to which $p$ has been moved. We have:

\begin{align*}
& ~ |\cost(\{p\}, C)-\cost(\{q_{p}\}, C)| 
\\
\leq & ~ \frac{1}{10} \cost(\{p\}, C)+11 \cdot \cost(\{p\}, C^{*})
\end{align*}

Taking all points in $P_{r} \setminus P_{r}^{*}$ into consideration,

\begin{align}\label{eq:cost_QC_cost_PC}
& ~ |\cost(P_{r} \setminus P_{r}^{*}, C)-\cost(Q_{r}, C)| \notag
\\
\leq & ~ \frac{1}{10} \cost(P_{r} \setminus P_{r}^{*}, C)+11 \cdot \cost(P_{r} \setminus P_{r}^{*}, C^{*})
\end{align}

Note that all points from $P_{r} \setminus P_{r}^{*}$ have been assigned to centers from $C \setminus\{r\}$. We turn to analyze the cost of moving the points back to the original location while maintaining the assignment. We use $Q_{r, i}$ to denote the points in $Q_{r}$ nearest to center $c_{i} \in C$ and $P_{r, i}$ to represent the set of their original locations. For $p \in P_{r, i}$ that was moved to $q_{p} \in Q_{r, i}$, we have:

\begin{align*}
& ~ |\cost(\{q_{p}\},\{c_{i}\})-\cost(\{p\},\{c_{i}\})| 
\\
\leq & ~ \frac{1}{10} \cost(\{q_{p}\},\{c_{i}\})+11 \cdot \cost(\{p\},\{q_{p}\})
\end{align*}

Taking all points in $P_{r}$ and the corresponding points in $Q_{r}$ into consideration,

\begin{align}\label{eq:cost_QC_cost_sum}
& ~ |\cost(Q_{r}, C)-\sum_{i=1}^{k} \cost(P_{r, i},\{c_{i}\})| \notag
\\
\leq & ~ \frac{1}{10} \cost(Q_{r}, C)+11 \cdot \cost(P_{r} \setminus P_{r}^{*}, C^{*}) \notag
\\
\leq & ~ \frac{1}{10}(\frac{11}{10} \cost(P_{r} \setminus P_{r}^{*}, C) + 11 \cdot \cost(P_{r} \setminus P_{r}^{*}, C^{*}))+11 \cdot \cost(P_{r} \setminus P_{r}^{*}, C^{*}) \notag
\\
\leq & ~ \frac{11}{100} \cost(P_{r}, C)+13 \cdot \cost(P_{r}, C^{*}) .
\end{align}
where the first step follows $\sum_{i=1}^{k} \cost(P_{r, i}, c_{i})=\cost(P_{r}\setminus P_{r}^{*}, C \setminus \{r\})$, the second step follows from Eq.~\eqref{eq:cost_QC_cost_PC} and the last step follows from $\cost(P_r\setminus P_r^*, C^*) \leq \cost(P_r, C^*)$.

Hence,
\begin{align*}
\mathrm{reassign}(P, C, c_{r}) = & ~ |\cost(P_{r} \setminus P_{r}^{*}, C) - \sum_{i=1}^k \cost(P_{r, i},\{c_{i}\})| 
\\
\leq & ~ |\cost(P_{r} \setminus P_{r}^{*}, C)-\cost(Q_{r}, C)| + |\cost(Q_{r}, C)-\sum_{i} \cost(P_{r, i}, c_{i})| 
\\
\leq & ~ \frac{21}{100} \cost(P_{r}, C)+24 \cost(P_{r}, C^{*})
\end{align*}
where the first step follows from definition of reassign and $\sum_{i=1}^{k} \cost(P_{r, i}, c_{i})=\cost(P_{r}\setminus P_{r}^{*}, C \setminus \{r\})$, the second step follows from inserting $\cost(Q_r, C)$ and the last step follows from Eq.~\eqref{eq:cost_QC_cost_PC} and Eq.~\eqref{eq:cost_QC_cost_sum}.

\end{proof}

\subsection{Good Cluster}
\label{sec:good_cluster}
In this section, we introduce the definition of "good" for an index in $H$ (Definition~\ref{def:reassign}).

\begin{definition}
\label{def:good_first}
We define a good cluster index $h \in H$ to be the one satisfying
\begin{align*}
& ~ \cost(P_{h}^{*}, C)-\mathrm{reassign}(P, C, c_{h})-9 \cost(P_{h}^{*},\{c_{h}^{*}\}) \\
> & ~
\frac{1}{100 k} \cdot \cost(P, C)
\end{align*}
\end{definition}

The following lemma shows a lower bound for the proportion of good clusters under condition $3 \sum_{h \in H} \cost(P_{h}^{*}, C)>\cost(P, C)$.
\begin{lemma}
\label{lem:sample_proportion_first}
Let $c_0 > 300$ be a constant. 
If $3 \sum_{h \in H} \cost(P_{h}^{*}, C)>\cost(P, C) \geq$ $500 \OPT_{k}$, then
\begin{align*}
\sum_{h \in H, h \mathrm{~is~good}} \cost(P_{h}^{*}, C) \geq (\frac{1}{9} - \frac{33}{c_0}) \cdot \cost(P, C)
\end{align*}
\end{lemma}

\begin{proof}
We already have $\sum_{h \in H} \cost(P_{h}^{*}, C) \geq \frac{1}{3} \cost(P, C)$. By the definition of good (Definition~\ref{def:good_first}) and Lemma~\ref{lem:reassign_upper_bound}

\begin{align*}
\sum_{h \in H, h \mathrm {~is~not~good}} \cost(P_{h}^{*}, C) \leq & ~ \sum_{h \in H} \mathrm { reassign }(P, C, c_{h})+9 \OPT_{k}+\frac{1}{100} \cost(P, C) 
\\
\leq & ~ \frac{22}{100} \cost(P, C)+33 \OPT_{k}
\end{align*}
where the first step follows from Definition~\ref{def:good_first} and the last step follows from Lemma~\ref{lem:reassign_upper_bound}.

Using $\cost(P, C) \geq 500 \OPT_{k}$, we obtain that
\begin{align*}
\sum_{h \in H, h \mathrm{~is~not~good}} \cost(P_{h}^{*}, C) \leq \frac{143}{500} \cdot \cost(P, C) .
\end{align*}

To the opposite, we have $\sum_{h \in H, h \mathrm{~is~good}} \cost(P_{h}^{*}, C) \geq \frac{23}{500} \cdot \cost(P, C)$. 

Thus, we complete the proof.
\end{proof}

\subsection{Lower Bound for Cost}
\label{sec:lower_bound_for_cost}
In this section, the following lemma provides a lower bound for the cost related to the mean center (Definition~\ref{def:mean_center})

\begin{lemma}
\label{lem:alpha_mu}
Given point set $Q \subset \mathbb{R}^{d}$, center set $C \subset \mathbb{R}^{d}$ of size $k$ and parameter $\alpha \geq 9$. Let $R \subset Q$ be the subset of $Q$ such that 
$\cost(R, \{\mu(Q)\}) = \frac{2}{|Q|} \cdot \cost(Q,\{\mu(Q)\})$ from $\mu(Q)$. If $\cost(Q, C) \geq \alpha \cdot \cost(Q,\{\mu(Q)\})$, we have
\begin{align*}
\cost(R, C) \geq(\frac{\alpha-1}{8}) \cdot \cost(Q,\{\mu(Q)\}),
\end{align*}
\end{lemma}

\begin{proof}
Lemma~\ref{lem:cost_mu} implies that the closest center in $C$ to $\mu(Q)$ has squared distance at least $\frac{\alpha-1}{|Q|} \cdot \cost(Q,\{\mu(Q)\})$. Hence, the squared distance of points in $R$ to $C$ is at least
\begin{align*}
(\sqrt{\alpha-1}-\sqrt{2})^{2} \cdot \cost(Q,\{\mu(Q)\}) /|Q| \geq (\alpha-1) / 4 \cdot \cost(Q,\{\mu(Q)\}) /|Q|
\end{align*}
where we use that $\alpha \geq 9$ so that $\sqrt{\alpha-1} \geq 2 \sqrt{2}$. By taking average, we get $|R| \geq|Q| / 2$. With the inequality above we obtain the result.
\end{proof}

\subsection{Good Cluster for General Cluster}
\label{sec:good_cluster_for_general_cluster}
In this section, we introduce definition of "good" for general cluster index.

\begin{definition}
\label{def:good_second}
Let $L$ be the subset of cluster centers $C=\{c_1, \cdots, c_k\}$ that doesn't capture (Definition~\ref{def:capture}) any optimal centers. For a general cluster index $i \in\{1, \cdots, k\}$, we say it is good if there exists a center $\ell \in L$ such that
\begin{align*}
& ~ \cost(P_{i}^{*}, C)-\mathrm{reassign}(P, C, \ell)-9 \cost(P_{i}^{*},\{c_{i}^{*}\}) \\
> & ~ 
\frac{1}{100 k} \cdot \cost(P, C)
\end{align*}
\end{definition}

We prove a lower bound similar to Lemma~\ref{lem:sample_proportion_first} in the case $3 \sum_{h \in H} \cost(P_{h}^{*}, C) \leq \cost(P, C)$.

\begin{lemma}
\label{lem:sample_proportion_second}
Let $c_0 > 300$ denote a constant. 
If $3 \sum_{h \in H} \cost(P_{h}^{*}, C) \leq \cost(P, C)$ and $\cost(P, C) \geq 500 \OPT_k$ we have
\begin{align*}
\sum_{r \in R, r \text { is good }} \cost(P_{r}^{*}, C) \geq (\frac{71}{300} - \frac{57}{c_0}) \cost(P, C) .    
\end{align*}
\end{lemma}
\begin{proof}
We already have $\sum_{r \in R} \cost(P_{r}^{*}, C) \geq 2 / 3 \cost(P, C)$. Note that $|R| \leq 2|L|$. By the definition of good (Definition~\ref{def:good_second}) and Lemma~\ref{lem:reassign_upper_bound} 
\begin{align*}
& ~ \sum_{r \in R, r \mathrm{~is~not~good}} \cost(P_r^*, C)
\\
\leq & ~ 2|L| \min _{\ell \in L} \mathrm{reassign}(P, C, \ell)+9 \OPT_{k}+\frac{1}{100} \cost(P, C) 
\\
\leq & ~ 2 \sum_{\ell \in L} \mathrm{ reassign }(P, C, \ell)+9 \OPT_{k}+\frac{1}{100} \cost(P, C) 
\\
\leq & ~ \frac{43}{100} \cost(P, C)+57 \OPT_{k} .
\end{align*}
where the first step follows from Definition~\ref{def:good_second}, the second step follows from $|L| \min _{\ell \in L} \mathrm{reassign}(P, C, \ell) \leq \sum_{\ell \in L} \mathrm{ reassign }(P, C, \ell)$ and the last step follows from Lemma~\ref{lem:reassign_upper_bound}.

Using $\sum_{i \in\{1, \cdots, k\}} \cost(P_{i}^{*}, C) \geq 500 \OPT_{k}$, we obtain that
\begin{align*}
\sum_{r \in R, r \mathrm {~is~not~good}} \cost(P_{r}^{*}, C) \leq \frac{11}{20} \cost(P, C)
\end{align*}
The bound follows from combining the previous inequality with $\sum_{r \in R} \cost(P_{r}^{*}, C) \geq 2 / 3 \cost(P, C)$
\end{proof}

\subsection{Cost Reduction}
\label{sec:cost_reduction}

Now, we can prove Lemma~\ref{lem:cost_decrease}. It suffices to show that using \textsc{DistanceOracle}, we can control the sampling error and give a constant lower bound for sampling probability.

\begin{lemma}\label{lem:cost_decrease}
Let $c_0 > 300$ denote a constant. 
Let $P$ be a point set in $\R^d$. Let $C$ center set satisfying  
$\cost(P, C)> c_0 \OPT_k$.  
Let $C'= \textsc{LocalSearch++}(P, C)$. With probability $\frac{1}{900}(1 - \delta)$, we have $\cost(P, C') \leq(1-\frac{1}{100 k}) \cost(P, C)$.
\end{lemma}

\begin{proof}
Let $R_h^*$ be the set $R$ defined in Lemma~\ref{lem:alpha_mu}. For the case $\sum_{h \in H} \cost(P_{h}^{*}, C) \geq \frac{1}{3} \cost(P, C)$, we conclude that 
\begin{align*}
    & ~ \sum_{h \in H, h \text { is good }} \cost(R_{h}^{*}, C) 
    \\
    \geq & ~ \frac{1}{9} \cdot (\frac{1}{9} - \frac{33}{c_0}) \cost(P, C) 
    \\
    \geq & ~ \frac{1 - \epsilon}{1 + \epsilon} (\frac{1}{81} - \frac{11}{3 c_0}) \cost(P, C) 
    \\
    \geq & ~ (\frac{1}{162} - \frac{11}{6 c_0}) \cost(P, C) 
\end{align*}
where the first step follows from Lemma~\ref{lem:sample_proportion_first} and Lemma~\ref{lem:alpha_mu}, the second step follows from  \textsc{Cost} part of Theorem~\ref{thm:distance_oracle} and the last step follows from $\frac{1 - \epsilon}{1 + \epsilon} \leq \frac{1}{2}$ when $\epsilon \in (0, 0.1)$,with probability at least $1-\delta$, $\delta \in (0, 0.1)$.

With probability $1 - \delta$, we obtain lower bound $\frac{1}{162} - \frac{11}{6c_0}$. Conditioned on this event happening, we have probability at least $\frac{1}{162} - \frac{11}{6c_0}$ to ensure the following analysis. The whole probability is $(\frac{1}{162} - \frac{11}{6c_0})(1 - \delta)$. Thus, we can sample a point from $\cup_{h \in H, h \mathrm{~is~good}}R_{h}^*$ with probability no less than $(\frac{1}{162} - \frac{11}{6 c_0})(1 - \delta)$.

 By Definition~\ref{def:good_first}, when we sample a point in $R_{h}^{*}$, we can replace it with $c_{h}$ to get an upper bound on the cost.
\begin{align*}
    & ~ \cost(P, C \setminus \{c_{h}\} \cup\{c\}) 
    \\
    \leq & ~ \cost(P, C)-\cost(P_{h}^{*}, C) \\
    & ~ +\mathrm{reassign}(P, C,\{c_{h}\})+\cost(P_{h}^{*},\{c\})
\end{align*}
which follows from Definition~\ref{def:good_first}.

By Lemma~\ref{lem:cost_mu} we have $\cost(P_{h}^{*},\{c\}) \leq 9 \cost(P_{h}^{*},\{c_{h}^{*}\})$. Thus, with probability at least $(\frac{1}{162} - \frac{11}{6 c_0})(1 - \delta)$, we have
\begin{align*}
& ~ \cost(P, C)-(\cost(P_{h}^{*}, C)\\
& ~ -\mathrm{reassign}(P, H, c_{h})-9 \cost(P_{h}^{*},\{c_{h}^{*}\}) 
\\
\leq & ~ (1-\frac{1}{100 k}) \cdot \cost(P, C)
\end{align*}

For the case $\sum_{h \in H} \cost(P_{h}^{*}, C) \leq \frac{1}{3} \cost(P, C)$, Lemma~\ref{lem:sample_proportion_second} gives a similar lower bound as Lemma~\ref{lem:sample_proportion_first}. Using a similar technique as above, we conclude the same result for this case.

Thus, we complete the proof.

\end{proof}

\subsection{Correctness of \textsc{FastKMeans++}}
\label{sec:correctness_k_means}
In this section, we show the correctness of our \textsc{FastKMeans++} algorithm using Lemma~\ref{lem:cost_decrease}. We remark that as long as we prove Lemma~\ref{lem:cost_decrease}, then we can achieve the following result  

\begin{lemma}[Correctness of \textsc{FastKMeans++}, correctness part of Theorem~\ref{thm:k_means_formal}
]\label{lem:correctness_k_means}
 
Given input data set $P \subset \R^d$, we set $Z = O(k \log \log k)$  
and use $C$ to denote the outcome of Algorithm~\ref{alg:k_means}. Let $C^*$ be the set of optimum centers. 
Then we have 
$\E[\cost(P, C)] = O(\cost(P, C^*)).$
\end{lemma}

\begin{proof}
Let $c_1 > 10^4$ be a constant.

Let $\hat{C}$ be the output of the first for-loop in Algorithm~\ref{alg:k_means}. Let $C$ be the final output of Algorithm~\ref{alg:k_means}. By Lemma~\ref{lem:cost_decrease}, conditioned on the fact that the cost of the centers is bigger than $5000 \OPT$ before the first call of LocalSearch++, 
we reduce the cost by a $(1-\frac{1}{100 k})$ multiplicative factor with probability $\frac{1}{1000}$.

We construct an auxiliary random process $X$ to model the evolution of our algorithm. It starts with initial value $\cost(P, \hat{C})$. Then, it goes through $Z=$ $10^5 k \log \log k$ iterations and reduces its value by a $(1-\frac{1}{100 k})$ multiplicative factor with probability $1 / 1000$ per iteration.

At the end of the process, it increases its value by adding quantity  
$500 \OPT_{k}$.

We note that the value of $X$ after $Z$ iterations stochastically dominates the cost of the clustering. Thus, we have

\begin{align*}
\E[X]= & ~ 500 \OPT_{k}+\cost(P, \hat{C}) \cdot \sum_{i=0}^{Z}{Z \choose i } (\frac{1}{1000})^{i} (\frac{999}{1000})^{Z-i}(1-\frac{1}{100 k})^{i}
\\
= & ~ \cost(P, \hat{C})(1-\frac{1}{10^5 k})^{10^5 k \log \log k} + 500 \OPT_{k} 
\\
\leq & ~ \frac{\cost(P, \hat{C})}{\log k}+500 \OPT_{k}
\end{align*}

where the first step follows from the expectation of binomial distribution, the second step follows from combining $(\frac{1}{1000})^i$ and $(1 - \frac{1}{100k})^i$ and the last step follows from $(1 - \frac{1}{n})^n \leq \frac{1}{e}$ when $n > 10^5$.

This implies that $\E[\cost(P, C) ~|~ \hat{C}] \leq \frac{\cost(P, \hat{C})}{\log k}+5000 \OPT_{k}$. In the meanwhile,

\begin{align}\label{eq:condition_expect_upper_bound}
\E[\cost(P, C)] = & ~ \sum_{\hat{C}} \E[\cost(P, C) ~|~ \hat{C}] P(\hat{C})\notag
\\
= & ~ \sum_{\hat{C}} P(\hat{C})(\frac{\cost(P, \hat{C})}{\log k}+500 \OPT_{k})\notag
\\
= & ~ \frac{\E[\cost(P, \hat{C})]}{\log k}+500 \OPT_{k}
\end{align}
where the first step follows from the definition of conditional expectation, the second step follows from Eq.~\eqref{eq:condition_expect_upper_bound} and the last step follows from $\sum_{\hat{C}}P(\hat{C}) = 1$.

Now the theorem follows from the following results in \cite{av06}
\begin{align*} 
\E[\cost(P, \hat{C})] \leq (8 \log k+2) \OPT_{k}.
\end{align*}

Hence,
\begin{align*} 
\E[\cost(P, C)] \leq 509 \OPT_{k}.
\end{align*}
\end{proof}

%% file: app_exp.tex
\section{More Experiments}\label{sec:app_exp}

Here we provide the links for the four real datasets.
\begin{itemize}
    \item SCADI Data Set \cite{zbd18,bz19} \footnote{\href{https://archive.ics.uci.edu/ml/datasets/SCADI}{https://archive.ics.uci.edu/ml/datasets/SCADI}} 
    \item Mturk User-Perceived Clusters over Images Data set \cite{clg+16}\footnote{\href{https://archive.ics.uci.edu/ml/datasets/Mturk+User-Perceived+Clusters+over+Images}{https://archive.ics.uci.edu/ml/datasets/Mturk+User-Perceived+Clusters+over+Images}} 
    \item Libras Movement Data set \cite{dmr+09}\footnote{\href{https://archive.ics.uci.edu/ml/datasets/Libras+Movement}{https://archive.ics.uci.edu/ml/datasets/Libras+Movement}} 
   \item Sales Transactions Dataset Weekly Data set \cite{ts14}\footnote{\href{https://archive.ics.uci.edu/ml/datasets/Sales\_Transactions\_Dataset\_Weekly}{https://archive.ics.uci.edu/ml/datasets/Sales\_Transactions\_Dataset\_Weekly}} 
\end{itemize}

Let $n$ be the number of points in the point set. Let $d$ denote the dimension of each node. We use $m$ to denote the dimension of each node after we process them with a sketching matrix. Let $k$ be the number of clusters and centers. We provide more results of experiments to show the influence of $n$, $d$, $m$, and $k$ in this part.
    
\begin{figure*}[!ht]
    \centering

    \subfloat[$d$ is changed]{ \includegraphics[width=0.3\textwidth]{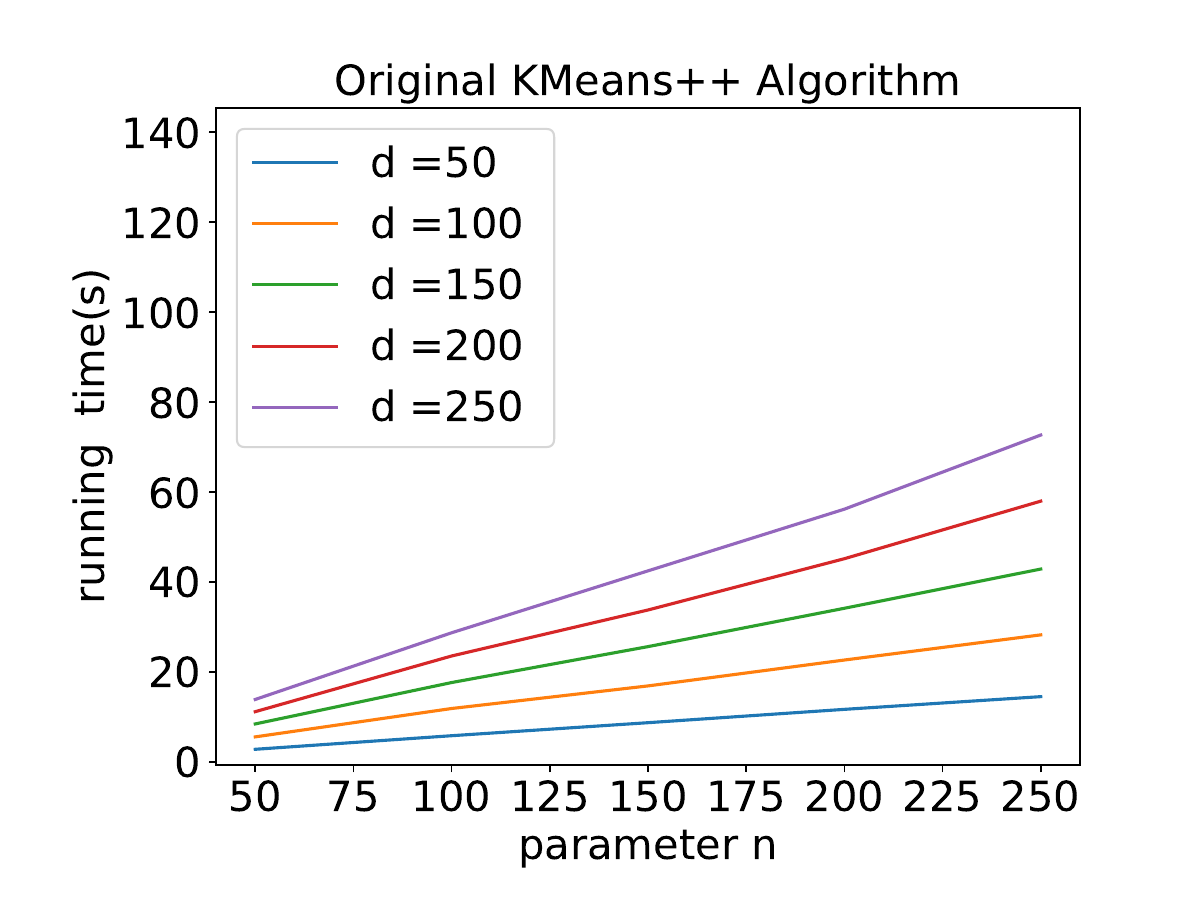}
   \label{fig:x_param_n_variable_d_original_total_time}
   } 
   \subfloat[$m$ is changed]{
    \includegraphics[width=0.3\textwidth]{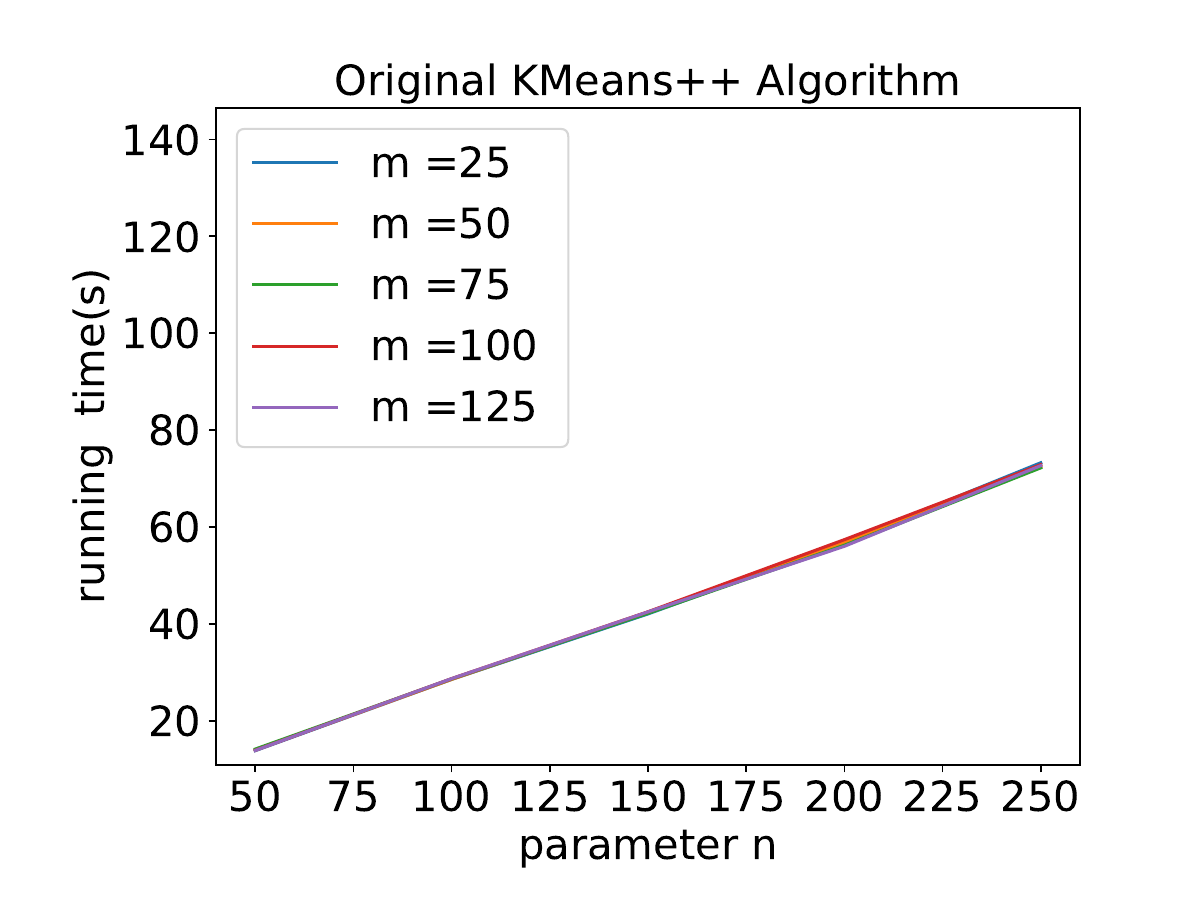}
   \label{fig:x_param_n_variable_m_original_total_time}
}
    \subfloat[$k$ is changed]{   
    \includegraphics[width=0.3\textwidth]{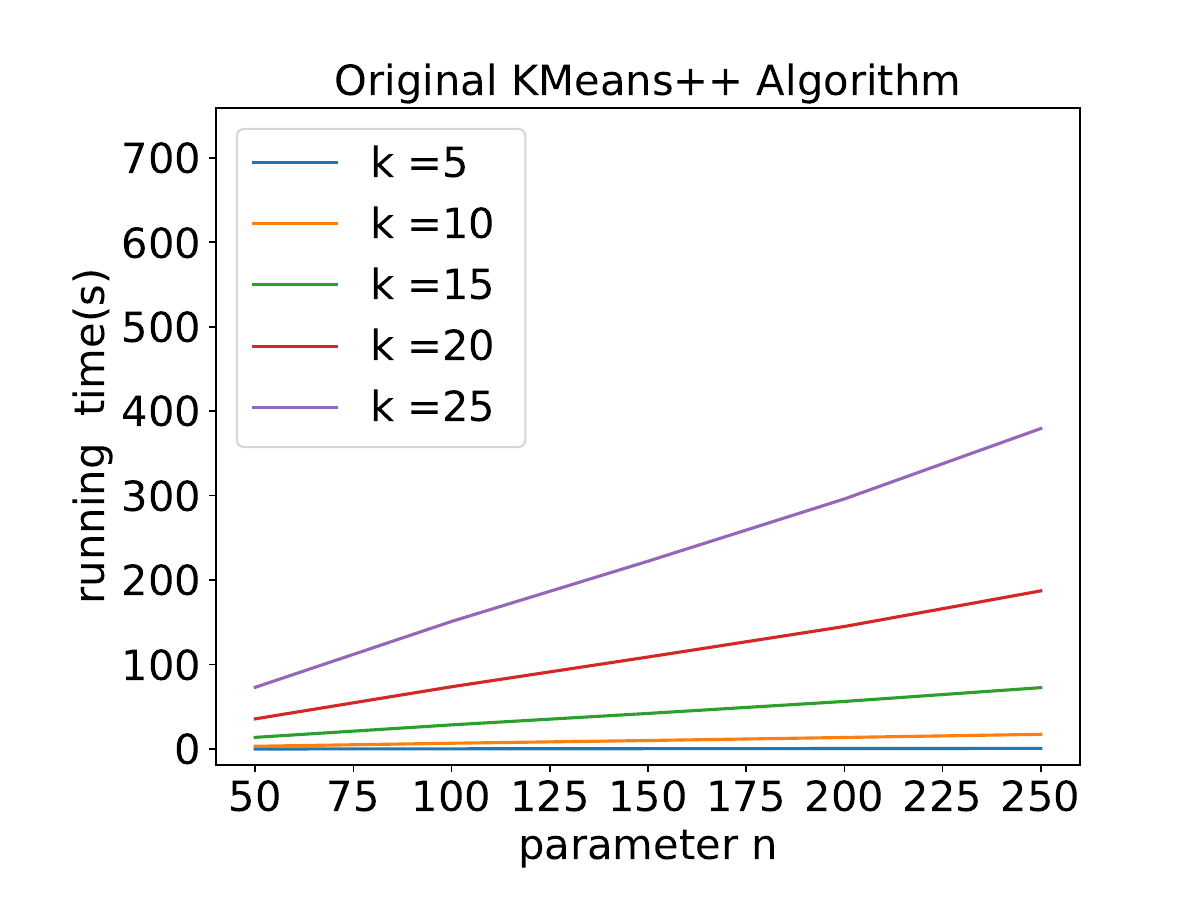}
   \label{fig:x_param_n_variable_k_original_total_time}
}
    
    \caption{The relationship between running time of original $k$-means++ algorithm and parameter $n$}
    \label{fig:the_relationship_between_running_time_of_original_algorithm_and_parameter_n}
\end{figure*}

\begin{figure*}[!ht]
    \centering

    \subfloat[$d$ is changed]{ \includegraphics[width=0.3\textwidth]{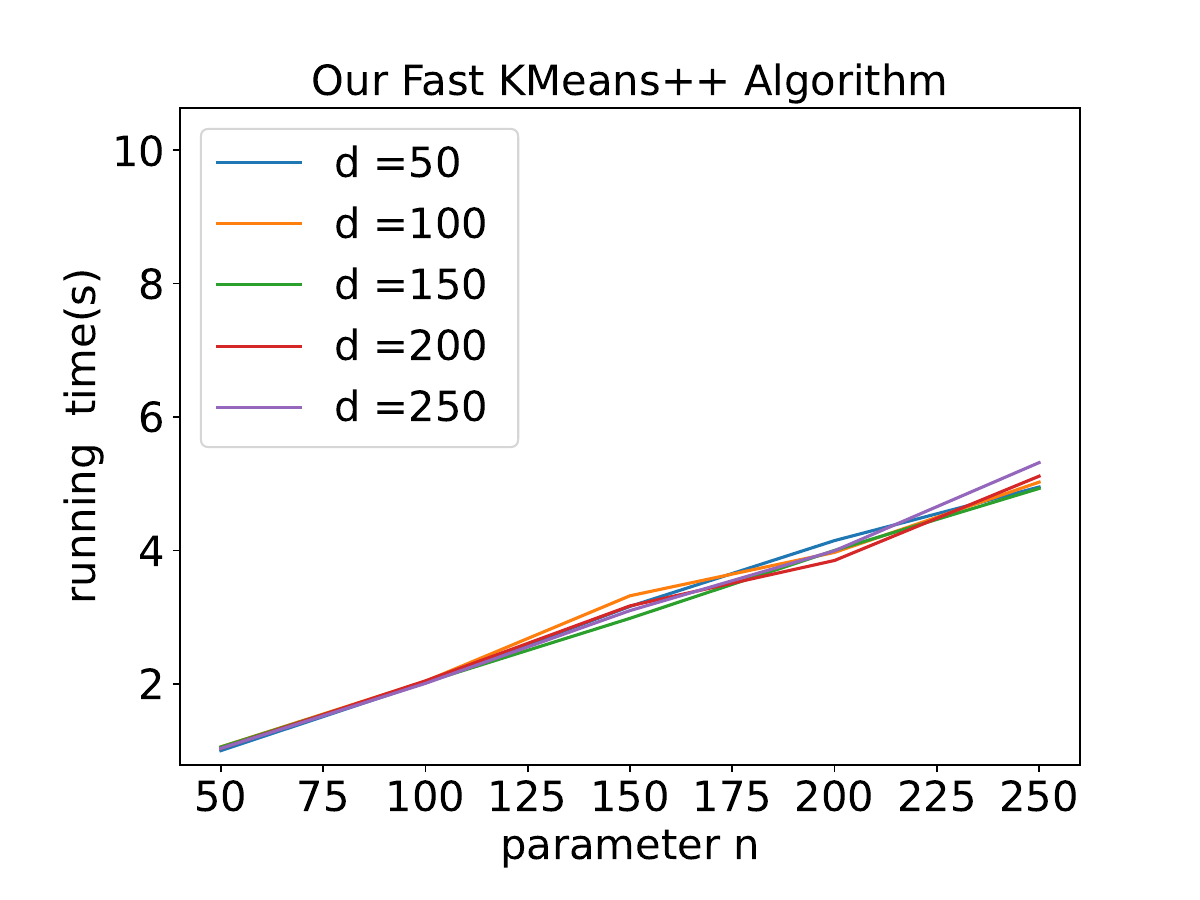}
   \label{fig:x_param_n_variable_d_our_total_time}
   } 
   \subfloat[$m$ is changed]{
    \includegraphics[width=0.3\textwidth]{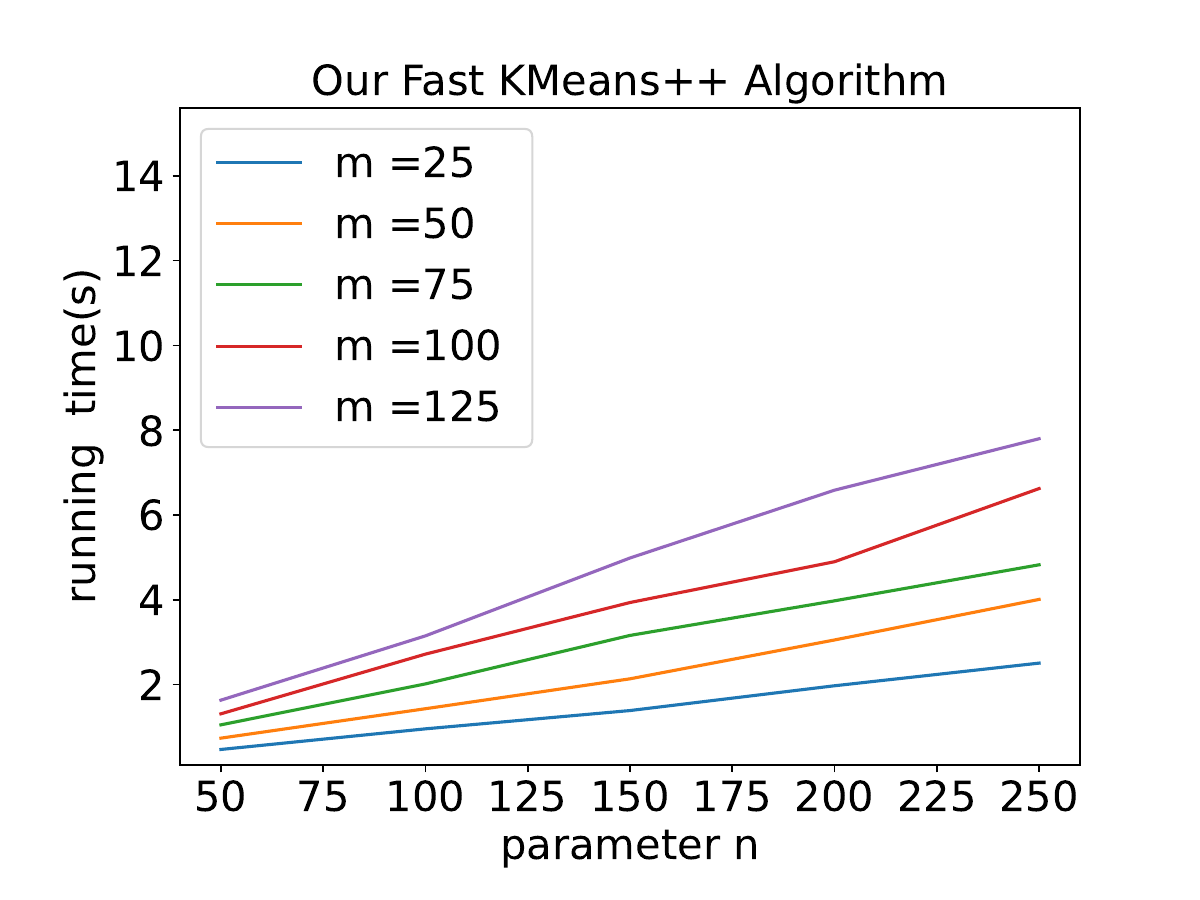}
   \label{fig:x_param_n_variable_m_our_total_time}
}
    \subfloat[$k$ is changed]{   
    \includegraphics[width=0.3\textwidth]{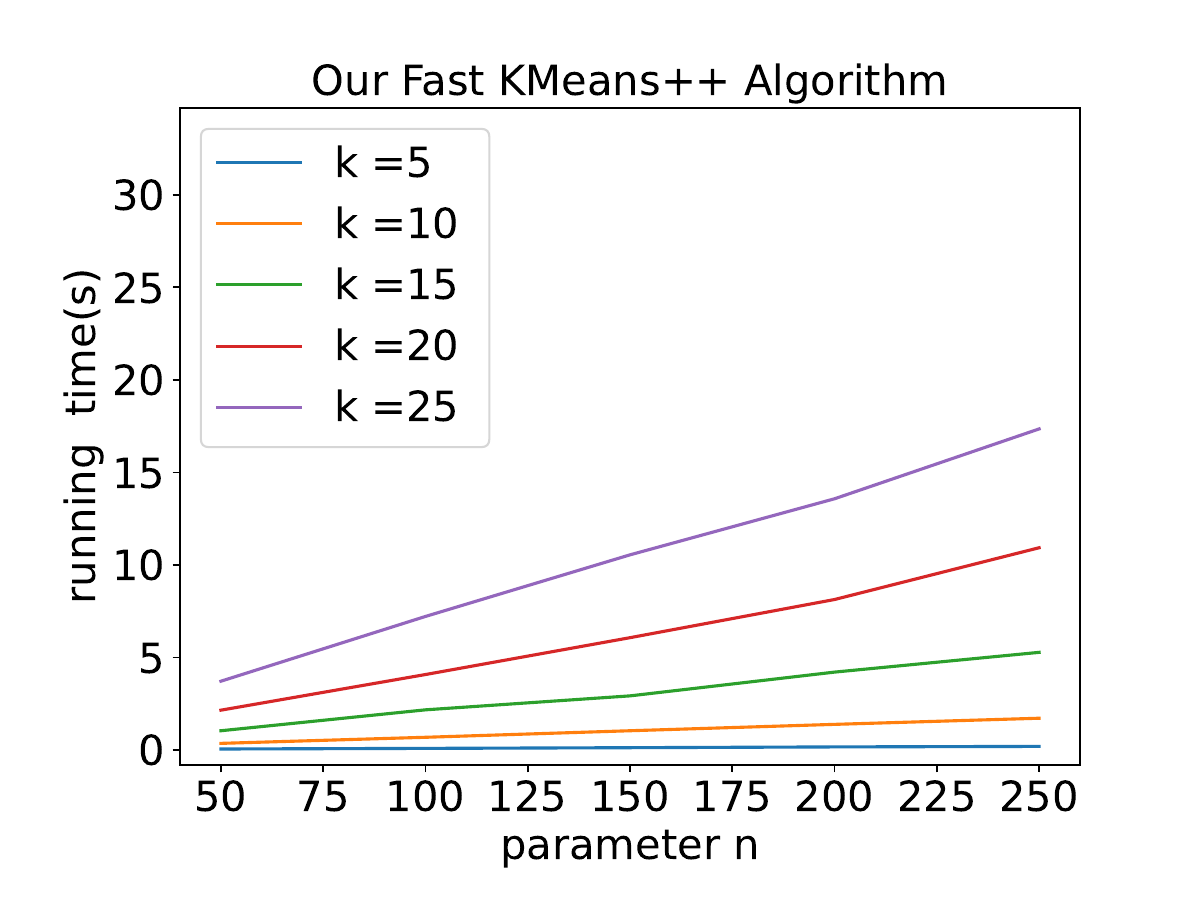}
   \label{fig:x_param_n_variable_k_our_total_time}
}
    
    \caption{The relationship between running time of our \textsc{FastKMeans++} algorithm and parameter $n$}
    \label{fig:the_relationship_between_running_time_of_our_algorithm_and_parameter_n}
\end{figure*}

    To evaluate if the influence of parameter $n$ will be affected by other parameters, we set different values for other parameters and run the original $k$-means++ algorithm and our \textsc{FastKMeans++} algorithm. Fig.~\ref{fig:the_relationship_between_running_time_of_original_algorithm_and_parameter_n} and Fig.~\ref{fig:the_relationship_between_running_time_of_our_algorithm_and_parameter_n} show that the running time of original $k$-means++ and \textsc{FastKMeans++} increase linearly as $n$ increases no matter what the value of other parameters is. 
    
\begin{figure*}[!ht]
    \centering

    \subfloat[$n$ is changed]{ \includegraphics[width=0.3\textwidth]{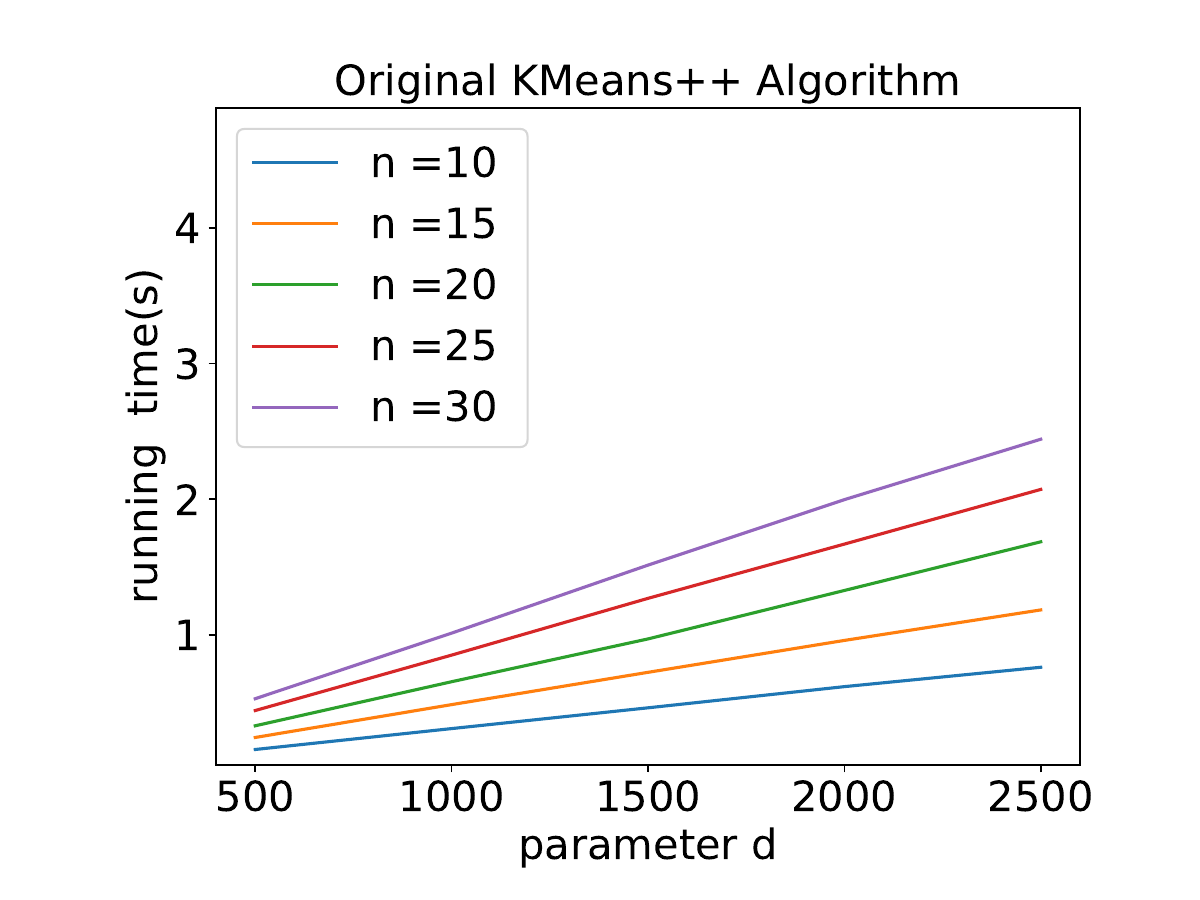}
   \label{fig:x_param_d_variable_n_original_total_time}
   } 
   \subfloat[$m$ is changed]{
    \includegraphics[width=0.3\textwidth]{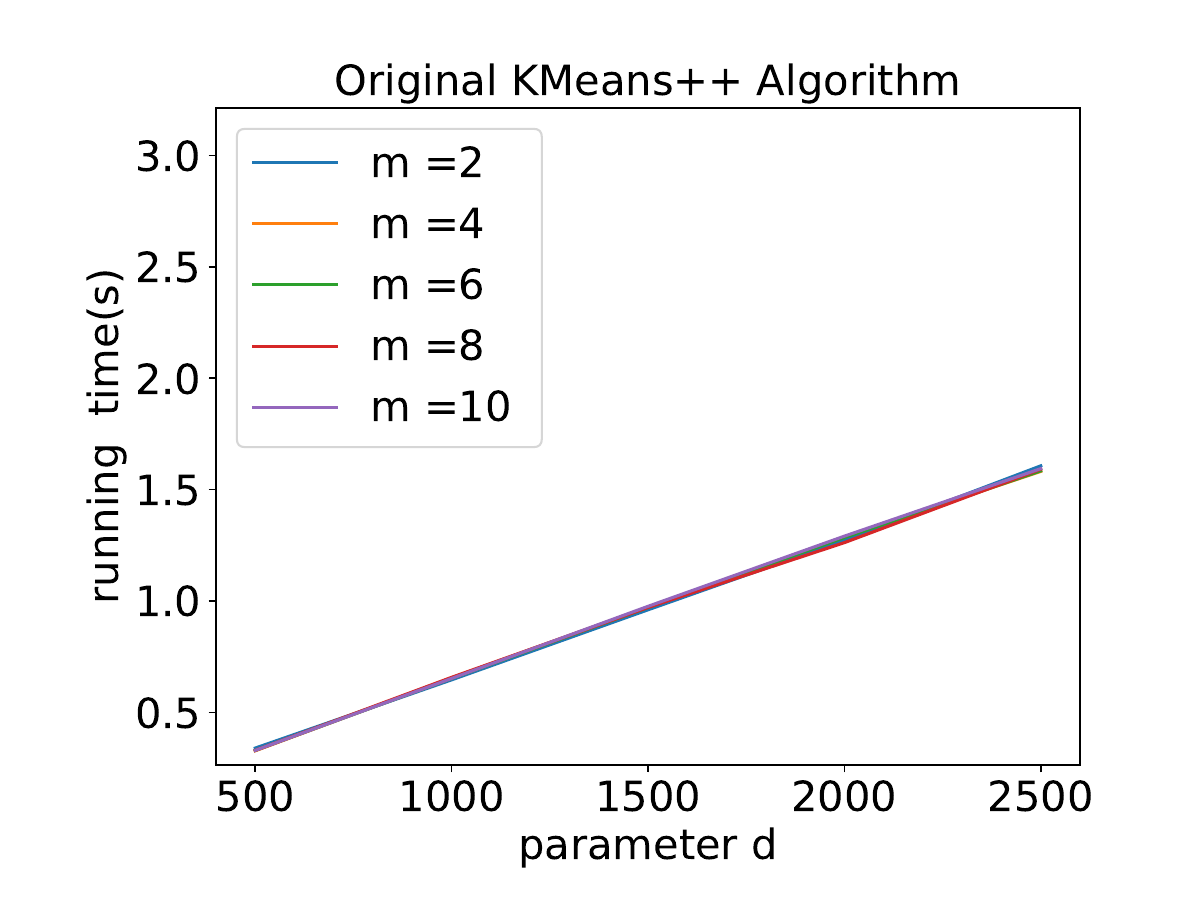}
   \label{fig:x_param_d_variable_m_original_total_time}
}
    \subfloat[$k$ is changed]{   
    \includegraphics[width=0.3\textwidth]{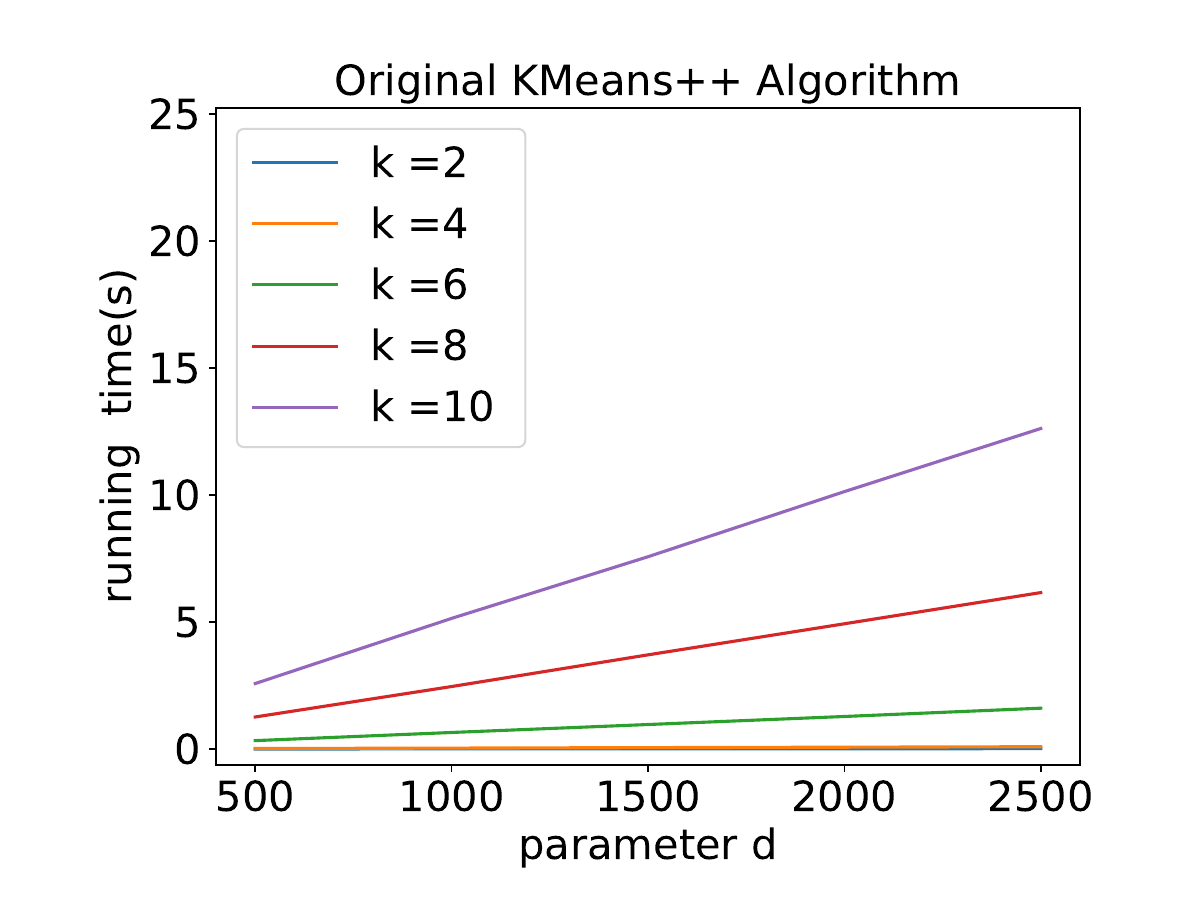}
   \label{fig:x_param_d_variable_k_original_total_time}
}
    
    \caption{The relationship between running time of original $k$-means++ algorithm and parameter $d$}
    \label{fig:the_relationship_between_running_time_of_original_algorithm_and_parameter_d}
\end{figure*}

\begin{figure*}[!ht]
    \centering

    \subfloat[$n$ is changed]{ \includegraphics[width=0.3\textwidth]{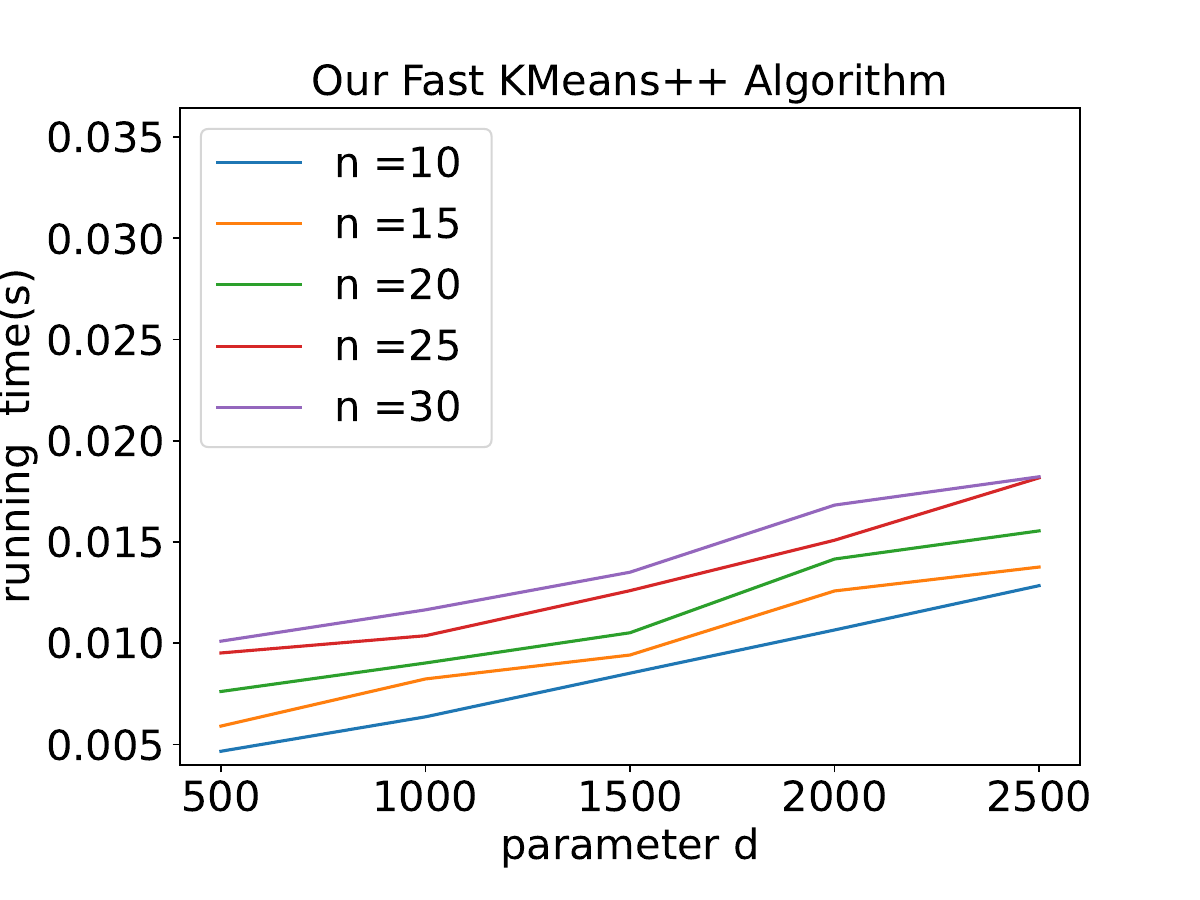}
   \label{fig:x_param_d_variable_n_our_total_time}
   } 
   \subfloat[$m$ is changed]{
    \includegraphics[width=0.3\textwidth]{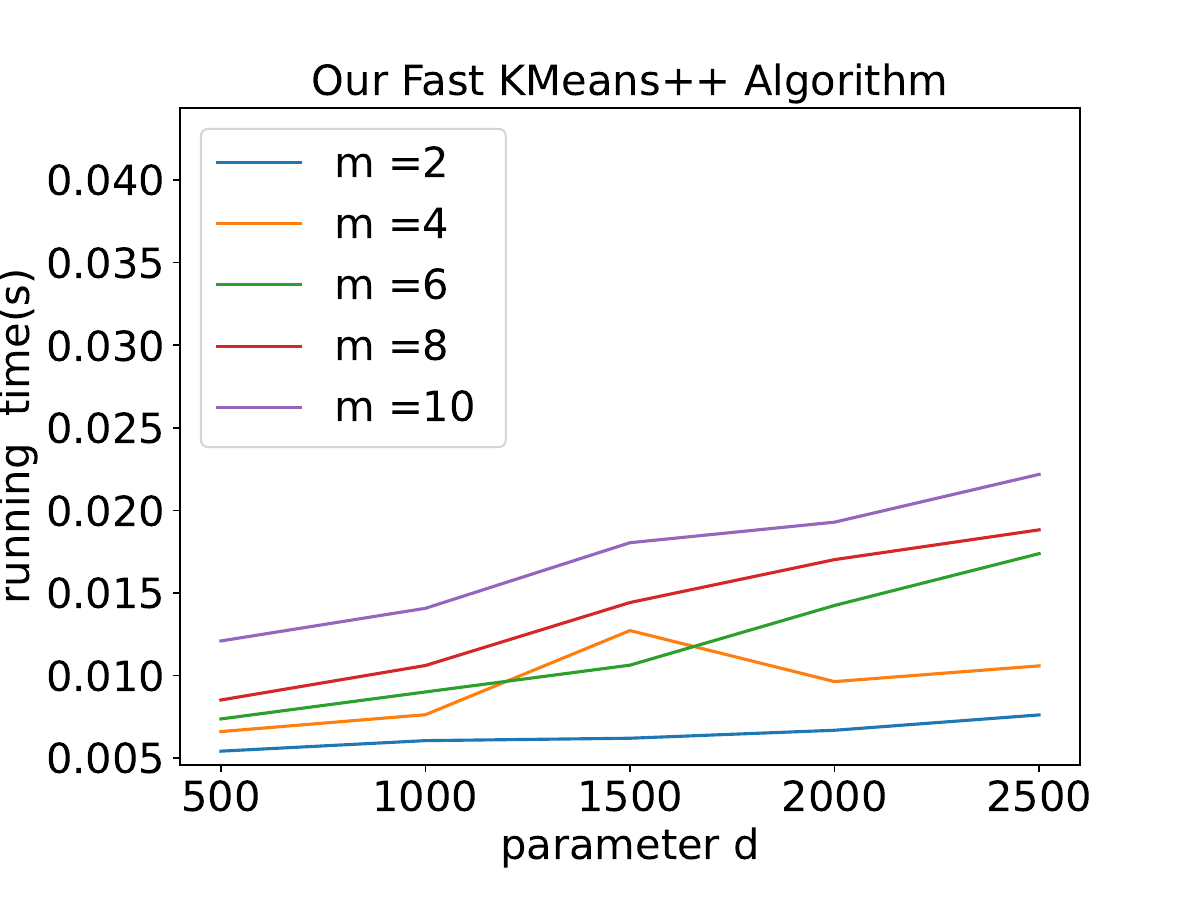}
   \label{fig:x_param_d_variable_m_our_total_time}
}
    \subfloat[$k$ is changed]{   
    \includegraphics[width=0.3\textwidth]{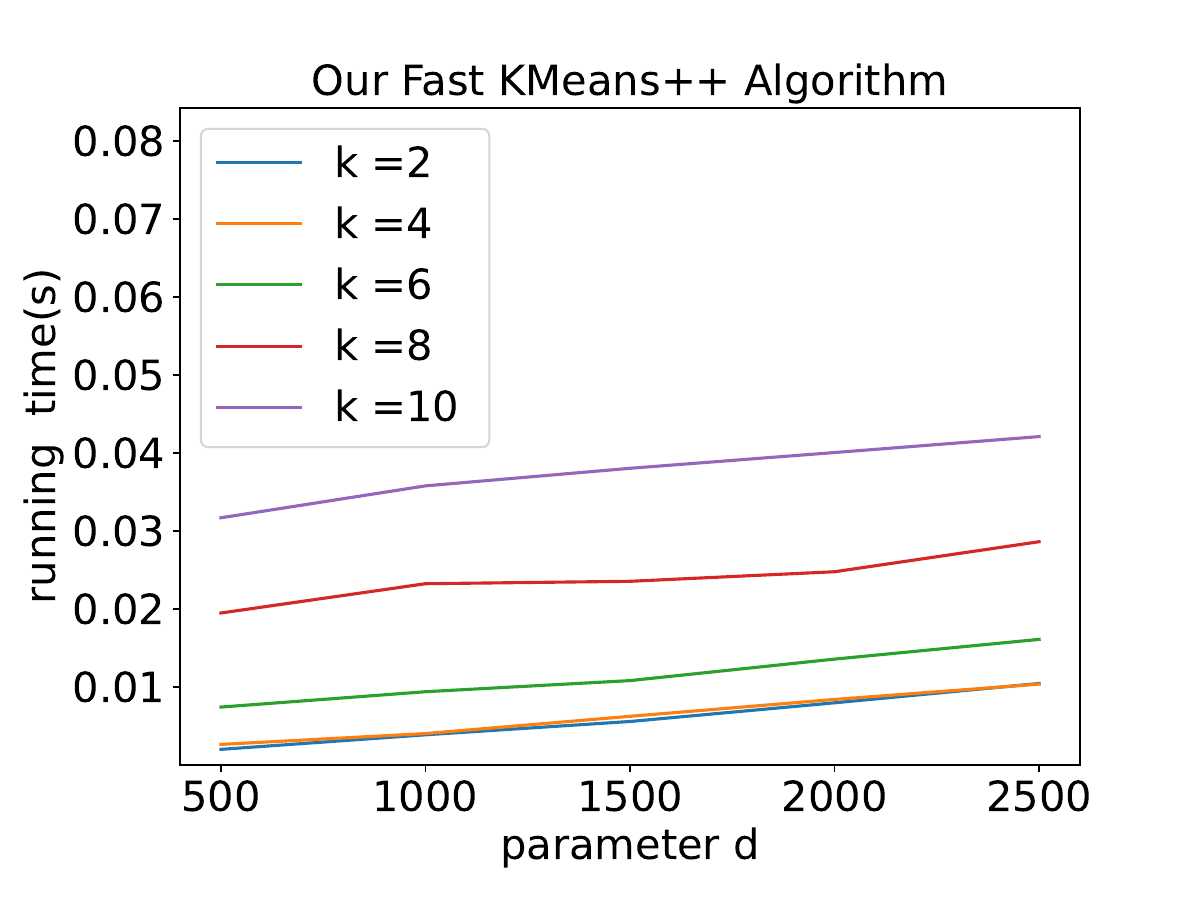}
   \label{fig:x_param_d_variable_k_our_total_time}
}
    
    \caption{The relationship between running time of our \textsc{FastKMeans++} algorithm and parameter $d$}
    \label{fig:the_relationship_between_running_time_of_our_algorithm_and_parameter_d}
\end{figure*}

    To evaluate if the influence of parameter $d$ will be affected by other parameters, we set different values for other parameters and run the original $k$-means++ algorithm and our \textsc{FastKMeans++} algorithm. To make the influence of $d$ on our \textsc{FastKMeans++} more clear, I set $d$ large enough compared with $n$, $m$, and $k$. Fig.~\ref{fig:the_relationship_between_running_time_of_original_algorithm_and_parameter_d} and 
    Fig.~\ref{fig:the_relationship_between_running_time_of_our_algorithm_and_parameter_d} show that the running time of original $k$-means++ and \textsc{FastKMeans++} increase linearly as $d$ increases no matter what the value of other parameters is. 
    
\begin{figure*}[!ht]
    \centering

    \subfloat[$n$ is changed]{ \includegraphics[width=0.3\textwidth]{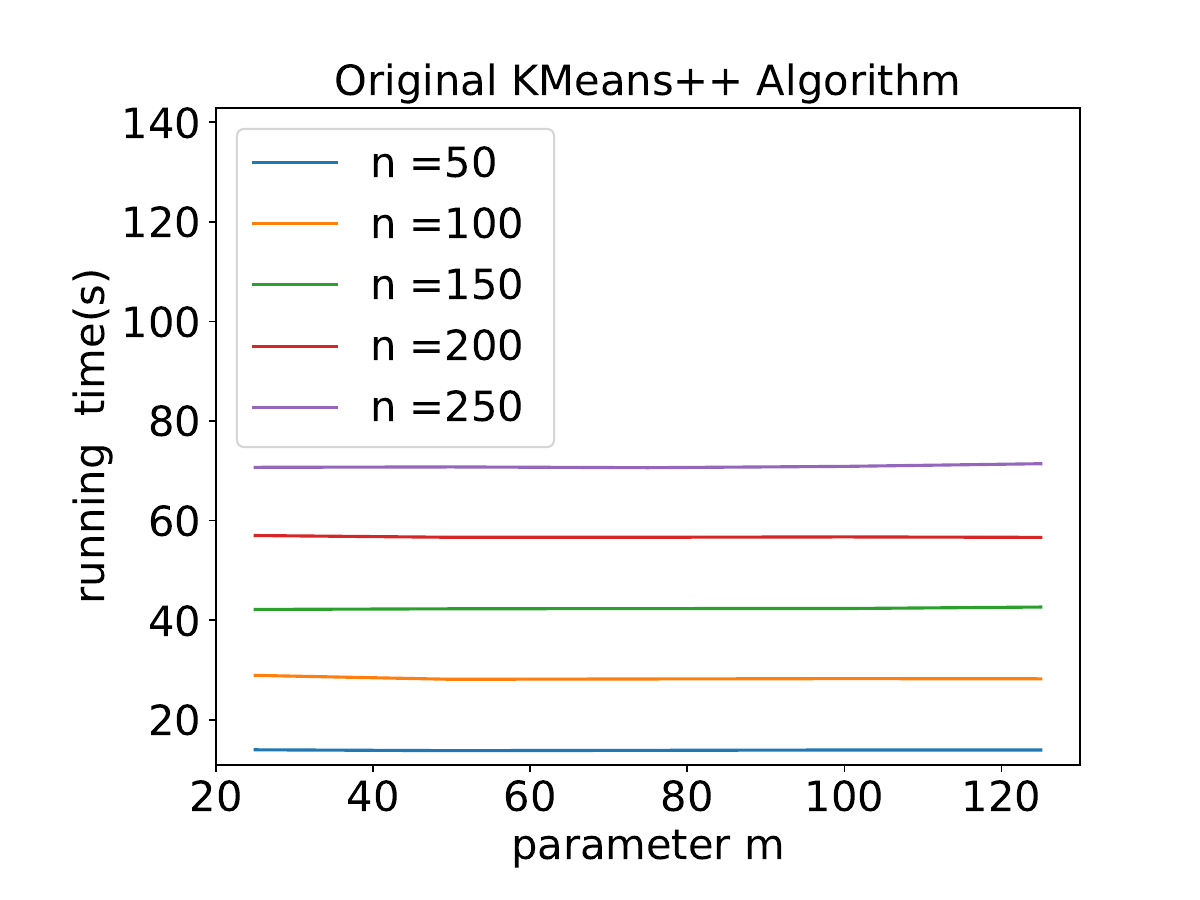}
   \label{fig:x_param_m_variable_n_original_total_time}
   } 
   \subfloat[$d$ is changed]{
    \includegraphics[width=0.3\textwidth]{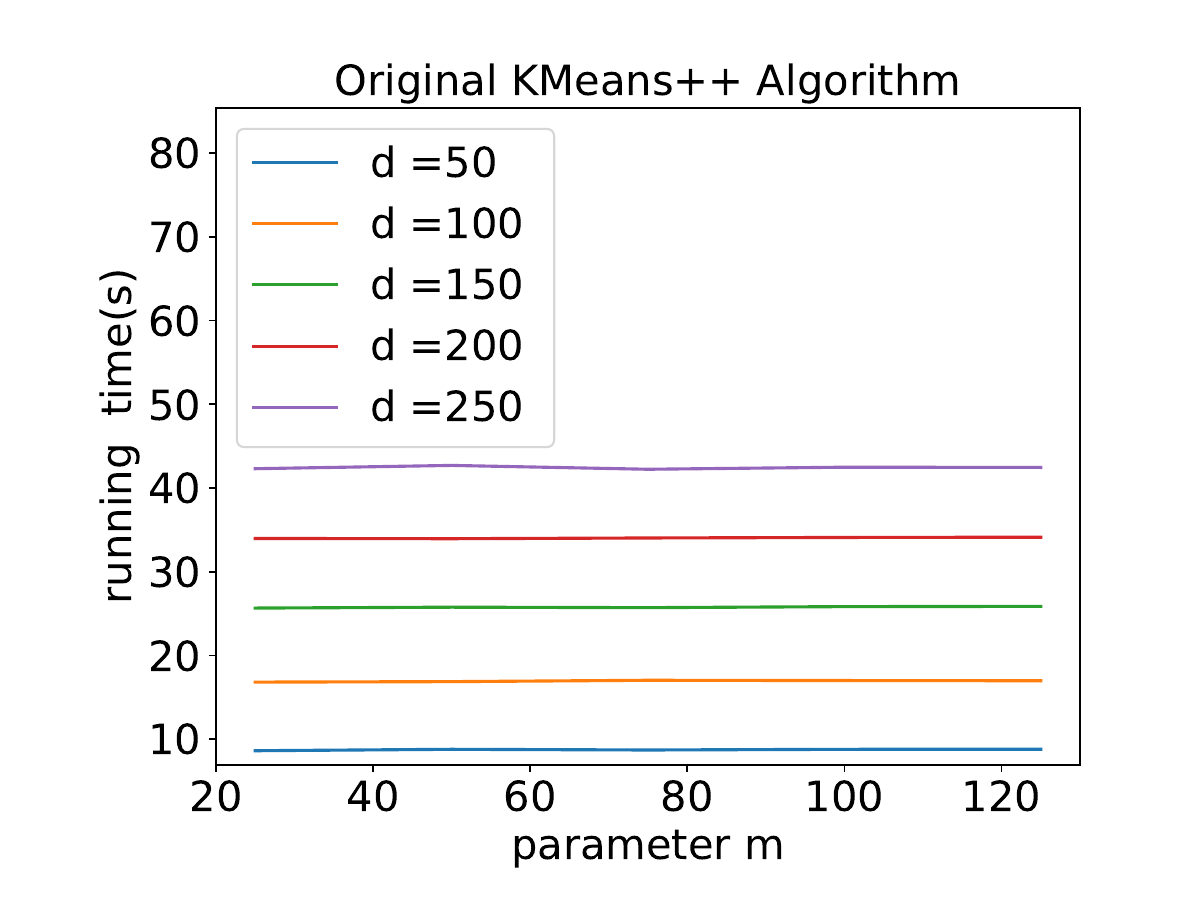}
   \label{fig:x_param_m_variable_d_original_total_time}
}
    \subfloat[$k$ is changed]{   
    \includegraphics[width=0.3\textwidth]{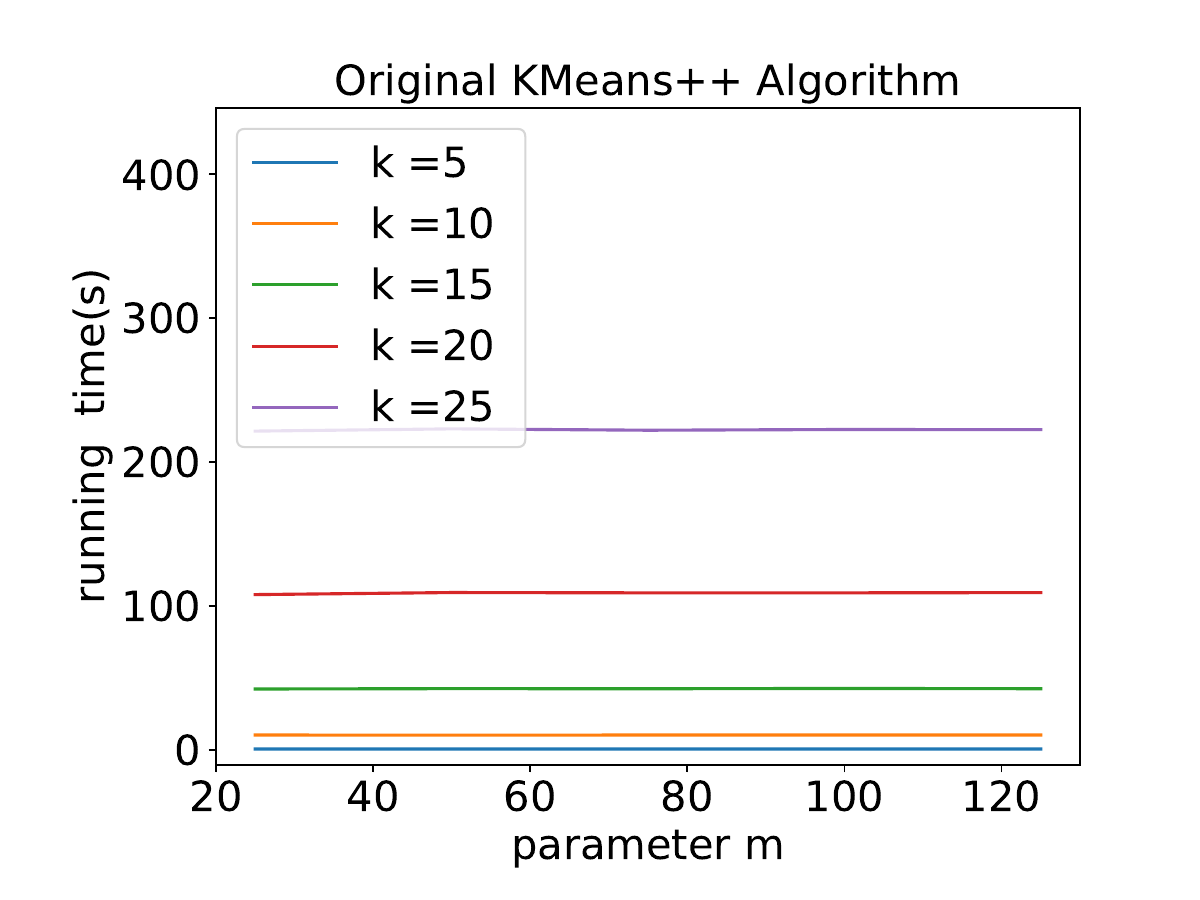}
   \label{fig:x_param_m_variable_k_original_total_time}
}
    
    \caption{The relationship between running time of original $k$-means++ algorithm and parameter $m$}
    \label{fig:the_relationship_between_running_time_of_original_algorithm_and_parameter_m}
\end{figure*}

\begin{figure*}[!ht]
    \centering

    \subfloat[$n$ is changed]{ \includegraphics[width=0.3\textwidth]{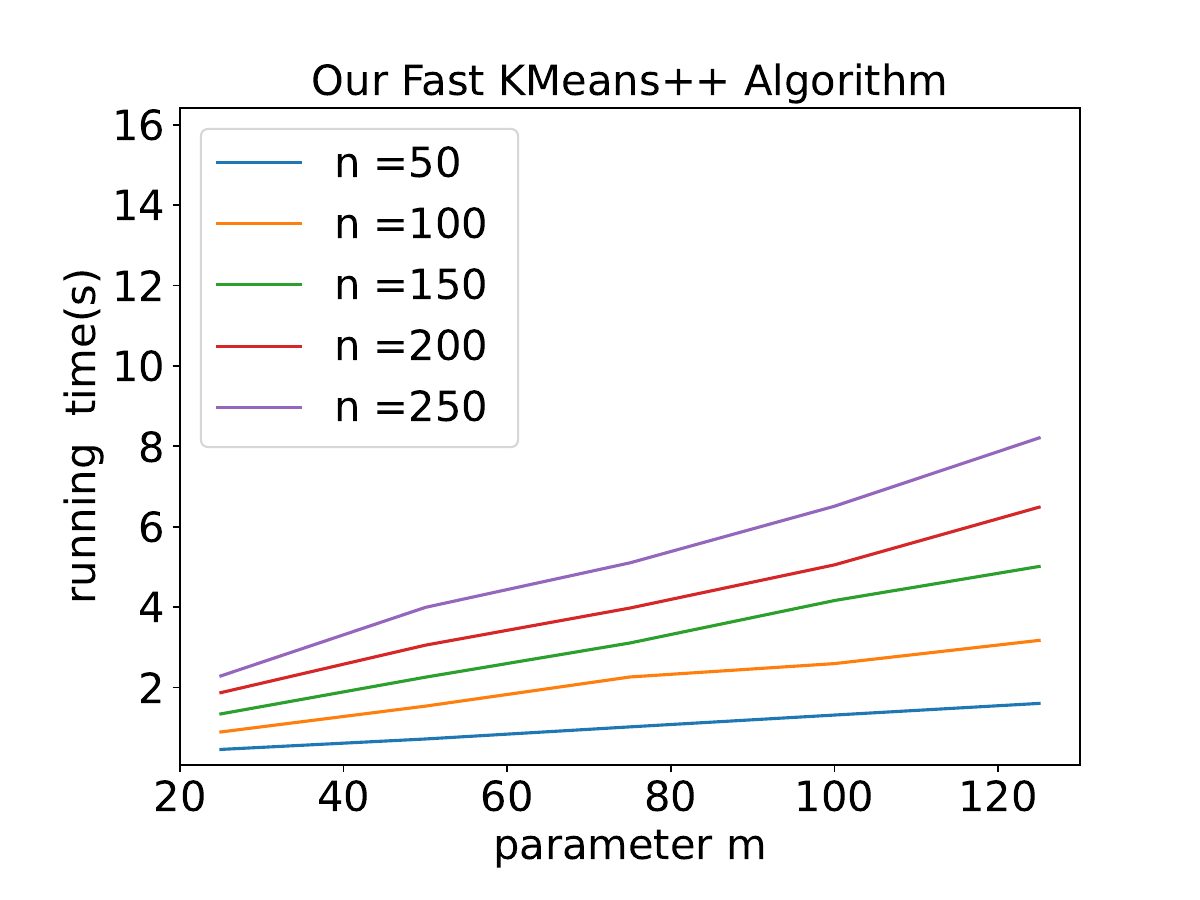}
   \label{fig:x_param_m_variable_n_our_total_time}
   } 
   \subfloat[$d$ is changed]{
    \includegraphics[width=0.3\textwidth]{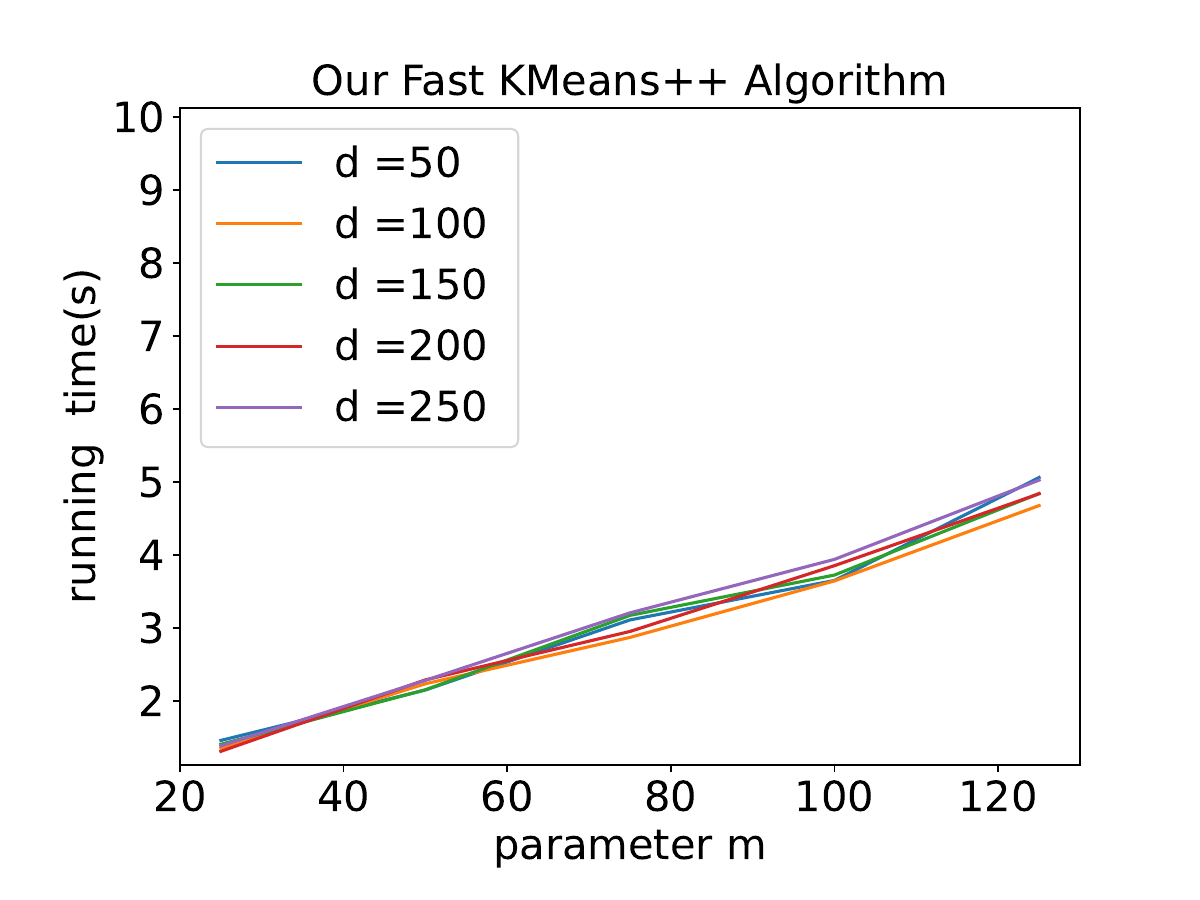}
   \label{fig:x_param_m_variable_d_our_total_time}
}
    \subfloat[$k$ is changed]{   
    \includegraphics[width=0.3\textwidth]{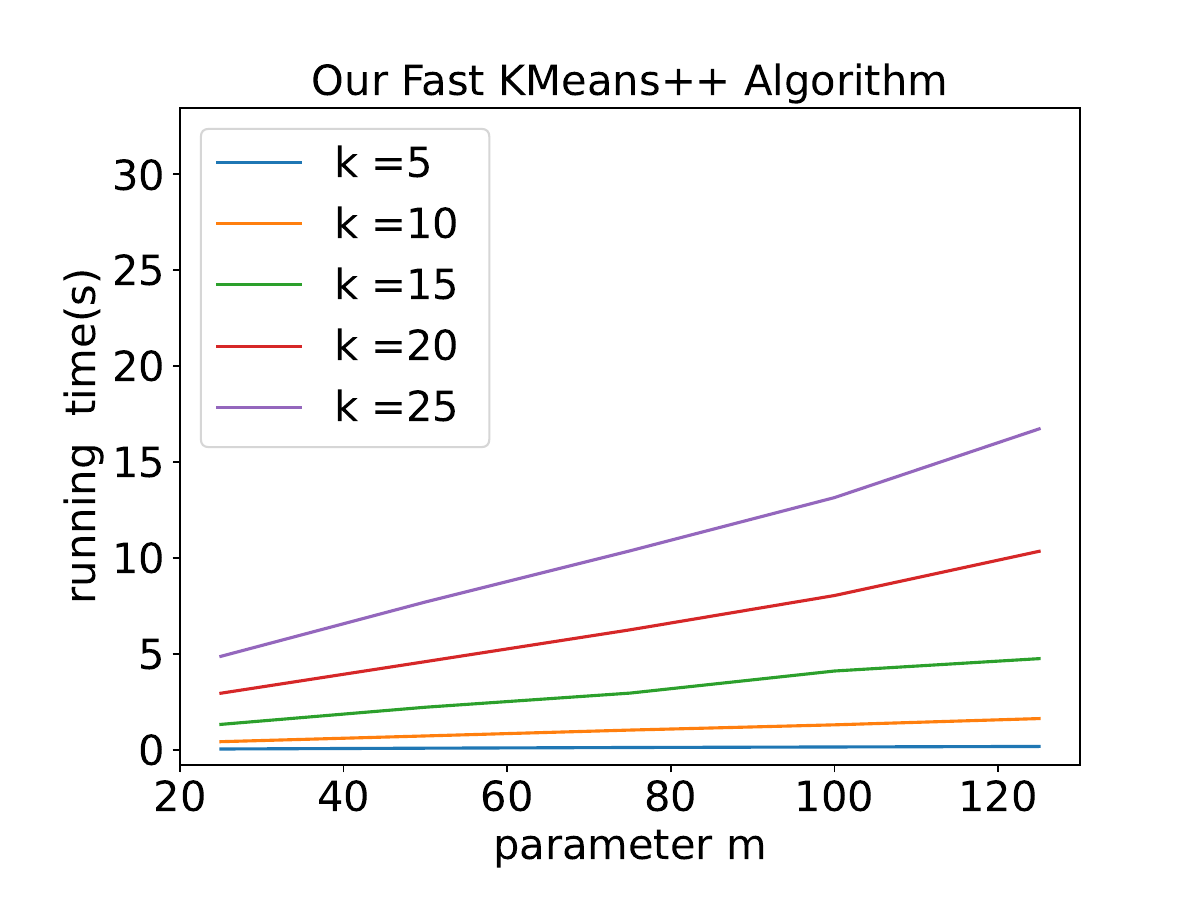}
   \label{fig:x_param_m_variable_k_our_total_time}
}
    
    \caption{The relationship between running time of our \textsc{FastKMeans++} algorithm and parameter $m$}
    \label{fig:the_relationship_between_running_time_of_our_algorithm_and_parameter_m}
\end{figure*}

    To evaluate if the influence of parameter $m$ will be affected by other parameters, we set different values for other parameters and run the original $k$-means++ algorithm and our \textsc{FastKMeans++} algorithm.  Fig.~\ref{fig:the_relationship_between_running_time_of_original_algorithm_and_parameter_m} shows that $m$ is irrelevant to the running time of the original $k$-means++ algorithm. Fig.~\ref{fig:the_relationship_between_running_time_of_our_algorithm_and_parameter_m} shows that the running time of \textsc{FastKMeans++} increases linearly as $m$ increases no matter what the value of other parameters is. 
    
\begin{figure*}[!ht]
    \centering

    \subfloat[$n$ is changed]{ \includegraphics[width=0.3\textwidth]{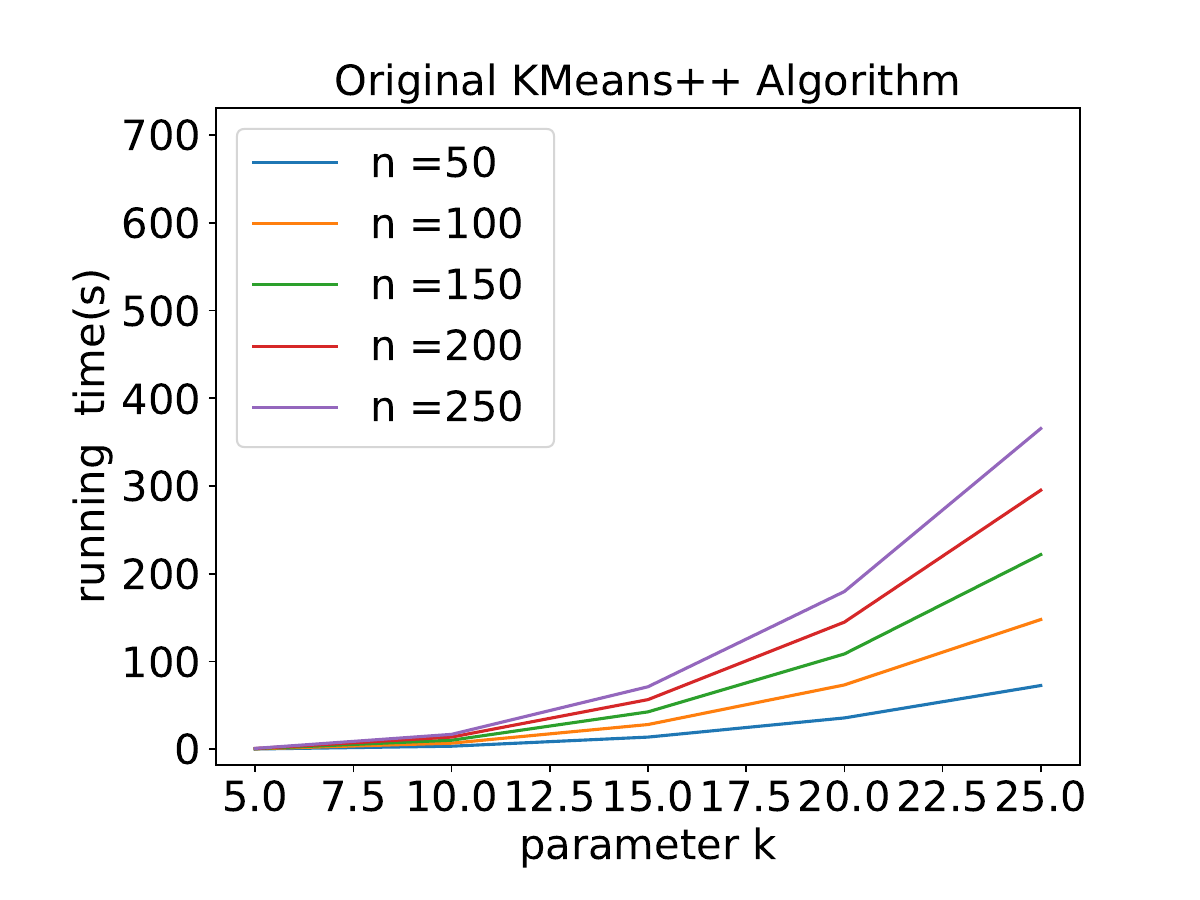}
   \label{fig:x_param_k_variable_n_original_total_time}
   } 
   \subfloat[$m$ is changed]{
    \includegraphics[width=0.3\textwidth]{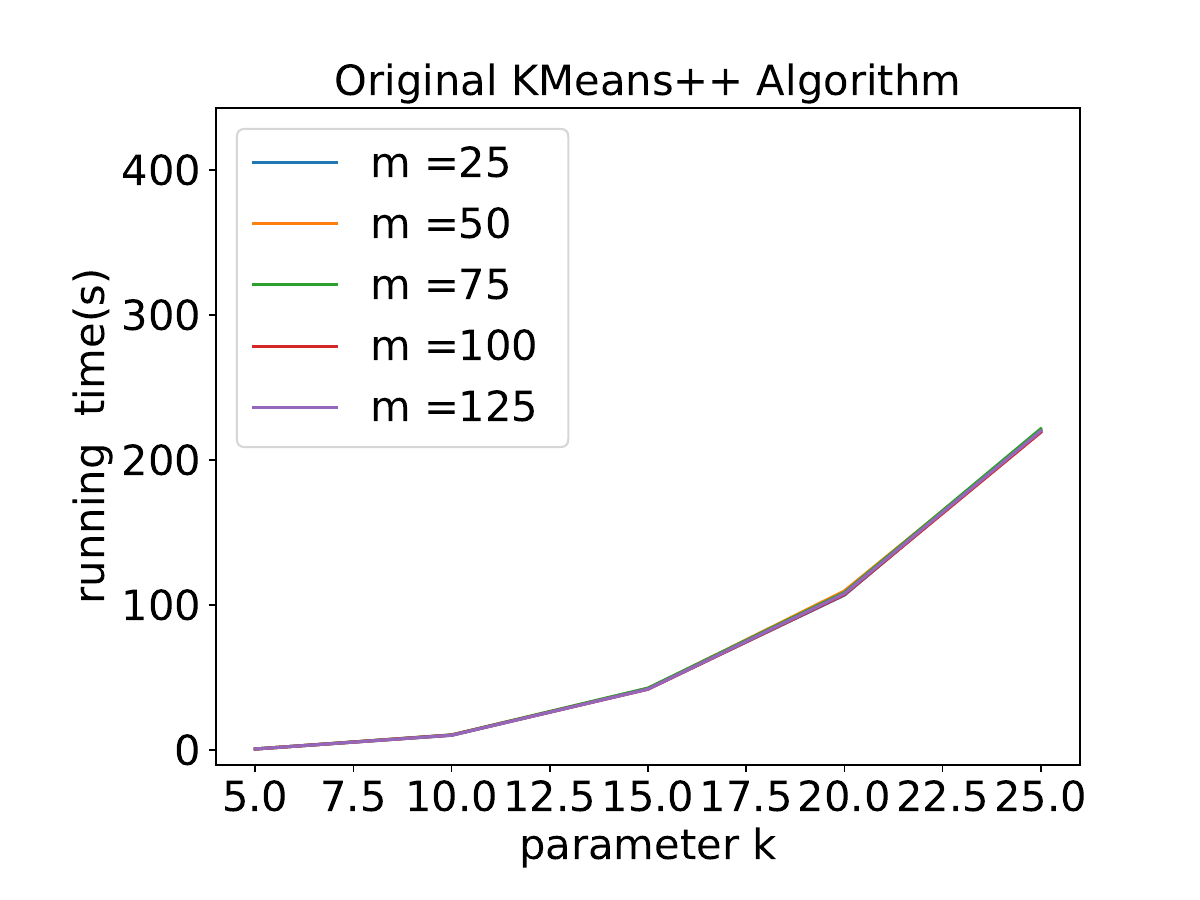}
   \label{fig:x_param_k_variable_m_original_total_time}
}
    \subfloat[$d$ is changed]{   
    \includegraphics[width=0.3\textwidth]{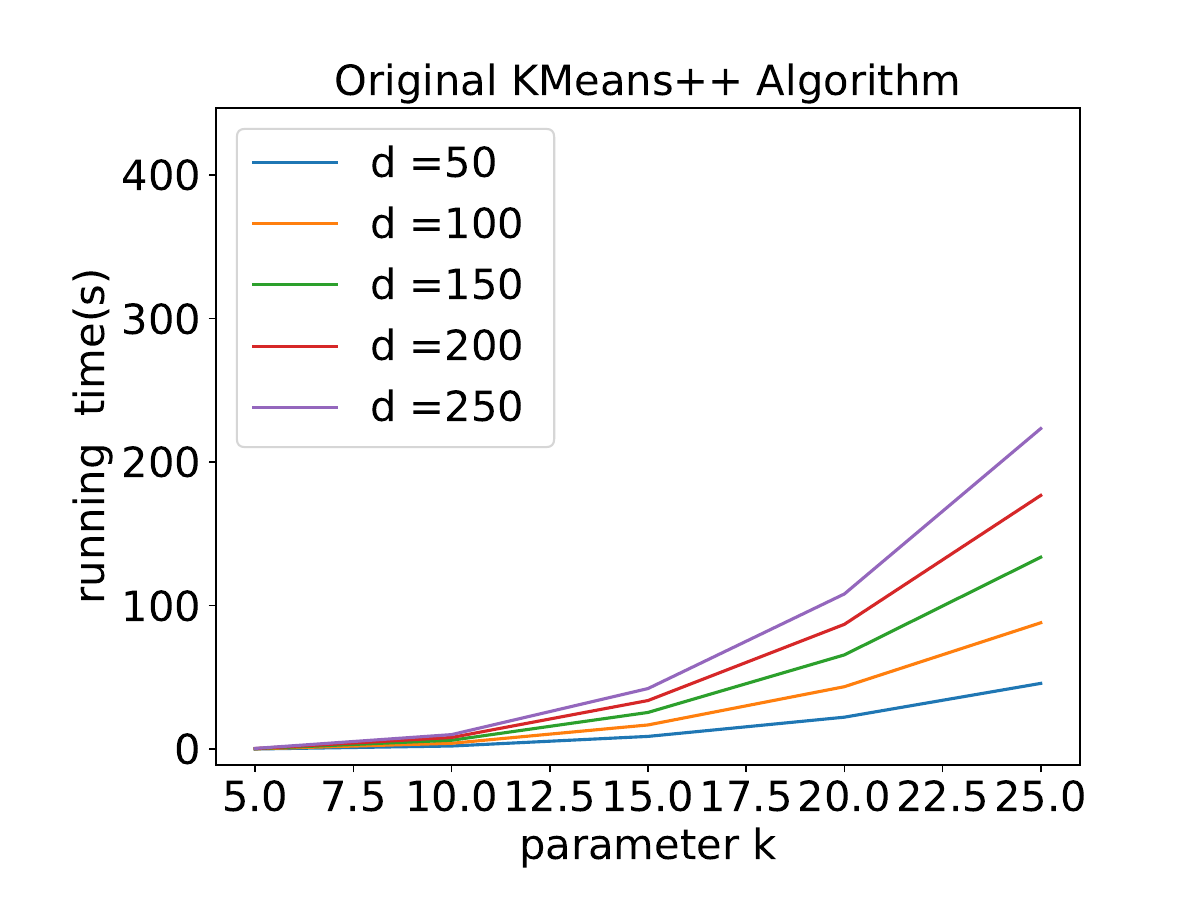}
   \label{fig:x_param_k_variable_d_original_total_time}
}
    
    \caption{The relationship between running time of original $k$-means++ algorithm and parameter $k$}
    \label{fig:the_relationship_between_running_time_of_original_algorithm_and_parameter_k}
\end{figure*}

\begin{figure*}[!ht]
    \centering

    \subfloat[$n$ is changed]{ \includegraphics[width=0.3\textwidth]{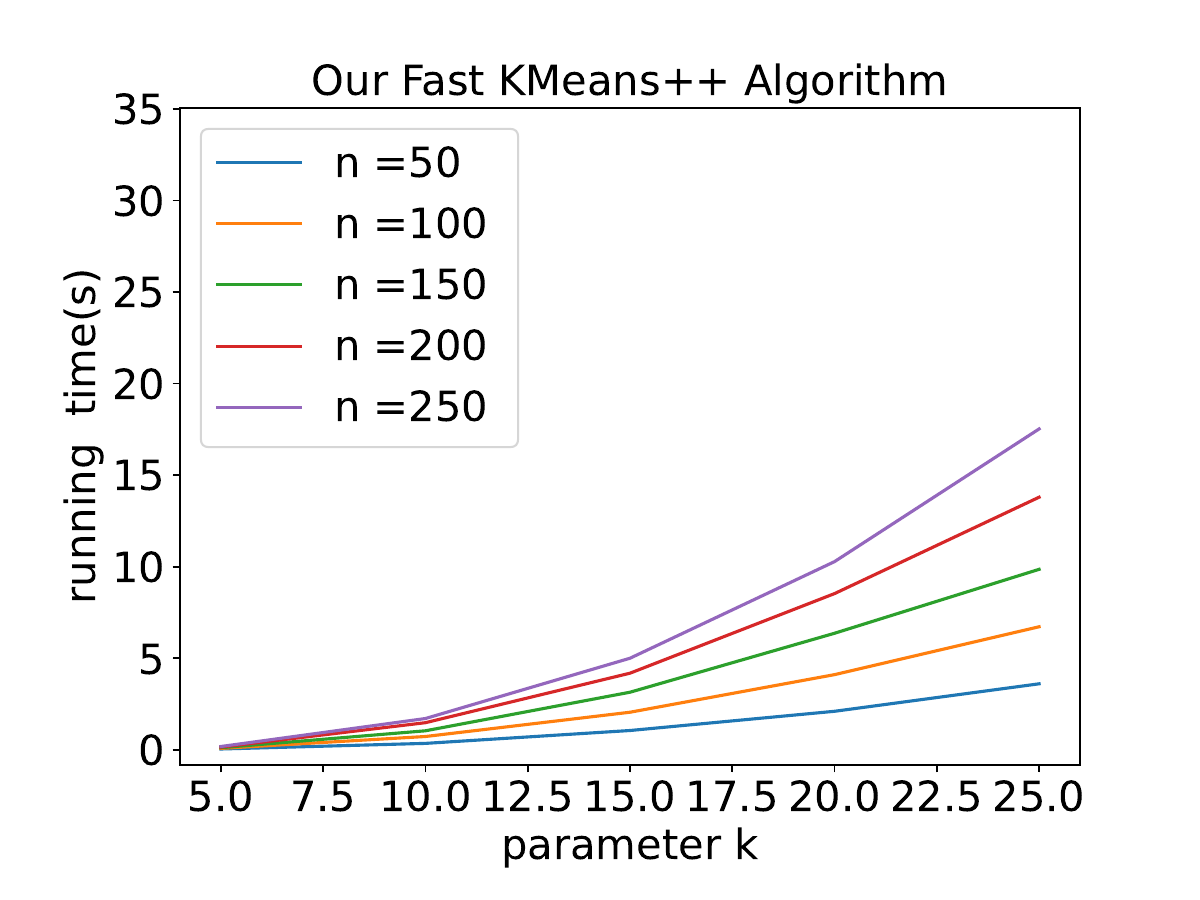}
   \label{fig:x_param_k_variable_n_our_total_time}
   } 
   \subfloat[$m$ is changed]{
    \includegraphics[width=0.3\textwidth]{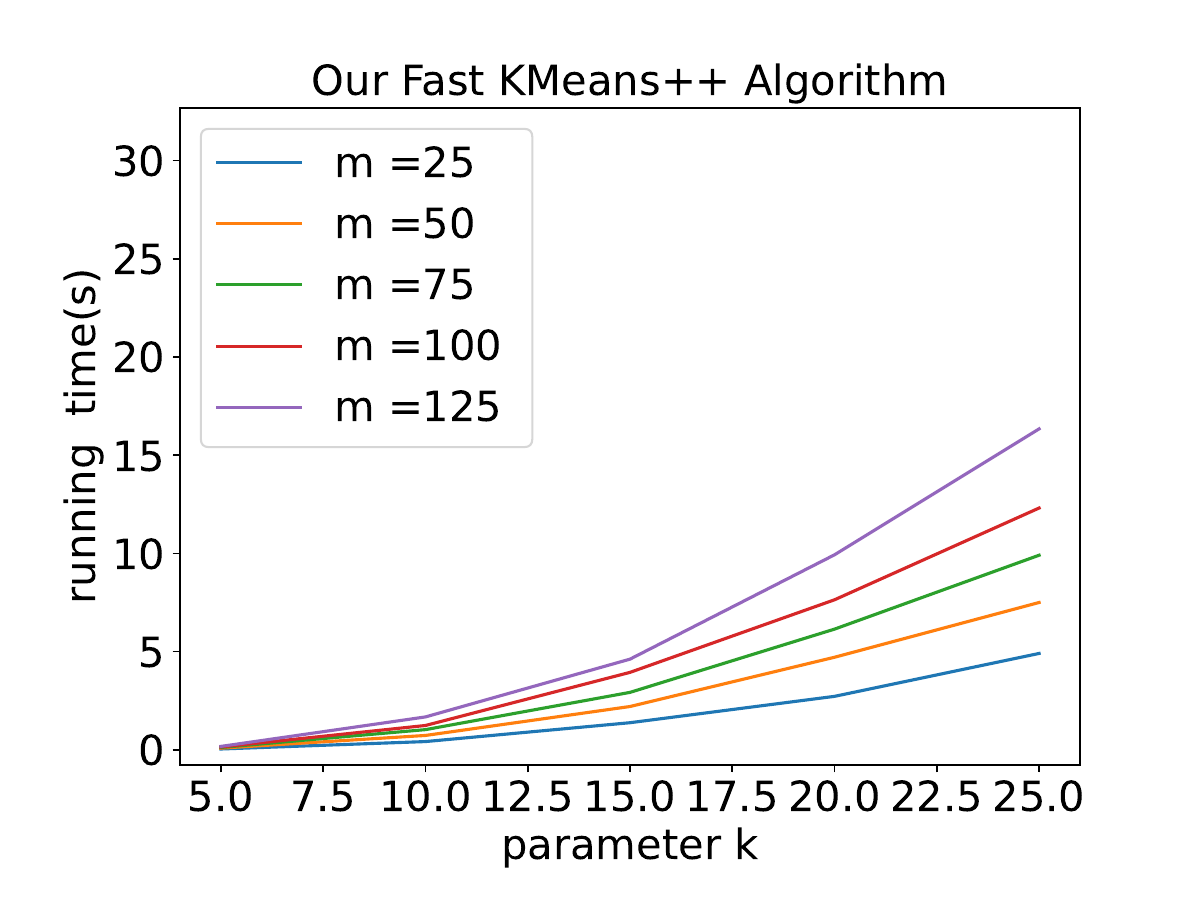}
   \label{fig:x_param_k_variable_m_our_total_time}
}
    \subfloat[$d$ is changed]{   
    \includegraphics[width=0.3\textwidth]{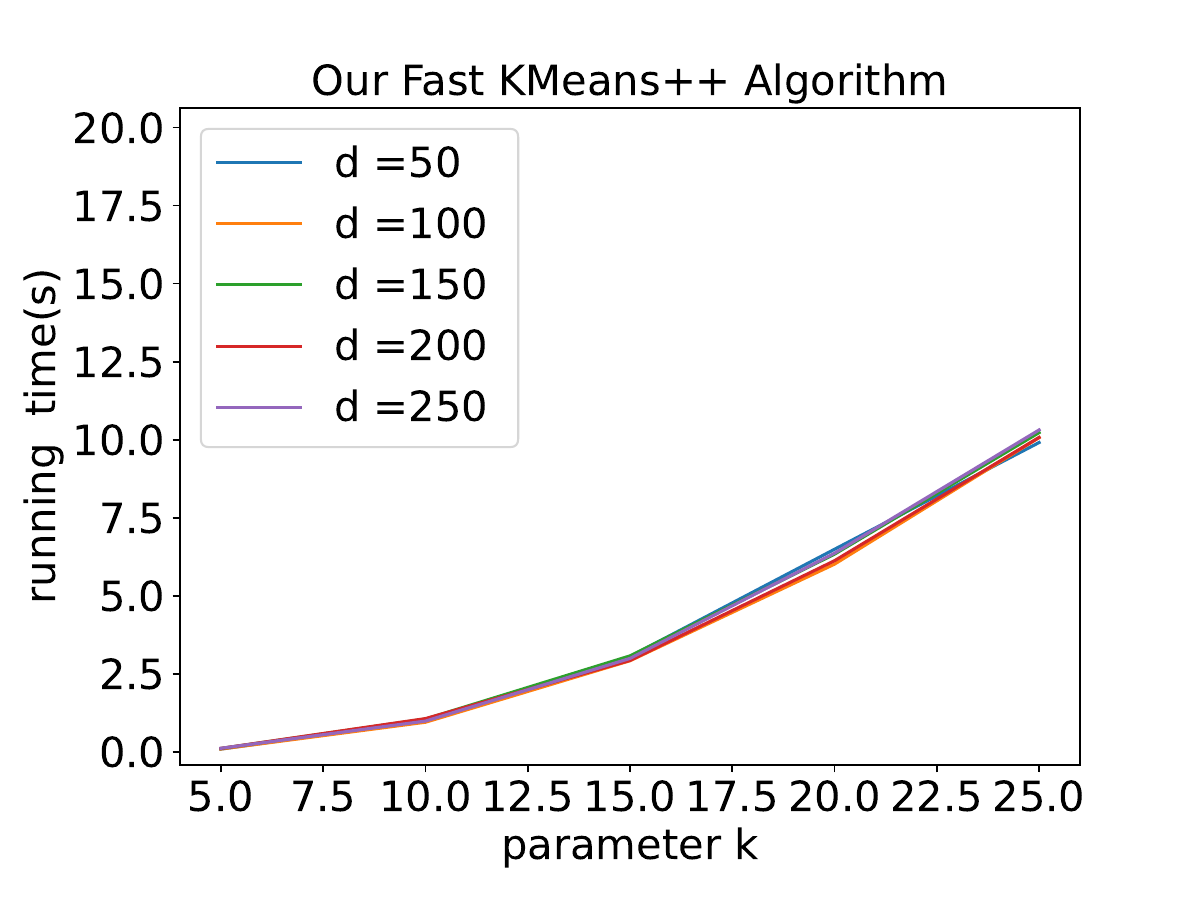}
   \label{fig:x_param_k_variable_d_our_total_time}
}
    
    \caption{The relationship between running time of our \textsc{FastKMeans++} algorithm and parameter $k$}
    \label{fig:the_relationship_between_running_time_of_our_algorithm_and_parameter_k}
\end{figure*}

    To evaluate if the influence of parameter $k$ will be affected by other parameters, we set different values for other parameters and run the original $k$-means++ algorithm and our \textsc{FastKMeans++} algorithm. Fig.~\ref{fig:the_relationship_between_running_time_of_original_algorithm_and_parameter_k} and 
    Fig.~\ref{fig:the_relationship_between_running_time_of_our_algorithm_and_parameter_k} show that the running time of original $k$-means++ and \textsc{FastKMeans++} increase squarely as $k$ increases no matter what the value of other parameters is.

    \textbf{When should we use \textsc{FastKMeans++} algorithm?} The result above demonstrates that our \textsc{FastKMeans++} algorithm much faster than original $k$-means++ algorithm. It also shows that if we set $m$ smaller, the running time of our algorithms will decrease too. Our \textsc{FastKMeans++} algorithm behaves better compared with original $k$-means++ algorithm especially when $d$ is very large. Therefore, our \textsc{FastKMeans++} algorithm can handle the $k$-means problem with high dimensional points. 

%% file: main.bbl
\newcommand{\etalchar}[1]{$^{#1}$}
\begin{thebibliography}{CALNF{\etalchar{+}}20}

\bibitem[ADK09]{adk09}
Ankit Aggarwal, Amit Deshpande, and Ravi Kannan.
\newblock Adaptive sampling for k-means clustering.
\newblock In {\em Approximation, Randomization, and Combinatorial Optimization.
  Algorithms and Techniques}, pages 15--28. Springer, 2009.

\bibitem[ALS{\etalchar{+}}22]{als+22}
Josh Alman, Jiehao Liang, Zhao Song, Ruizhe Zhang, and Danyang Zhuo.
\newblock Bypass exponential time preprocessing: Fast neural network training
  via weight-data correlation preprocessing.
\newblock {\em arXiv preprint arXiv:2211.14227}, 2022.

\bibitem[ANFSW19]{ajos19}
Sara Ahmadian, Ashkan Norouzi-Fard, Ola Svensson, and Justin Ward.
\newblock Better guarantees for k-means and euclidean k-median by primal-dual
  algorithms.
\newblock {\em SIAM Journal on Computing}, 49(4):FOCS17--97, 2019.

\bibitem[AV06]{av06}
David Arthur and Sergei Vassilvitskii.
\newblock k-means++: The advantages of careful seeding.
\newblock Technical report, Stanford, 2006.

\bibitem[BPSW21]{bpsw21}
Jan van~den Brand, Binghui Peng, Zhao Song, and Omri Weinstein.
\newblock Training (overparametrized) neural networks in near-linear time.
\newblock In {\em ITCS}, 2021.

\bibitem[BSY23]{bsy23}
Song Bian, Zhao Song, and Junze Yin.
\newblock Federated empirical risk minimization via second-order method.
\newblock {\em arXiv preprint arXiv:2305.17482}, 2023.

\bibitem[BV15]{bv15}
Sayan Bandyapadhyay and Kasturi Varadarajan.
\newblock On variants of k-means clustering.
\newblock {\em arXiv preprint arXiv:1512.02985}, 2015.

\bibitem[BZ19]{bz19}
SMM~Fatemi Bushehri and Mohsen~Sardari Zarchi.
\newblock An expert model for self-care problems classification using
  probabilistic neural network and feature selection approach.
\newblock {\em Applied Soft Computing}, 82:105545, 2019.

\bibitem[CA18]{ca18}
Vincent Cohen-Addad.
\newblock A fast approximation scheme for low-dimensional k-means.
\newblock In {\em Proceedings of the Twenty-Ninth Annual ACM-SIAM Symposium on
  Discrete Algorithms}, pages 430--440. SIAM, 2018.

\bibitem[CAKM19]{cakm19}
Vincent Cohen-Addad, Philip~N Klein, and Claire Mathieu.
\newblock Local search yields approximation schemes for k-means and k-median in
  euclidean and minor-free metrics.
\newblock {\em SIAM Journal on Computing}, 48(2):644--667, 2019.

\bibitem[CALNF{\etalchar{+}}20]{cln+20}
Vincent Cohen-Addad, Silvio Lattanzi, Ashkan Norouzi-Fard, Christian Sohler,
  and Ola Svensson.
\newblock Fast and accurate $ k $-means++ via rejection sampling.
\newblock {\em Advances in Neural Information Processing Systems},
  33:16235--16245, 2020.

\bibitem[CCLY19]{ccly19}
Michael~B Cohen, Ben Cousins, Yin~Tat Lee, and Xin Yang.
\newblock A near-optimal algorithm for approximating the john ellipsoid.
\newblock In {\em Conference on Learning Theory}, pages 849--873. PMLR, 2019.

\bibitem[Che09]{che09}
Ke~Chen.
\newblock On coresets for k-median and k-means clustering in metric and
  euclidean spaces and their applications.
\newblock {\em SIAM Journal on Computing}, 39(3):923--947, 2009.

\bibitem[CLg{\etalchar{+}}16]{clg+16}
Ting-Yu Cheng, Guiguan Lin, xinyang gong, Kang-Jun Liu, and Shan-Hung~(Brandon)
  Wu.
\newblock Learning user perceived clusters with feature-level supervision.
\newblock In D.~Lee, M.~Sugiyama, U.~Luxburg, I.~Guyon, and R.~Garnett,
  editors, {\em Advances in Neural Information Processing Systems}, volume~29.
  Curran Associates, Inc., 2016.

\bibitem[CLS19]{cls19}
Michael~B Cohen, Yin~Tat Lee, and Zhao Song.
\newblock Solving linear programs in the current matrix multiplication time.
\newblock In {\em STOC}, 2019.

\bibitem[CN12]{cn12}
Adam Coates and Andrew~Y Ng.
\newblock Learning feature representations with k-means.
\newblock In {\em Neural networks: Tricks of the trade}, pages 561--580.
  Springer, 2012.

\bibitem[CW13]{cw13}
Kenneth~L Clarkson and David~P Woodruff.
\newblock Low-rank approximation and regression in input sparsity time.
\newblock In {\em Journal of the ACM (JACM), A Preliminary version of this
  paper is appeared at STOC}, 2013.

\bibitem[DG17]{dg19}
Dheeru Dua and Casey Graff.
\newblock {UCI} machine learning repository, 2017.

\bibitem[DH04]{dh04}
Chris Ding and Xiaofeng He.
\newblock K-means clustering via principal component analysis.
\newblock In {\em Proceedings of the twenty-first international conference on
  Machine learning}, page~29, 2004.

\bibitem[DLS23]{dls23}
Yichuan Deng, Zhihang Li, and Zhao Song.
\newblock An improved sample complexity for rank-1 matrix sensing.
\newblock {\em arXiv preprint arXiv:2303.06895}, 2023.

\bibitem[DMC15]{dmc15}
Nameirakpam Dhanachandra, Khumanthem Manglem, and Yambem~Jina Chanu.
\newblock Image segmentation using k-means clustering algorithm and subtractive
  clustering algorithm.
\newblock {\em Procedia Computer Science}, 54:764--771, 2015.

\bibitem[DMR{\etalchar{+}}09]{dmr+09}
Daniel~B Dias, Renata~CB Madeo, Thiago Rocha, Helton~H Biscaro, and Sarajane~M
  Peres.
\newblock Hand movement recognition for brazilian sign language: a study using
  distance-based neural networks.
\newblock In {\em 2009 international joint conference on neural networks},
  pages 697--704. IEEE, 2009.

\bibitem[DSY23]{dsy23}
Yichuan Deng, Zhao Song, and Junze Yin.
\newblock Faster robust tensor power method for arbitrary order.
\newblock {\em arXiv preprint arXiv:2306.00406}, 2023.

\bibitem[FMS07]{fms07}
Dan Feldman, Morteza Monemizadeh, and Christian Sohler.
\newblock A ptas for k-means clustering based on weak coresets.
\newblock In {\em Proceedings of the twenty-third annual symposium on
  Computational geometry}, pages 11--18, 2007.

\bibitem[FRS19]{frs19}
Zachary Friggstad, Mohsen Rezapour, and Mohammad~R Salavatipour.
\newblock Local search yields a ptas for k-means in doubling metrics.
\newblock {\em SIAM Journal on Computing}, 48(2):452--480, 2019.

\bibitem[FS08]{fs08}
Gereon Frahling and Christian Sohler.
\newblock A fast k-means implementation using coresets.
\newblock {\em International Journal of Computational Geometry \&
  Applications}, 18(06):605--625, 2008.

\bibitem[FSS20]{cdm20}
Dan Feldman, Melanie Schmidt, and Christian Sohler.
\newblock Turning big data into tiny data: Constant-size coresets for k-means,
  pca, and projective clustering.
\newblock {\em SIAM Journal on Computing}, 49(3):601--657, 2020.

\bibitem[GG12]{gg12}
Allen Gersho and Robert~M Gray.
\newblock {\em Vector quantization and signal compression}, volume 159.
\newblock Springer Science \& Business Media, 2012.

\bibitem[GS22]{gs22}
Yuzhou Gu and Zhao Song.
\newblock A faster small treewidth sdp solver.
\newblock {\em arXiv preprint arXiv:2211.06033}, 2022.

\bibitem[GSWY23]{gswy23}
Yeqi Gao, Zhao Song, Weixin Wang, and Junze Yin.
\newblock A fast optimization view: Reformulating single layer attention in llm
  based on tensor and svm trick, and solving it in matrix multiplication time.
\newblock {\em arXiv preprint arXiv:2309.07418}, 2023.

\bibitem[GSY23]{gsy23_hyper}
Yeqi Gao, Zhao Song, and Junze Yin.
\newblock An iterative algorithm for rescaled hyperbolic functions regression.
\newblock {\em arXiv preprint arXiv:2305.00660}, 2023.

\bibitem[GSYZ23]{gsyz23}
Yuzhou Gu, Zhao Song, Junze Yin, and Lichen Zhang.
\newblock Low rank matrix completion via robust alternating minimization in
  nearly linear time.
\newblock {\em arXiv preprint arXiv:2302.11068}, 2023.

\bibitem[HPK05]{as05}
Sariel Har-Peled and Akash Kushal.
\newblock Smaller coresets for k-median and k-means clustering.
\newblock In {\em Proceedings of the twenty-first annual symposium on
  Computational geometry}, pages 126--134, 2005.

\bibitem[HPM04]{hpm04}
Sariel Har-Peled and Soham Mazumdar.
\newblock On coresets for k-means and k-median clustering.
\newblock In {\em Proceedings of the thirty-sixth annual ACM symposium on
  Theory of computing}, pages 291--300, 2004.

\bibitem[JL84]{jl84}
William~B Johnson and Joram Lindenstrauss.
\newblock Extensions of lipschitz mappings into a hilbert space.
\newblock {\em Contemporary mathematics}, 26(189-206):1, 1984.

\bibitem[JLSW20]{jlsw20}
Haotian Jiang, Yin~Tat Lee, Zhao Song, and Sam Chiu-wai Wong.
\newblock An improved cutting plane method for convex optimization,
  convex-concave games and its applications.
\newblock In {\em STOC}, 2020.

\bibitem[JLSZ23]{jlsz23}
Haotian Jiang, Yin~Tat Lee, Zhao Song, and Lichen Zhang.
\newblock Convex minimization with integer minima in $\widetilde{O}(n^4)$ time.
\newblock {\em arXiv preprint arXiv:2304.03426}, 2023.

\bibitem[JSWZ21]{jswz21}
Shunhua Jiang, Zhao Song, Omri Weinstein, and Hengjie Zhang.
\newblock Faster dynamic matrix inverse for faster lps.
\newblock In {\em STOC}, 2021.

\bibitem[JV01]{jv01}
Kamal Jain and Vijay~V Vazirani.
\newblock Approximation algorithms for metric facility location and k-median
  problems using the primal-dual schema and lagrangian relaxation.
\newblock {\em Journal of the ACM (JACM)}, 48(2):274--296, 2001.

\bibitem[KMN{\etalchar{+}}04]{kmn+04}
Tapas Kanungo, David~M Mount, Nathan~S Netanyahu, Christine~D Piatko, Ruth
  Silverman, and Angela~Y Wu.
\newblock A local search approximation algorithm for k-means clustering.
\newblock {\em Computational Geometry}, 28(2-3):89--112, 2004.

\bibitem[KSS10]{kss10}
Amit Kumar, Yogish Sabharwal, and Sandeep Sen.
\newblock Linear-time approximation schemes for clustering problems in any
  dimensions.
\newblock {\em Journal of the ACM (JACM)}, 57(2):1--32, 2010.

\bibitem[Llo82]{llo82}
Stuart Lloyd.
\newblock Least squares quantization in pcm.
\newblock {\em IEEE transactions on information theory}, 28(2):129--137, 1982.

\bibitem[LS19]{ls19}
Silvio Lattanzi and Christian Sohler.
\newblock A better k-means++ algorithm via local search.
\newblock In {\em International Conference on Machine Learning}, pages
  3662--3671. PMLR, 2019.

\bibitem[LSW17]{lsw17}
Euiwoong Lee, Melanie Schmidt, and John Wright.
\newblock Improved and simplified inapproximability for k-means.
\newblock {\em Information Processing Letters}, 120:40--43, 2017.

\bibitem[LSZ19]{lsz19}
Yin~Tat Lee, Zhao Song, and Qiuyi Zhang.
\newblock Solving empirical risk minimization in the current matrix
  multiplication time.
\newblock In {\em Conference on Learning Theory (COLT)}, pages 2140--2157.
  PMLR, 2019.

\bibitem[NN13]{nn13}
Jelani Nelson and Huy~L Nguy{\^e}n.
\newblock Osnap: Faster numerical linear algebra algorithms via sparser
  subspace embeddings.
\newblock In {\em 2013 ieee 54th annual symposium on foundations of computer
  science}, pages 117--126. IEEE, 2013.

\bibitem[QSZZ23]{qszz23}
Lianke Qin, Zhao Song, Lichen Zhang, and Danyang Zhuo.
\newblock An online and unified algorithm for projection matrix vector
  multiplication with application to empirical risk minimization.
\newblock In {\em AISTATS}, 2023.

\bibitem[SY21]{sy21}
Zhao Song and Zheng Yu.
\newblock Oblivious sketching-based central path method for linear programming.
\newblock In {\em International Conference on Machine Learning}, pages
  9835--9847. PMLR, 2021.

\bibitem[SYYZ22]{syyz22}
Zhao Song, Xin Yang, Yuanyuan Yang, and Tianyi Zhou.
\newblock Faster algorithm for structured john ellipsoid computation.
\newblock {\em arXiv preprint arXiv:2211.14407}, 2022.

\bibitem[SYYZ23a]{syyz23_weighted}
Zhao Song, Mingquan Ye, Junze Yin, and Lichen Zhang.
\newblock Efficient alternating minimization with applications to weighted low
  rank approximation.
\newblock {\em arXiv preprint arXiv:2306.04169}, 2023.

\bibitem[SYYZ23b]{syyz23_ellinf}
Zhao Song, Mingquan Ye, Junze Yin, and Lichen Zhang.
\newblock A nearly-optimal bound for fast regression with $\ell_\infty$
  guarantee.
\newblock In {\em International Conference on Machine Learning}, pages
  32463--32482. PMLR, 2023.

\bibitem[SYZ23a]{syz23_atten}
Zhao Song, Junze Yin, and Lichen Zhang.
\newblock Solving attention kernel regression problem via pre-conditioner.
\newblock {\em arXiv preprint arXiv:2308.14304}, 2023.

\bibitem[SYZ23b]{syz23}
Zhao Song, Junze Yin, and Ruizhe Zhang.
\newblock Revisiting quantum algorithms for linear regressions: Quadratic
  speedups without data-dependent parameters.
\newblock {\em arXiv preprint arXiv:2311.14823}, 2023.

\bibitem[SZZ21]{szz21}
Zhao Song, Lichen Zhang, and Ruizhe Zhang.
\newblock Training multi-layer over-parametrized neural network in subquadratic
  time.
\newblock {\em arXiv preprint arXiv:2112.07628}, 2021.

\bibitem[TSL14]{ts14}
Swee~Chuan Tan and Jess~Pei San~Lau.
\newblock Time series clustering: A superior alternative for market basket
  analysis.
\newblock In {\em Proceedings of the First International Conference on Advanced
  Data and Information Engineering (DaEng-2013)}, pages 241--248. Springer,
  Singapore, 2014.

\bibitem[Wei16]{wei16}
Dennis Wei.
\newblock A constant-factor bi-criteria approximation guarantee for k-means++.
\newblock {\em Advances in Neural Information Processing Systems}, 29, 2016.

\bibitem[ZBD18]{zbd18}
MS~Zarchi, SMM~Fatemi Bushehri, and M~Dehghanizadeh.
\newblock Scadi: A standard dataset for self-care problems classification of
  children with physical and motor disability.
\newblock {\em International Journal of Medical Informatics}, 2018.

\bibitem[Zha22]{z22}
Lichen Zhang.
\newblock Speeding up optimizations via data structures: Faster search, sample
  and maintenance.
\newblock Master's thesis, Carnegie Mellon University, 2022.

\end{thebibliography}
